\documentclass[5p,authoryear]{elsarticle}

\usepackage{amssymb}
\usepackage{amsthm}
\usepackage{amsmath}
\usepackage{aas_macros}
\usepackage{rotating}
\usepackage{url}

\usepackage{lineno}

\journal{Icarus}

\newcommand{\ud}{\mathrm{d}}
\newcommand{\vect}[1]{\boldsymbol{\mathrm{#1}}}

\renewcommand{\deg}{^\circ}

\newcommand{\au}{\,\mathrm{au}}
\newcommand{\km}{\,\mathrm{km}}
\newcommand{\meter}{\,\mathrm{m}}

\newcommand{\Myr}{\,\mathrm{Myr}}
\newcommand{\yr}{\,\mathrm{yr}}

\newcommand{\second}{\,\mathrm{s}}
\newcommand{\magnitude}{\,\mathrm{mag}}

\newcommand{\kg}{\,\mathrm{kg}}

\newcommand{\logl}{\log L}

\begin{document}

\begin{frontmatter}

\title{Debiased orbit and absolute-magnitude distributions for near-Earth objects}

\author[label1]{Mikael Granvik\corref{cor1}}
\address[label1]{Department of Physics, P.O. Box 64, 00014 University of Helsinki, Finland}
\cortext[cor1]{Corresponding author (email: mgranvik@iki.fi)}

\author[label2]{Alessandro Morbidelli}
\address[label2]{Observatoire de la Cote d'Azur, Boulevard de l'Observatoire, F 06304 Nice Cedex 4, France}

\author[label3]{Robert Jedicke}
\address[label3]{Institute for Astronomy, 2680 Woodlawn Dr, HI 96822, USA}

\author[label2,label3,label4,label5,fn1]{Bryce Bolin}
\address[label4]{B612 Asteroid Institute, 20 Sunnyside Ave, Suite 427, Mill Valley, CA 94941, USA}
\address[label5]{Department of Astronomy, University of Washington, 3910 15th Ave NE, Seattle, WA 98195, USA}
\fntext[fn1]{B612 Asteroid Institute and DIRAC Institute Postdoctoral Fellow}

\author[label6]{William F. Bottke}
\address[label6]{Southwest Research Institute, 1050 Walnut Street, Suite 300, Boulder, CO 80302, USA}

\author[label7]{Edward Beshore}
\address[label7]{University of Arizona, 933 North Cherry Avenue, Tucson, Arizona 85721-0065, USA}

\author[label8]{David Vokrouhlick\'y}
\address[label8]{Institute of Astronomy, Charles University, V Hole\v{s}ovi\v{c}k\'ach 2, CZ 18000 Prague 8, Czech Republic}

\author[label6]{David Nesvorn\'y}

\author[label2]{Patrick Michel}

\begin{abstract}
The debiased absolute-magnitude and orbit distributions as well as
source regions for near-Earth objects (NEOs) provide a fundamental
frame of reference for studies of individual NEOs and more complex
population-level questions. We present a new four-dimensional model of
the NEO population that describes debiased steady-state distributions
of semimajor axis, eccentricity, inclination, and absolute magnitude
$H$ in the range $17<H<25$. The modeling approach improves upon the
methodology originally developed by \citet[][Science 288,
  2190--2194]{bot2000a} in that it is, for example, based on more
realistic orbit distributions and uses source-specific
absolute-magnitude distributions that allow for a power-law slope that
varies with $H$.  We divide the main asteroid belt into six different
entrance routes or regions (ER) to the NEO region: the $\nu_6$, 3:1J,
5:2J and 2:1J resonance complexes as well as Hungarias and
Phocaeas. In addition we include the Jupiter-family comets as the
primary cometary source of NEOs. We calibrate the model against NEO
detections by Catalina Sky Surveys' stations 703 and G96 during
2005--2012, and utilize the complementary nature of these two systems
to quantify the systematic uncertainties associated to the resulting
model. We find that the (fitted) $H$ distributions have significant
differences, although most of them show a minimum power-law slope at
$H\sim20$. As a consequence of the differences between the ER-specific
$H$ distributions we find significant variations in, for example, the
NEO orbit distribution, average lifetime, and the relative
contribution of different ERs as a function of $H$. The most important
ERs are the $\nu_6$ and 3:1J resonance complexes with JFCs
contributing a few percent of NEOs on average. A significant
contribution from the Hungaria group leads to notable changes compared
to the predictions by Bottke et al.\ in, for example, the orbit
distribution and average lifetime of NEOs. We predict that there are
$962^{+52}_{-56}$ ($802^{+48}_{-42}\times10^3$) NEOs with $H<17.75$
($H<25$) and these numbers are in agreement with the most recent
estimates found in the literature (the uncertainty estimates only
account for the random component). Based on our model we find that
relative shares between different NEO groups (Amor, Apollo, Aten,
Atira, Vatira) are (39.4,54.4,3.5,1.2,0.3)\%, respectively, for the
considered $H$ range and that these ratios have a negligible
dependence on $H$.  Finally, we find an agreement between our estimate
for the rate of Earth impacts by NEOs and recent estimates in the
literature, but there remains a potentially significant discrepancy in
the frequency of Tunguska-sized and Chelyabinsk-sized impacts.
\end{abstract}


\end{frontmatter}


\section{Introduction}
\label{sec:intro}

Understanding the orbital and size distributions as well as the source
regions for near-Earth objects (NEOs; for a glossary of acronyms and
terms, see Table \ref{table:acronyms}) is one of the key topics in
contemporary planetary science
\citep{2015aste.book..243B,2015aste.book..835H,2015aste.book..855A}. Here
we present a new model describing the debiased absolute-magnitude
($H$) and orbital (semimajor axis $a$, eccentricity $e$, inclination
$i$) distributions for NEOs. The model also enables a probabilistic
assessment of source regions for individual NEOs.

\begin{table}
    \centering
    \caption{Glossary of acronyms and terms.}
    \label{table:acronyms}
    \begin{tabular}{ll}
    \hline
    Acronym/ & Definition \\ 
    term & \\ 
    \hline
    703 & Catalina Sky Survey (telescope) \\
    AICc & corrected Akaike Information Criteria \\
    CSS & Catalina Sky Survey (703 and G96) \\
    ER & escape/entrance route/region \\
    G96 & Mt.~Lemmon Survey (part of CSS) \\
    HFD & $H$-frequency distribution \\
    IMC & intermediate Mars-crosser \\
    JFC & Jupiter-family comet \\
    MAB & main asteroid belt \\
    MBO & main-belt object \\
    ML & maximum likelihood \\
    MMR & mean-motion resonance \\
    MPC & Minor Planet Center \\
    MOID & minimum orbital intersection distance \\
    NEO & near-Earth object (here asteroid or \\
     & comet with $q<1.3\au$ and $a<4.2\au$) \\
    PHO & NEO with MOID$<0.05\au$ and $H<22$ \\
    RMS & root-mean-square \\
    SR & secular resonance \\
    YORP & Yarkovsky-O'Keefe-Radzievskii-Paddack \\
     & effect \\
    Amor & NEO with $1.017\au<q<1.3\au$ \\
    Apollo & NEO with $a>1.0\au$ and $q<1.017\au$ \\
    Aten & NEO with $a<1.0\au$ and $Q>0.983\au$ \\
    Atira & NEO with $0.718\au<Q<0.983\au$ \\
    Vatira & NEO with $0.307\au<Q<0.718\au$ \\
    $a$ & semimajor axis \\
    $e$ & eccentricity \\
    $i$ & inclination \\
    $\Omega$ & longitude of ascending node \\
    $\omega$ & argument of perihelion \\
    $M_0$ & mean anomaly \\
    $H$ & absolute magnitude in V band \\
    $D$ & diameter \\
    $q$ & perihelion distance \\
    $Q$ & aphelion distance \\
    \hline
    \end{tabular}
\end{table}

We follow the conventional notation and define an NEO as an asteroid
or comet (active, dormant or extinct) with perihelion distance
$q<1.3\au$ and semimajor axis $a<4.2\au$. The latter requirement is
not part of the official definition, which has no limit on $a$, but it
limits NEOs to the inner solar system and makes comparisons to the
exisiting literature easier \citep[cf.][]{bot2002a}. The population of
transneptunian objects may contain a substantial number of objects
with $q<1.3\au$ that are thus not considered in this work. NEOs are
further divided into the Amors ($1.017\au<q<1.3\au$), Apollos
($a>1.0\au$ and $q<1.017\au$), Atens ($a<1.0\au$ and aphelion distance
$Q>0.983\au$), Atiras ($0.718\au<Q<0.983\au$) that are detached from
the Earth, and the so-called Vatiras ($0.307\au<Q<0.718\au$) that are
detached from Venus \citep{gre2012a}. NEOs are also classified as
potentially hazardous objects (PHOs) when their minimum orbital
intersection distance (MOID) with respect to the Earth is less than
$0.05\au$ and $H<22$.

Several papers have reported estimates for the debiased orbit
distribution and/or the $H$-frequency distribution (HFD) for NEOs over
the past 25 years or so. The basic equation that underlies most of the
studies describes the relationship between the known NEO population
$n$, the discovery efficiency $\epsilon$, and the true population $N$
as functions of $a$, $e$, $i$, and $H$:
\begin{equation}
  n(a,e,i,H) = \epsilon(a,e,i,H)\,N(a,e,i,H)\,. \label{eq:baseff}
\end{equation}

\citet{rab1993a} derived the first debiased orbit distribution and HFD
for NEOs. The model was calibrated by using only 23 asteroids
discovered by the Spacewatch Telescope between September 1990 and
December 1991. The model is valid in the diameter range $10\meter
\lesssim D \lesssim10\km$, and the estimated rate of Earth impacts by
NEOs is about 100 times larger than the current best estimates at
diameters $D\sim10\meter$
\citep{2013Natur.503..238B}. \citet{rab1993a} concluded that the small
NEOs have a different HFD slope compared to NEOs with
$D\gtrsim100\meter$ and suggested that future studies should ``assess
the effect of a size-dependent orbit distribution on the derived size
distribution.'' \citet{rab2000a} used methods similar to those
employed by \citet{rab1993a} to estimate the HFD based on 45 NEOs
detected by the Jet Propulsion Laboratory's Near-Earth-Asteroid
Tracking (NEAT) program. They concluded that there should be
$700\pm230$ NEOs with $H<18$.

Although the work by \citet{rab1993a} showed that it is possible to
derive a reasonable estimate for the true population, the relatively
small number of known NEOs ($\sim10^2$--$10^4$) implies that the
maximum resolution of the resulting four-dimensional model is poor and
a scientifically useful resolution is limited to marginalized
distributions in one dimension. Therefore additional constraints had
to be found to derive more useful four-dimensional models of the true
population. \citet{bot2000a} devised a methodology which utilizes the
fact that objects originating in different parts of the main asteroid
belt (MAB) or the cometary region will have statistically distinct
orbital histories in the NEO region. Assuming that there is no
correlation between $H$ and ($a,e,i$), \citet{bot2000a} decomposed the
true population $N(a,e,i,H)$ into $N(H)\,\sum R_s(a,e,i)$, where
$R_s(a,e,i)$ denotes the steady-state orbit distribution for NEOs
entering the NEO region through entrance route $s$. The primary
dynamical mechanisms responsible for delivering objects from the MAB
and cometary region into the NEO region were already understood at
that time, and models for $R_s(a,e,i)$ could therefore be obtained
through direct orbital integration of test particles placed in, or in
the vicinity of, escape routes from the MAB. The parameters left to be
fitted described the relative importance of the steady-state orbit
distributions and the overall NEO HFD. Fitting a model with three
escape routes from the MAB, that is, the $\nu_6$ secular resonance
(SR), the intermediate Mars crossers (IMC), and the 3:1J mean-motion
resonance (MMR) with Jupiter, to 138 NEOs detected by the Spacewatch
survey, \citet{bot2000a} estimated that there are $910^{+100}_{-120}$
NEOs with $H<18$. Their estimates for the contributions from the
different escape routes had large uncertainties that left the relative
contributions from the different escape routes statistically
indistinguishable. \citet{bot2002a} extended the model by also
accounting for objects from the outer MAB and the Jupiter-family-comet
(JFC) population, but their contribution turned out to be only about
15\% combined whereas the contributions by the inner MAB escape routes
were, again, statistically indistinguishable. The second column in
Table~\ref{table:b02numbers} provides the numbers predicted by
\citet{bot2002a} for all NEOs as well as NEO subgroups.

\begin{table}[h!]
  \centering
  \small
  \caption{The \citet{bot2002a} estimate for the number of NEOs with
    $H < 18$ and $a<7.2\au$, and the known population
    (ASTORB\protect\footnotemark{} 2018-01-30) with $H<18$, $H<17$ and
    $H<16$ as a function of NEO subgroup.}\label{table:b02numbers}
  \begin{tabular}{lrrrr}
    \hline
    Group & B02 & Known & Known & Known \\
    & $H<18$ & $H<18$ & $H<17$ & $H<16$ \\
    \hline
    Amor & $310\pm38$ & 504 & 218 & 81 \\ 
    Apollo & $590\pm71$ & 532 & 226 & 84 \\ 
    Aten & $58\pm9$ & 36 & 17 & 5 \\ 
    Atira & $20\pm3$ & 3 & 2 & 0 \\ 
    Vatira & -- & 0 & 0 & 0 \\ 
    NEO & $960\pm120$ & 1,075 & 463 & 170 \\ 
    \hline
  \end{tabular}
\end{table}

\footnotetext{\url{ftp://ftp.lowell.edu/pub/elgb/astorb.dat.gz}}
\citet{2001Icar..153..214D} presented an alternative method for
estimating the HFD that is based on the re-detection ratio, that is,
the fraction of objects that are re-detections of known objects rather
than new discoveries. They based their analysis on 784 NEOs detected
by the Lincoln Near-Earth Asteroid Research (LINEAR) project during
1999--2000. Based on the resulting HFD, that was valid for $13.5 \le H
\le 20.0$, they estimated that there should exist $855\pm101$ NEOs
with $H<18$. \citet{2015Icar..257..302H} extended the method and redid
the analysis with 11,132 NEOs discovered by multiple surveys. They
produced an HFD that is valid for $9<H<30.5$ and estimated that there
should be $1230\pm27$ ($990\pm20$) NEOs with $H<18$ ($H<17.75$). Later
an error was discovered in the treatment of absolute magnitudes that
the Minor Planet Center (MPC) reports to only a tenth of a
magnitude. Correcting for the rounding error reduced the number of
NEOs by 5\% \citep{stokes2017}.

\citet{2001Sci...294.1691S} used 1,343 detections of 606 different
NEOs by the LINEAR project to estimate, in practice, one-dimensional
debiased distributions for $a$, $e$, $i$, and $H$ by using a technique
relying on the $n(a,e,i,H)/\epsilon(a,e,i,H)$ ratio, where
$N(a,e,i,H)$ had to be marginalized over three of the parameters to
provide a useful estimate for the fourth parameter. They estimated
that there are $1227^{+170}_{-90}$ NEOs with $H<18$ and, in terms of
orbital elements, the most prominent difference compared to
\citet{bot2000a} was a predicted excess of NEOs with $i \gtrsim
20\deg$. \citet{2004Icar..170..295S} extended the model by
\citet{2001Sci...294.1691S} to include the taxonomic and albedo
distributions. They estimated that there would be $1090\pm180$ NEOs
with diameter $D>1\km$, and that 60\% of all NEOs should be dark, that
is, they should belong to the C, D, and X taxonomic complexes.

\citet{2011ApJ...743..156M} estimated that there are $981\pm19$ NEOs
with $D>1\km$ and $20,500\pm3,000$ with $D>100\meter$ based on
infrared (IR) observations obtained by the Widefield Infrared Survey
Explorer (WISE) mission. They also observed that the fraction of dark
NEOs with geometric albedo $p_V < 0.1$ is about 40\%, which should be
close to the debiased estimate given that IR surveys are essentially
unbiased with respect to $p_V$. \citet{2012ApJ...752..110M} extended
the analysis of WISE observations to NEO subpopulations and estimated
that there are $4,700\pm1,450$ PHOs with $D>100\meter$. In addition,
they found that the albedos of Atens are typically larger than the
albedos for Amors.

\citet{gre2012a} improved the steady-state orbit distributions that
were used by \citet{bot2002a} by using six times more test asteroids
and four times shorter timesteps for the orbital integrations. The new
orbit model focused on the $a<1\au$ region and discussed, for the
first time, the so-called Vatira population with orbits entirely
inside the orbit of Venus. The new integrations revealed that NEAs can
evolve to retrograde orbits even in the absence of close encounters
with planets \citep{gre2012b}. Similar retrograde orbits must have
existed in the integrations carried out by \citet{bot2000a,bot2002a},
but have apparently been overlooked. The fraction of NEAs on
retrograde orbits was estimated at about 0.1\% (within a factor of
two) of the entire NEO population. We note that the model by
\citet{gre2012a} did not attempt a re-calibration of the model
parameters but used the best-fit parameters found by \citet{bot2002a}.

\citet{2016Natur.530..303G} used an approach similar to
\citet{bot2002a} to derive a debiased four-dimensional model of the
HFD and orbit distribution. The key result of that paper was the
identification of a previously unknown sink for NEOs, most likely
caused by the intense solar radiation experienced by NEOs on orbits
with small perihelion distances. They also showed that dark NEOs
disrupt more easily than bright NEOs, and concluded that this explains
why the Aten asteroids have higher albedos than other NEOs
\citep{2012ApJ...752..110M}. Based on 7,952 serendipitious detections
of 3,632 distinct NEOs by the Catalina Sky Survey (CSS) they predicted
that there exists $1,008\pm45$ NEOs with $H<17.75$, which is in
agreement with other recent estimates.

\citet{Tricarico2017} analyzed the data obtained by the 9 most
prolific asteroid surveys over the past two decades and predicted that
there should exist $1096.6\pm13.7$ ($920\pm10$) NEOs with $H<18$
($H<17.75$). Their method relied, again, on computing the
$n(a,e,i,H)/\epsilon(a,e,i,H)$ ratio. The chosen approach implied that
the resulting population estimate is systematically too low, because
only bins that contain one or more known NEOs contribute to the
overall population regardless of the value of
$\epsilon(a,e,i,H)$. Detailed tests showed that the problem caused by
empty bins remains moderate when optimizing the bin
sizes. \citet{Tricarico2017} also showed that the cumulative HFDs
based on individual surveys were similar and this lends further
credibility to their results.

\citet{2017Icar..284..114S} derived the NEO HFD based on NEO
detections obtained with the Panoramic Survey Telescope and Rapid
Response System 1 (Pan-STARRS 1). Their methodology was, again, based
on computing the ratio $n(a,e,i,H)/\epsilon(a,e,i,H)$, where the
observational bias was obtained using a realistic survey simulation
\citep{2013PASP..125..357D}. Marginalizing over the orbital parameters
to provide a useful estimate for the HFD, \citet{2017Icar..284..114S}
found a distribution that agrees with \citet{2016Natur.530..303G} and
\citet{2015Icar..257..302H}.

Estimates for the number of km-scale and larger NEOs have thus
converged to about 900--1000 objects but there still remains
significant variation at the smaller sizes ($H\gtrsim23$). In what
follows we therefore primarily focus on the sub-km-scale NEOs.

Of the models described above only \citet{bot2000a}, \citet{bot2002a}
and \citet{2016Natur.530..303G} are four-dimensional models, that is,
they simultaneously and explicitly describe the correlations between
all four parameters throughout the considered $H$ range. These are
also the only models that provide information on the source regions
for NEOs although \citet{2016Natur.530..303G} did not explicitly
report this information. Although the model by \citet{bot2002a} has
been very popular and also able to reproduce the known NEO population
surprisingly well, it has some known shortcomings. The most obvious
problem is that the number of currently \emph{known} $H<18$ Amors
exceeds the \emph{predicted} number of $H<18$ Amors by more than
$5\sigma$ (Table~\ref{table:b02numbers}). \citet{bot2002a} are also
unable to reproduce the NEO, and in particular Aten, inclination
distribution \citep{2013ApJ...767L..18G}. These shortcomings are most
readily explained by the limited number of detections that the model
was calibrated with, but may also be explained by an unrealistic
initial inclination distribution for the test asteroids which were
used for computing the orbital steady-state distributions, or by not
accounting for Yarkovsky drift when populating the so-called
intermediate source regions in the MAB. An {\it intermediate source
  region} refers to the escape route from, e.g., the MAB and into the
NEO region whereas a {\it source region} refers to the region where an
object originates. In what follows we do not explicitly differentiate
between the two but refer to both with the term {\it
  entrance/escape route/region} (ER) for the sake of simplicity.
\citet{bot2002a} also used a single power law to describe the NEO HFD,
that is, neither variation in the HFD between NEOs from different ERs
nor deviations from the power-law form of the HFD were allowed. The
ERs, such as the intermediate Mars-crossers (IMCs) in \citet{bot2000a}
and \citet{bot2002a}, have been perceived to be artificial because
they are not the actual sources in the MAB. The IMC source also added
to the degeneracy of the model because the steady-state orbit
distribution for asteroids escaping the MAB overlaps with the
steady-state orbit distributions for asteroids escaping the MAB
through both the 3:1J MMR and the $\nu_6$ SR. On the other hand,
objects can and do escape out of a myriad of tiny resonances in the
inner MAB, feeding a substantial population of Mars-crossing
objects. Objects traveling relatively rapidly by the Yarkovsky effect
are more susceptible to jumping across these tiny resonances, while
those moving more slowly can become trapped \citep{bot2002b}. Modeling
this portion of the planet-crossing population correctly is therefore
computationally challenging. In this paper, we employ certain
compromises on how asteroids evolve in the MAB rather than invoke
time-consuming full-up models of Yarkovsky/YORP evolution
\citep[e.g.,][]{2015aste.book..509V}. The penalty is that we may miss
bodies that escape the MAB via tiny resonances. Our methods to deal
with this complicated issue are discussed in
\citet{2016Natur.530..303G,gra2017a} and below. We stress that the
observational data used for the \citet{bot2002a} model, 138 NEOs
observed by the Spacewatch survey, was well described with the model
they developed. The shortcomings described above have only become
apparent with the $>100$ times larger sample of known NEOs that is
available today (15,624 as of 2017-02-09). The most notable
shortcoming of the \citet{bot2002a} model in terms of application is
that it is strictly valid only for NEOs with $H<22$, roughly
equivalent to a diameter of $\gtrsim100\meter$. In addition, the
resolution of the steady-state orbit distribution limits the utility
of models that are based on it \citep[see, e.g.,][]{gra2012a}.

The improvements presented in this work compared to \citet{bot2002a}
are possible through the availability of roughly a factor of 30 more
observational data than used by \citet{bot2002a}, using more accurate
orbital integrations with more test asteroids and a shorter time step,
using more ERs (7 vs 5), and by using different and more flexible
absolute-magnitude distributions for different ERs.

The (incomplete) list of questions we will answer are:
\begin{itemize}

\item What is the total number of Amors, Apollos, Atens, Atiras, and
  Vatiras in a given size-range?

\item What is the origin for the observed excess (as compared to
  prediction \citet{bot2002a}) of NEOs with $20\deg \lesssim i
  \lesssim 40\deg$? Is there a particular source or are these orbits
  in a particular phase of their dynamical evolution, like the Kozai
  cycle?

\item What is the relative importance of each of the ERs in the MAB?

\item What is the fraction of comets in the NEO population?

\item Is there a measurable difference in the orbit distribution
  between small and large NEOs?

\item Are there differences in the HFDs of NEOs from different ERs?
  What are the differences?

\item What is the implication of these results for our understanding
  of the asteroid-Earth impact risk?

\item How does the predicted impact rate compare with the observed
  bolide rate?

\item What is the HFD for NEOs on retrograde orbits?

\item How does the resulting NEO HFD compare with independent
  estimates obtained through, for example, crater counting?

\item Is the NEO population in a steady state?

\end{itemize}

\section{Theory and methods}
\label{sec:theory}

Let us, for a moment, assume that we could correctly model all the
size-dependent, orbit-dependent, dynamical pathways from the MAB to
the NEO region, and we knew the orbit and size distributions of
objects in the MAB. In that case we could estimate the population
statistics for NEOs by carrying out direct integrations of test
asteroids from the MAB through the NEO region until they reach a
sink. While it has been shown that such a direct modeling is
reasonably accurate for km-scale and larger objects, it breaks down
for smaller objects \citep{gra2017a}. The most obvious missing piece
is that we do not know the orbit and size distributions of small
main-belt objects (MBOs) --- the current best estimates suggest that
the inventory is complete for diameters $D\gtrsim 1.5\km$
\citep{2015aste.book..795J}.

Instead we take another approach that can cope with our imperfect
knowledge and gives us a physically meaningful set of knobs to fit the
observations.  We build upon the methodology originally developed by
\citet{bot2000a} by using ER-dependent HFDs that allow for a smoothly
changing slope as a function of $H$. Equation (\ref{eq:baseff}) can
therefore be rewritten as
\begin{eqnarray}
  n(a,e,i,H) = \epsilon(a,e,i,H)\,\times \nonumber \\ 
   \sum_{s=1}^{N_{\rm ER}} N_s(H; N_{0, s}, \alpha_{{\rm min},s}, H_{{\rm min},s}, c_s)\,R_s(a,e,i)\,, \label{eq:impeff}
\end{eqnarray}
where $N_{\rm ER}$ is the number of ERs in the model, and the equation
for the differential $H$ distribution allows for a smooth,
second-degree variation of the slope:
\begin{eqnarray}
  N_s(H; N_{0, s}, \alpha_{{\rm min},s}, H_{{\rm min},s}, c_s) = \nonumber \\ 
  N_{0,s}\,10^{\int_{H_0}^{H}\left[\alpha_{{\rm min},s} + c_s(H'-H_{{\rm min},s})^2\right]\,dH'} = \nonumber \\
  N_{0,s}\,10^{\alpha_{{\rm min},s}(H-H_0) + \frac{c_s}{3}\left[(H-H_{{\rm min},s})^3 - (H_0-H_{{\rm min},s})^3\right]} \,. \label{eq:hdistr}
\end{eqnarray}
The steady-state orbital distributions, $R_s(a,e,i)$, are estimated
numerically by carrying out orbital integrations of numerous test
asteroids in the NEO region and recording the time that the test
asteroids spend in various parts of the ($a,e,i$) space (see
Sects.~\ref{sec:integrationmethods} and \ref{sec:integrations}). The
orbit distributions are normalized so that for each ER $s$
\begin{equation}
  \iiint R_s(a,e,i) \,\ud a \,\ud e \,\ud i = 1\,. \label{eq:resnorm}
\end{equation}
In practice the integration over the corresponding NEO orbital space
in Eq.~(\ref{eq:resnorm}) is replaced with a simple summation over a
grid of finite cells.

With the orbit distributions $R_s(a,e,i)$ fixed, the free parameters
to be fitted describe the HFDs for the different ERs: the number
density $N_{0,s}$ at the reference magnitude $H_0$ (common to all
sources and chosen to be $H_0=17$), the minimum slope $\alpha_{{\rm
    min},s}$ of the absolute magnitude distribution, the curvature
$c_s$ of the absolute-magnitude-slope relation, and the absolute
magnitude $H_{{\rm min},s}$ corresponding to the minimum slope.  Note
that at $H=H_0$ in our parametrization, the (unnormalized) $N_{0,s}$
effectively take the same role as the (normalized) weighting factors
by which different ERs contribute to the NEO population in the
\citet[][2002]{bot2000a} models.  Because HFDs are ER-resolved in our
approach, the relative weighting at $H \neq H_0$ is not explicitly
available but has to be computed separately.

As shown by \citet{2016Natur.530..303G} it is impossible to find an
acceptable fit to NEOs with small perihelion distances,
$q=q(a,e)=a(1-e)$, when assuming that the sinks for NEOs are
collisions with the Sun or planets, or an escape from the inner solar
system. The model is able to reproduce the observed NEO distribution
only when assuming that NEOs are completely destroyed at small, yet
nontrivial distances from the Sun. In addition to the challenges with
numerical models of such a complex disruption event in a detailed
physical sense, it is also computationally challenging to merely
\emph{fit} for an average disruption
distance. \citet{2016Natur.530..303G} performed an incremental fit to
an accuracy of 0.001~au. The incremental fit was facilitated by
constructing multiple different steady-state orbit distributions, each
with a different assumption for the average disruption distance, and
then identifying the orbit distribution which leads to the best
agreement with the observations. Each of the steady-state orbit
distributions were constructed so that the test asteroids did not
contribute to the orbit distribution after they crossed the assumed
average disruption distance. \citet{2016Natur.530..303G} used the
perihelion distance $q$ as the distance metric. While it is clear that
super-catastrophic disruption can explain the lack of NEOs on
small-$q$ orbits, such a simple model does not allow for an accurate
reproduction of the observed $q$ distribution. For instance,
\citet{2016Natur.530..303G} explicitly showed that the disruption
distance depends on asteroid diameter and geometric albedo.

Here we take an alternative and non-physical route to fit the
small-perihelion-distance part of the NEO population to improve the
quality of the fit: we use orbit distributions that do \emph{not}
account for disruptions at small $q$ and instead fit a linear penalty
function, $p(a,e)$, with an increasing penalty against orbits with
smaller $q$. Equation (\ref{eq:baseff}) now reads
\begin{eqnarray}
  && n(a,e,i,H) = \epsilon(a,e,i,H)\times \nonumber \\
  && \sum_{s=1}^{N_{\rm ER}} N_s(H; N_{0,s}, \alpha_{{\rm min},s}, H_{{\rm min},s}, c_s)\times \nonumber \\
  && \frac{\left[1-p(a,e)\right]\,R_s(a,e,i)}{\sum_{a,e,i}\left[1-p(a,e)\right]\,R_s(a,e,i)}\,, \label{eq:empeff}
\end{eqnarray}
where
\begin{eqnarray*}
  p(a,e) = \left\{
  \begin{array}{cl}
    k(q_0 - q(a,e)) & \mathrm{for}\; q \le q_0,\\
    0 & \mathrm{for}\; q > q_0,\\
  \end{array}\right.
\end{eqnarray*}
and we solve for two additional parameters---the linear slope, $k$, of
the penalty function and the maximum perihelion distance where the
penalty is applied, $q_0$. Note that the penalty function does not
have a dependence on $H$ although it has been shown that small NEOs
disrupt at larger distances compared to large NEOs
\citep{2016Natur.530..303G}.  We chose to use a functional form
independent of $H$ to limit the number of free parameters.

In total we need to solve for $N_\mathrm{par} = 4N_\mathrm{ER}+2$
parameters. In the following three subsections we will describe the
methods used to estimate the orbital-element steady-state
distributions and discovery efficiencies as well as to solve the
efficiency equation.

\subsection{Estimation of observational selection effects}
\label{sec:biasmethods}

All asteroid surveys are affected by observational selection effects
in the sense that the detected population needs to be corrected in
order to find the true population. The known distribution of asteroid
orbits is not representative of their actual distribution because
asteroid discovery and detection is affected by an object's size,
lightcurve amplitude, rotation period, apparent rate of motion, color,
and albedo, and the detection system's limiting magnitude, survey
pattern, exposure time, the sky-plane density of stars, and other
secondary factors. Combining the observed orbit distributions from
surveys with different detection characteristics further complicates
the problem unless the population under consideration is essentially
`complete', i.e., all objects in the sub-population are
known. \citet{2016Icar..266..173J} provide a detailed description of
the methods employed for estimating selection effects in this work.
Their technique builds upon earlier methods
\citep{1998Icar..131..245J,jed2002a,gra2012a} and takes advantages of
the increased availability of computing power to calculate a fast and
accurate estimate of the observational bias.

The ultimate calculation of the observational bias would provide the
efficiency of detecting an object as a function of all the parameters
listed above but this calculation is far too complicated,
computationally expensive, and unjustified for understanding the NEO
orbit distribution and HFD at the current time.  Instead, we invoke
many assumptions about unknown or unmeasurable parameters and average
over the underlying system and asteroid properties to estimate the
selection effects.

The fundamental unit of an asteroid observation for our purposes is a
`tracklet' composed of individual detections of the asteroid in
multiple images on a single night \citep{kub2007a}.  At the mean time
of the detections the tracklet has a position and rate of motion ($w$)
on the sky and an apparent magnitude ($m$; perhaps in a particular
band). Note that \citet{2016Icar..266..173J} use a different
notation. The tracklet detection efficiency depends on all these
parameters and can be sensitive to the detection efficiency in a
single image due to sky transparency, optical effects, and the
background of, e.g., stars, galaxies, and nebulae.  We average these
effects over an entire night and calculate the detection efficiency
($\bar\epsilon(m)$) as a function of apparent magnitude for the CSS
images using the system's automated detection of known MBOs.  To
correct for the difference in apparent rates of motion between NEOs
and MBOs we used the results of \citet{zav2008a} who measured the
detection efficiency of stars that were artificially trailed in CSS
images at known rates.  Thus, we calculated the average nightly NEO
detection efficiency as a function of the observable tracklet
parameters: $\bar\epsilon(m,w)$.

The determination of the observational bias as a function of the
orbital parameters ($\epsilon(a,e,i,H)$) involved convolving an
object's observable parameters ($m,w$) with its ($a,e,i,H$)
averaged over the orbital angular elements (longitude of ascending
node $\Omega$, argument of perihelion $\omega$, mean anomaly $M_0$) 
that can appear in the
fields from which a tracklet is composed.  For each image we step
through the range of allowed topocentric distances ($\Delta$) and
determine the range of angular orbital elements that could have been
detected for each $(a,e,i,H)$ combination. Since the location of the
image is known ((R.A.,Dec.)=$(\alpha,\delta)$) and the topocentric
location of the observer is known, then, given $\Delta$ and $(a,e,i)$
it is possible to calculate the range of values of the other orbital
elements that can appear in the field. Under the assumption that the
distributions of the angular orbital elements are flat it is then
possible to calculate $\epsilon(a,e,i,H)$ in a field and then in all
possible fields using the appropriate probabilistic
combinatorics. While \citet{2014Icar..229..236J} have shown that the
argument of perihelion, longitude of ascending node and longitude of
perihelion distributions for NEOs have modest but
statistically-significant non-uniformities, we consider them to be
negligible for our purposes compared to the other sources for
systematics.

\subsection{Orbit integrator}
\label{sec:integrationmethods}

The orbital integrations to obtain the NEO steady-state orbit
distributions are carried out with an augmented version of the SWIFT
RMVS4 integrator \citep{1994Icar..108...18L}. The numerical methods,
in particular those related to Yarkovsky modeling (not used in the
main simulations of this work but only in some control simulations to
attest its importance), are detailed in \citet{gra2017a}. The only
additional feature implemented in the software was the capability to
ingest test asteroids (with different initial epochs) on the fly 
as the integration progresses, and
this was done solely to reduce the computing time required.

\subsection{Estimation of model parameters}

We employ an extended maximum-likelihood (EML) scheme \citep{cow1998a}
and the simplex optimization algorithm \citep{nel1965a} when solving
Eqs. (\ref{eq:impeff}) and (\ref{eq:empeff}) for the parameters $\vect
P$ that describe the model.

Let $(n_1,n_2,\ldots,n_{N_\mathrm{bin}})$ be the non-zero bins in the
binned version of $n(a,e,i,H)$, and
$(\nu_1,\nu_2,\ldots,\nu_{N_\mathrm{bin}})$ be the corresponding bins
containing the expectation values, that is, the model prediction for
the number of observations in each bin. The joint probability-density
function (PDF) for the distribution of observations
$(n_1,n_2,\ldots,n_{N_\mathrm{bin}})$ is given by the multinomial
distribution:
\begin{equation}\label{eq:likelihood}
p_\mathrm{joint} = n_\mathrm{tot}!\,\prod_{k=1}^{N_\mathrm{bin}}
\frac{1}{n_k!}\left(\frac{\nu_k}{n_\mathrm{tot}}\right)^{n_k},
\end{equation}
where $\nu_k/n_\mathrm{tot}$ gives the probability to be in bin $k$.
In EML the measurement is defined to consist of first determining
\begin{equation}\label{eq:ntot}
n_\mathrm{tot}=\sum_{k=1}^{N_\mathrm{bin}} n_k
\end{equation}
observations from a Poisson distribution with mean $\nu_\mathrm{tot}$
and then distributing those observations in the histogram
$(n_1,n_2,\ldots,n_{N_\mathrm{bin}})$. That is, the total number of
detections is regarded as an additional constraint. The extended
likelihood function $L$ is defined as the joint PDF for the total
number of observations $n_\mathrm{tot}$ and their distribution in the
histogram $(n_1,n_2,\ldots,n_{N_\mathrm{bin}})$. The joint PDF is
therefore obtained by multiplying Eq.~(\ref{eq:likelihood}) with a
Poisson distribution with mean
\begin{equation}\label{eq:nutot}
\nu_\mathrm{tot} = \sum_{k=1}^{N_\mathrm{bin}} \nu_k
\end{equation}
and accounting for the fact that the probability for being in bin $k$
is now $\nu_k/\nu_\mathrm{tot}$:
\begin{eqnarray}\label{eq:elikelihood}
p_\mathrm{joint}' & = &
\frac{\nu_\mathrm{tot}^{n_\mathrm{tot}}\,\exp(-\nu_\mathrm{tot})}{n_\mathrm{tot}!}\,n_\mathrm{tot}!\,\prod_{k=1}^{N_\mathrm{bin}}
\frac{1}{n_k!}\left(\frac{\nu_k}{\nu_\mathrm{tot}}\right)^{n_k} \nonumber \\
& = &
\nu_\mathrm{tot}^{n_\mathrm{tot}}\,\exp(-\nu_\mathrm{tot})\,\prod_{k=1}^{N_\mathrm{bin}}
\frac{1}{n_k!}\left(\frac{\nu_k}{\nu_\mathrm{tot}}\right)^{n_k} \nonumber \\
& = & \exp(-\nu_\mathrm{tot})\,\prod_{k=1}^{N_\mathrm{bin}}
\frac{1}{n_k!}\nu_k^{n_k}
\end{eqnarray}

Neglecting variables that do not depend on the parameters that are
solved for, the logarithm of Eq.\ (\ref{eq:elikelihood}), that is, the
log-likelihood function, can be written as
\begin{equation}\label{eq:logel}
\logl = -\nu_\mathrm{tot} + \sum_{k=1}^{N_\mathrm{bin}} n_k\,\log \nu_k \,,
\end{equation}
where the first term on the right hand side emerges as a consequence
of accounting for the total number of detections.

The optimum solution, in the sense of maximum log-likelihood,
$\logl_\mathrm{max}$, is obtained using the simplex algorithm
\citep{nel1965a} which starts with $N_\mathrm{par}+1$ random
$N_\mathrm{par}$-dimensional solution vectors $\vect P_l$
($l=1,N_\mathrm{par}+1$) where $N_\mathrm{par}$ is the number of
parameters to be solved for. The simplex crawls towards the optimum
solution in the $N_\mathrm{par}$-dimensional phase space by improving,
at each iteration step, the parameter values of the worst solution
$\logl_\mathrm{min}$ towards the parameter values of the best solution
$\logl_\mathrm{max}$ according to the predefined sequence of simplex
steps. The optimization ends when $\logl_\mathrm{max} -
\logl_\mathrm{min} < \epsilon$ and all $|\vect P_{i,m} - \vect
P_{j,m}| < \epsilon_1$, where $i,j$ refer to different solution
vectors, $m$ is the index for a given parameter, and $\epsilon_1 \sim
2\times10^{-15}$. To ensure that an optimum solution has been found we
repeat the simplex optimization using the current best solution and
$N$ random solution vectors until $\logl_\mathrm{max}$ changes by less
than $\epsilon_2$ in subsequent runs, where $\epsilon_2 \sim
10^{-10}$. We found suitable values for $\epsilon_1$ and $\epsilon_2$
empirically. Larger values would prevent the optimum solutions to be
found and smaller values would not notably change the results.
Finally, we employ 10 separate simplex chains to verify that different
initial conditions lead to the same optimum solution.

As an additional constraint we force the fitted parameters $\vect P$
to be non-negative. The reasoning behind this choice is that negative
parameter values are either unphysical ($N_{0,s}$, $\alpha_{{\rm
    min},s}$, $c_s$, $k$, $q_0$) or unconstrained ($H_{{\rm
    min},s}$). A minimum slope occuring for $H_{{\rm min},s}<0$ is
meaningless because all NEOs have $H\gtrsim9.4$ and we fit for
$17<H<25$. Hence $H_{{\rm min},s}\gtrsim0$ is an acceptable
approximation in the hypothetical case that the simplex algorithm
would prefer $H_{{\rm min},s}<0$.

In what follows we use low-resolution orbit distributions ($\delta
a=0.1\au$, $\delta e=0.04$, $\delta i=4\deg$) to fit for the model
parameters, because it was substantially faster than using the default
resolution of the steady-state orbit distributions ($\delta
a=0.05\au$, $\delta e=0.02$, $\delta i=2\deg$). In both cases we use a
resolution of $\delta H=0.25\magnitude$ for the absolute magnitude.
We combine the best-fit parameters obtained in low resolution with the
orbit distributions in default resolution to provide our final
model. We think this is a reasonable approach because the orbit
distributions are fairly smooth regardless of resolution and the
difference between fitting in low or default resolution leads to
negligible differences in the resulting models.

Although \citet{gra2017a} identified about two dozen different ERs,
concerns about degeneracy issues prevented us from including all the
ERs separately in the final model. Instead we made educated decisions
in combining the steady-state orbit distributions into larger
complexes by, e.g., minimizing Akaike's Information Criteria
\citep[AIC;][]{aka1974a} with a correction for multinomial data and
sample size \citep[Eq.~7.91 in][]{burnham-anderson2002}:
\begin{equation}
 {\rm AICc} = 2N_\mathrm{par} - 2\log L_\mathrm{max} +
  \frac{2N_\mathrm{par}(N_\mathrm{par}+4)}{4N_\mathrm{bin}-N_\mathrm{par}-4}\,. \label{eq:aicc}
\end{equation}

\section{Distribution of NEOs as observed by CSS}
\label{sec:cssobs}

The Mt.\ Lemmon (IAU code G96) and Catalina (703) stations of the
Catalina Sky Survey \citep[CSS;][]{Christensen2012} discovered or
accidentally rediscovered 4035 and 2858 NEOs, respectively, during the
8-year period 2005--2012. The motivation for using the data from these
telescopes during this time period is that one of these two telescopes
was the top PHO discovery system from 2005 through 2011 and the two
systems have a long track record of consistent, well-monitored
operations. The combination of these two factors provided us reliable,
high-statistics discoveries of NEOs suited to the debiasing procedure
employed in this work.

Details of CSS operations and performance can be found elsewhere
\citep[e.g.][]{Christensen2012,2016Icar..266..173J} but generally, the
G96 site with its 1.5-m telescope can be considered a narrow-field
`deep' survey whereas the 0.7-m 703 Schmidt telescope is a wide-field
but `shallow' survey.  The different capabilities provide an excellent
complementarity for this work to validate our methods as described
below.  To ensure good quality data we used NEO detections only on
nights that met our criteria \citep{2016Icar..266..173J} for tracklet
detection efficiency ($\epsilon_0$), limiting magnitude ($V_{\rm
  lim}$), and a parameter related to the stability of the limiting
magnitude on a night ($V_{\rm width}$).  About 80\% of all 703 fields
and nearly 88\% of the G96 fields passed our requirements.  The
average tracklet detection efficiency for the fields that passed the
requirements were 75\% and 88\% for 703 and G96, respectively, while
the limiting magnitudes were $V=19.44$ and $V=21.15$
\citep{2016Icar..266..173J}.

\begin{figure*}
  \centering
  \includegraphics[width=\columnwidth]{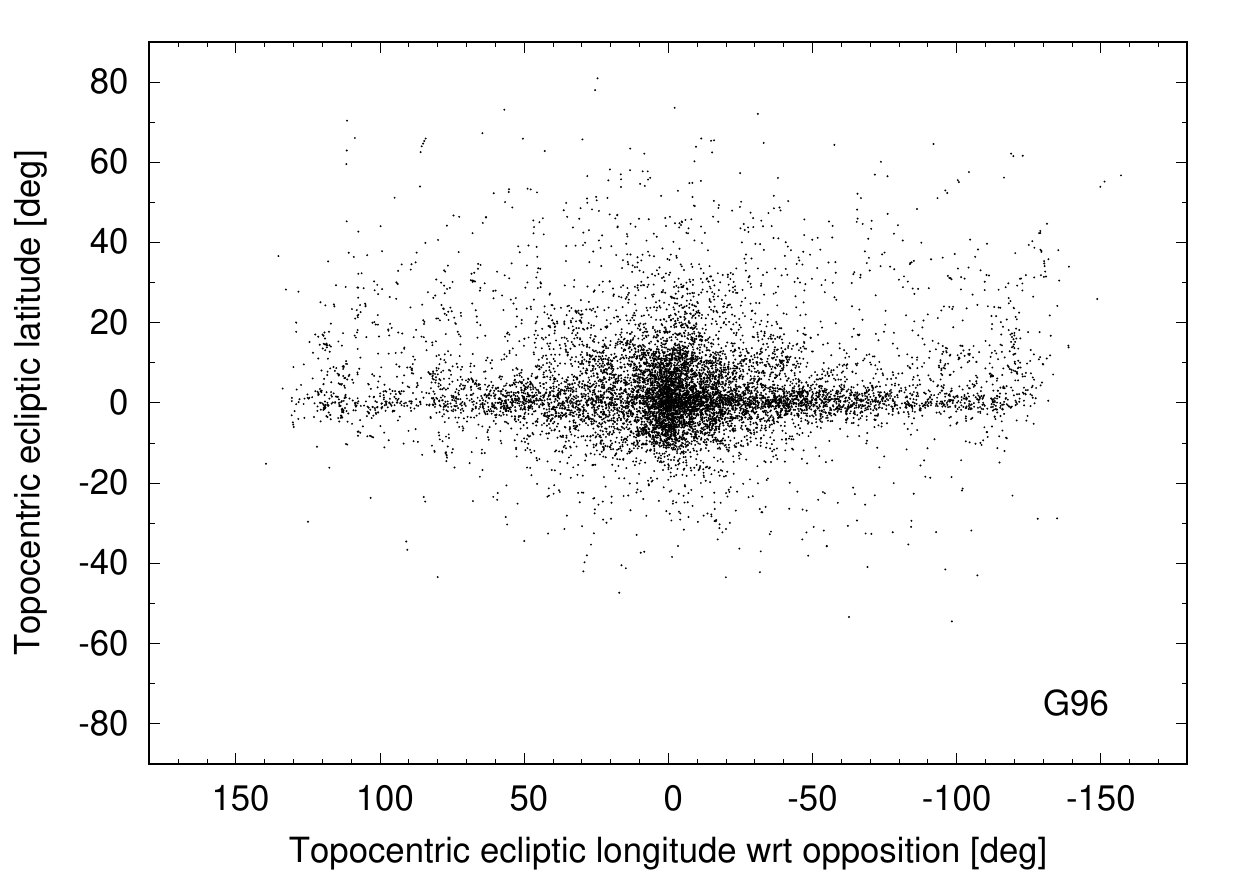}
  \includegraphics[width=\columnwidth]{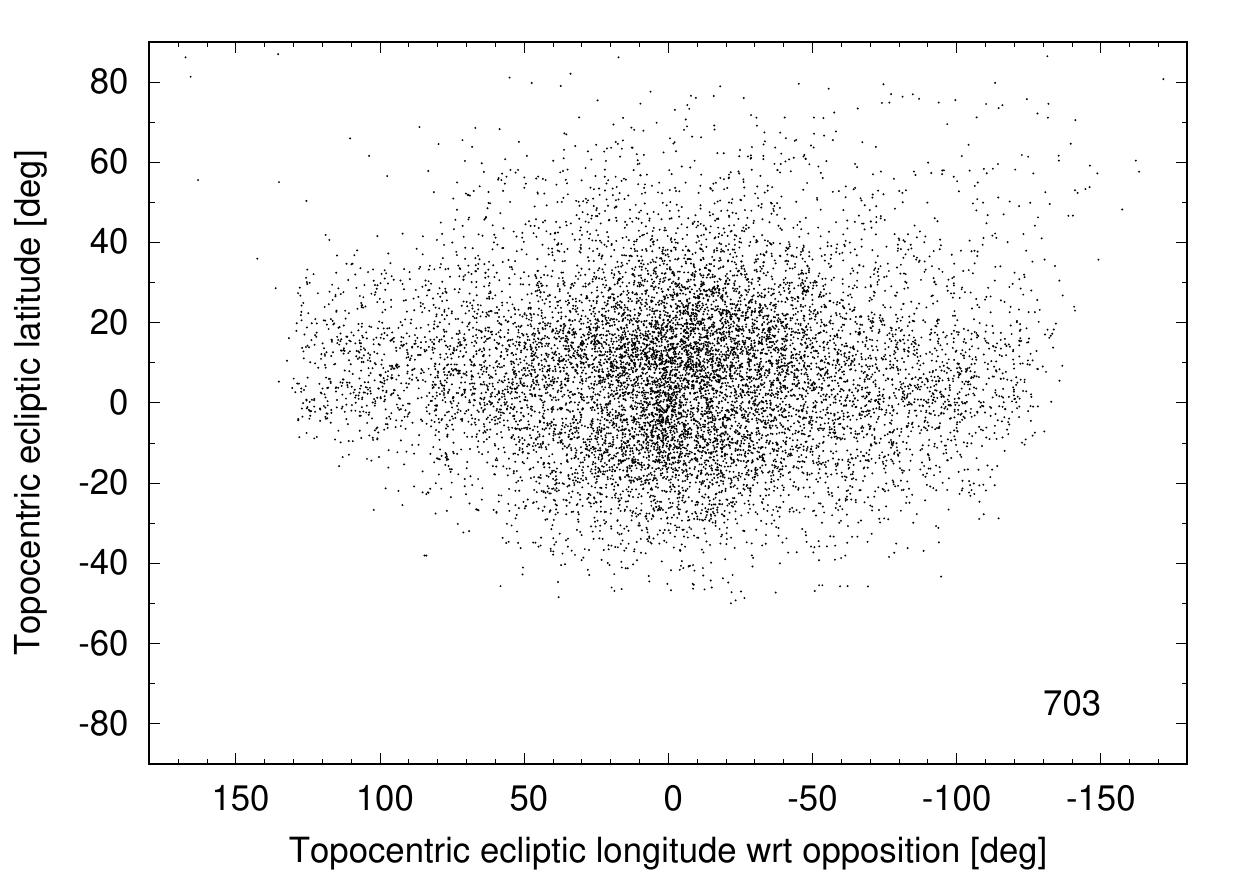}
  \includegraphics[width=\columnwidth]{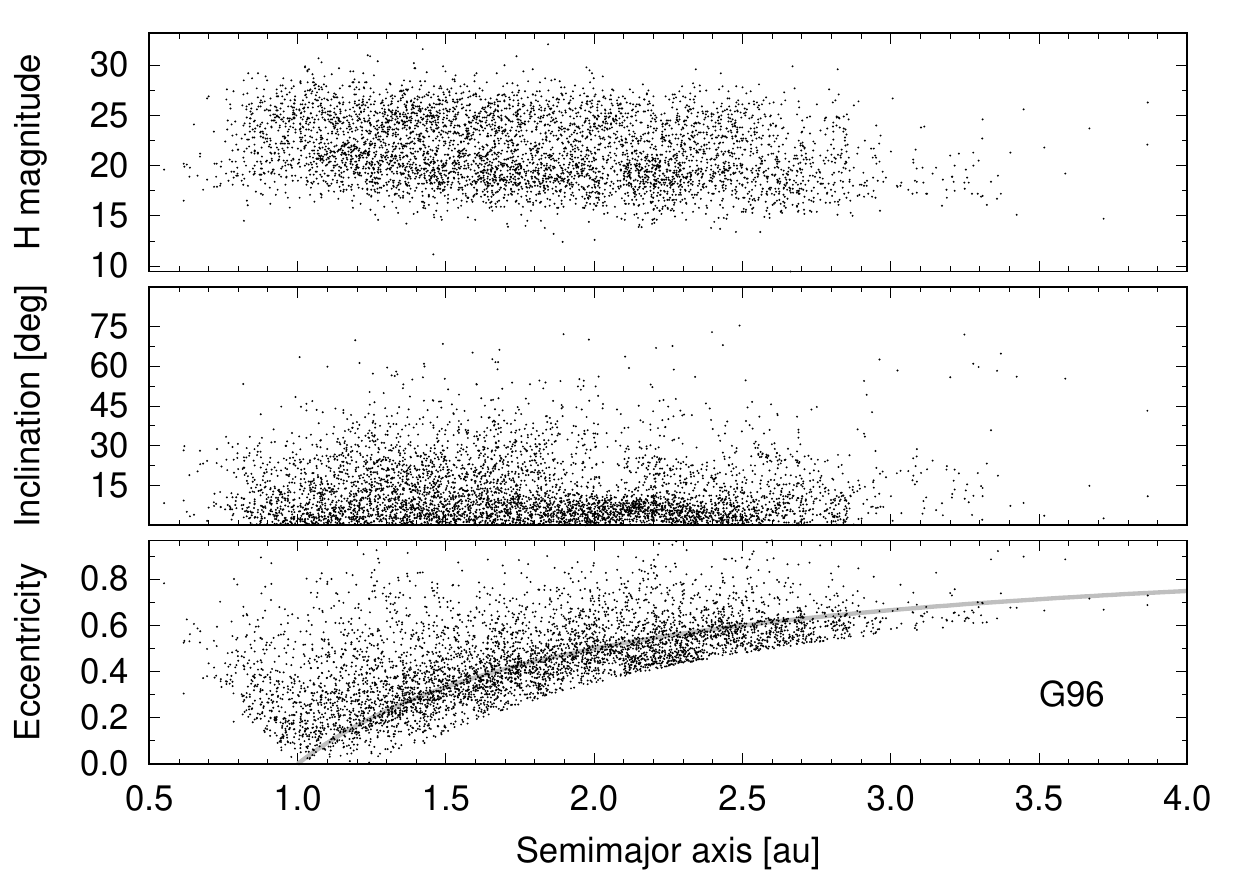}
  \includegraphics[width=\columnwidth]{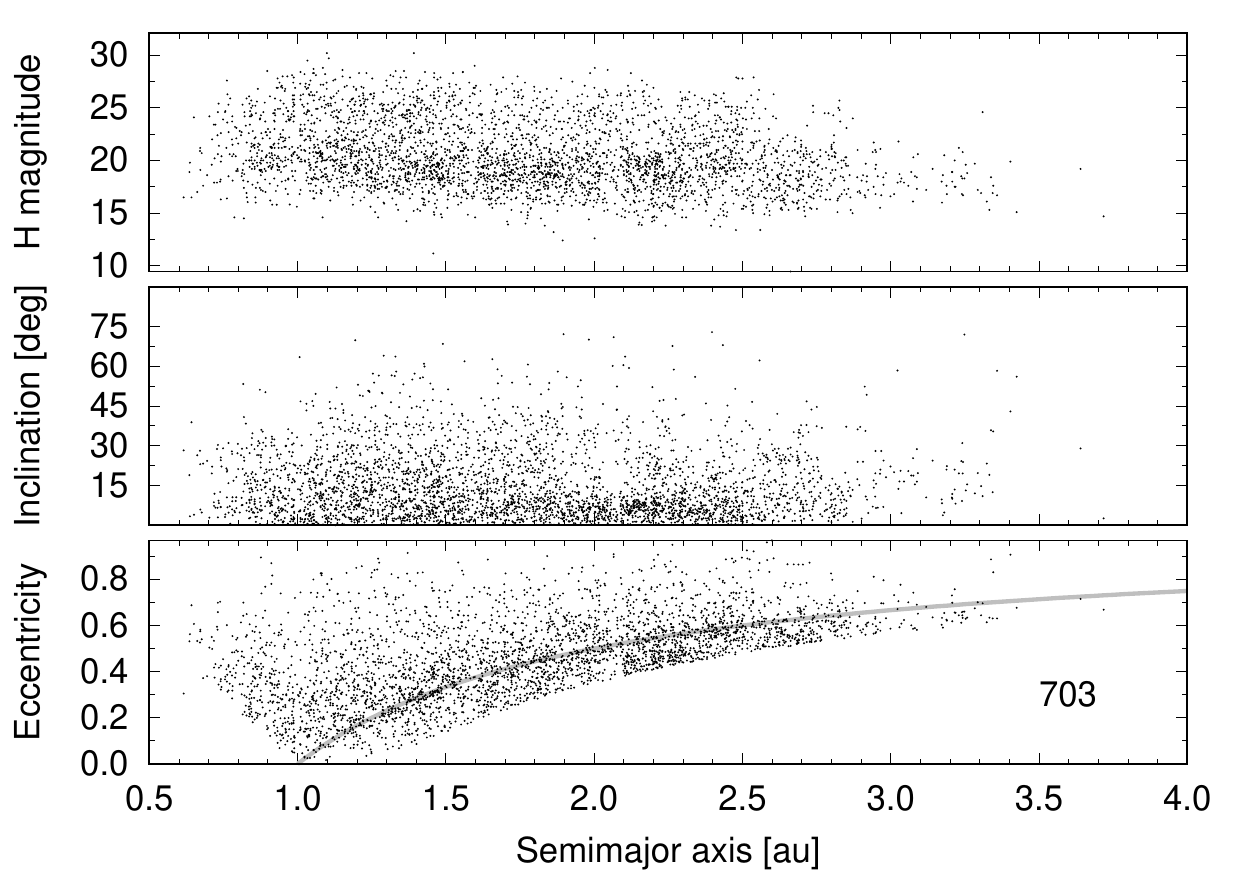}
  \includegraphics[width=\columnwidth]{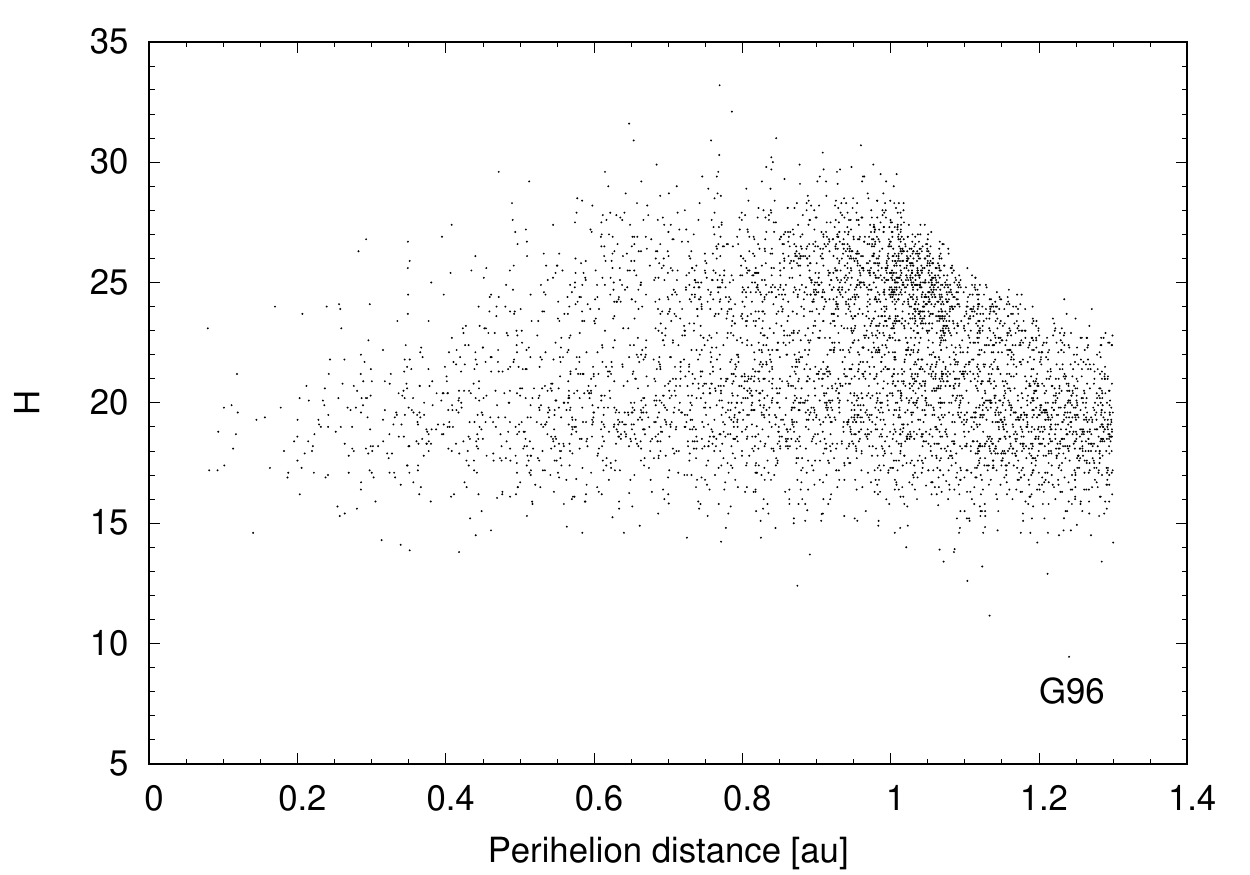}
  \includegraphics[width=\columnwidth]{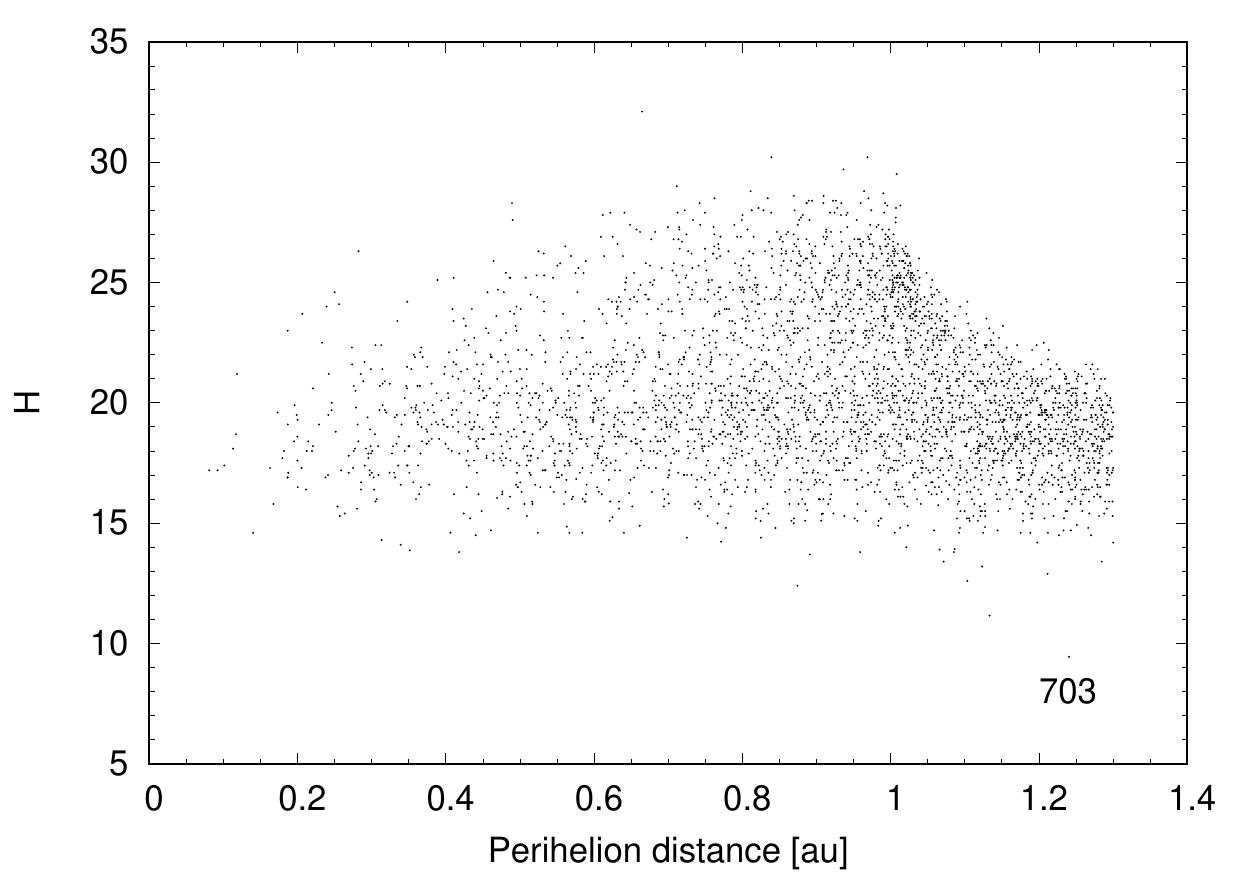}
  \caption{Ecliptic coordinates at discovery for NEOs detected by (top
    left) G96 and (top right) 703. Observed ($a$,$i$,$e$,$H$)
    distributions for NEOs detected by (middle left) G96 and (middle
    right) 703. The gray line in the ($a$,$e$) panels corresponds to
    $q=1\au$. Observed ($q$,$H$) distributions for NEOs detected by
    (bottom left) G96 and (bottom right) 703.}
  \label{fig:obsaeiH}
\end{figure*}

All NEOs that were identified in tracklets in fields acquired on
nights that met our criteria were included in this analysis. It is
important to note that the selection of fields and nights was in no
way based on NEO discoveries. The list of NEOs includes new
discoveries and previously known objects that were independently
re-detected by the surveys. The ecliptic coordinates of the CSS
detections at the time of detection show that the G96 survey
concentrates primarily on the ecliptic whereas the wide-field 703
survey images over a much broader region of the sky
(Fig.\ \ref{fig:obsaeiH}, top 2 panels). It is also clear from these
distributions that both surveys are located in the northern hemisphere
as no NEOs were discovered with ecliptic latitudes $<-50\deg$.

The detected NEOs' $a$, $e$, $i$, and $H$ distributions are also shown
in Fig.\ \ref{fig:obsaeiH} and display similar distributions for both
stations. The enhancement near the $q=1\au$ line in the ($a$,$e$)
plots is partly caused by observational biases. The small NEOs that
can be detected by ground-based surveys must be close to Earth to be
brighter than a system's limiting magnitude, and objects on orbits
with perihelia near the Earth's orbit spend more time near the Earth,
thereby enhancing the number of detected objects with
$q\sim1\au$. This effect is obvious in the bottom panels in
Fig.\ \ref{fig:obsaeiH} in which it is clear that detections of small
NEOs ($H\gtrsim25$) are completely lacking for $q>1.1\au$ ($q>1.05\au$
for 703); in other words, very small objects can be detected only when
they approach close to the Earth. Thus, the NEO model described herein
is not constrained by observational data in that region of ($q,H$)
space. Instead the constraints derive from our understanding of NEO
orbital dynamics.

It is also worth noting the clear depletion of objects with $H\sim22$
in the two bottom plots of Fig.\ \ref{fig:obsaeiH} (also visible in
the $H$-$a$ panels in the middle of the figure).  The depletion band
in this absolute magnitude range is clearly not an observational
artifact because there is no reason to think that objects in this size
range are more challenging to detect than slightly bigger and slightly
smaller objects. The explanation is that the HFD cannot be reproduced
with a simple power-law function but has a plateau around
$H\sim22$. This plateau reduces their number statistics simultaneously
and combined with their small sizes reduces their likelihood of
detection. Going to slightly smaller objects will increase the number
statistics and they are therefore detected in greater numbers than
their larger counterparts.

\section{CSS observational selection effects}
\label{sec:cssbias}

\begin{figure}
  \centering
  \includegraphics[width=\columnwidth]{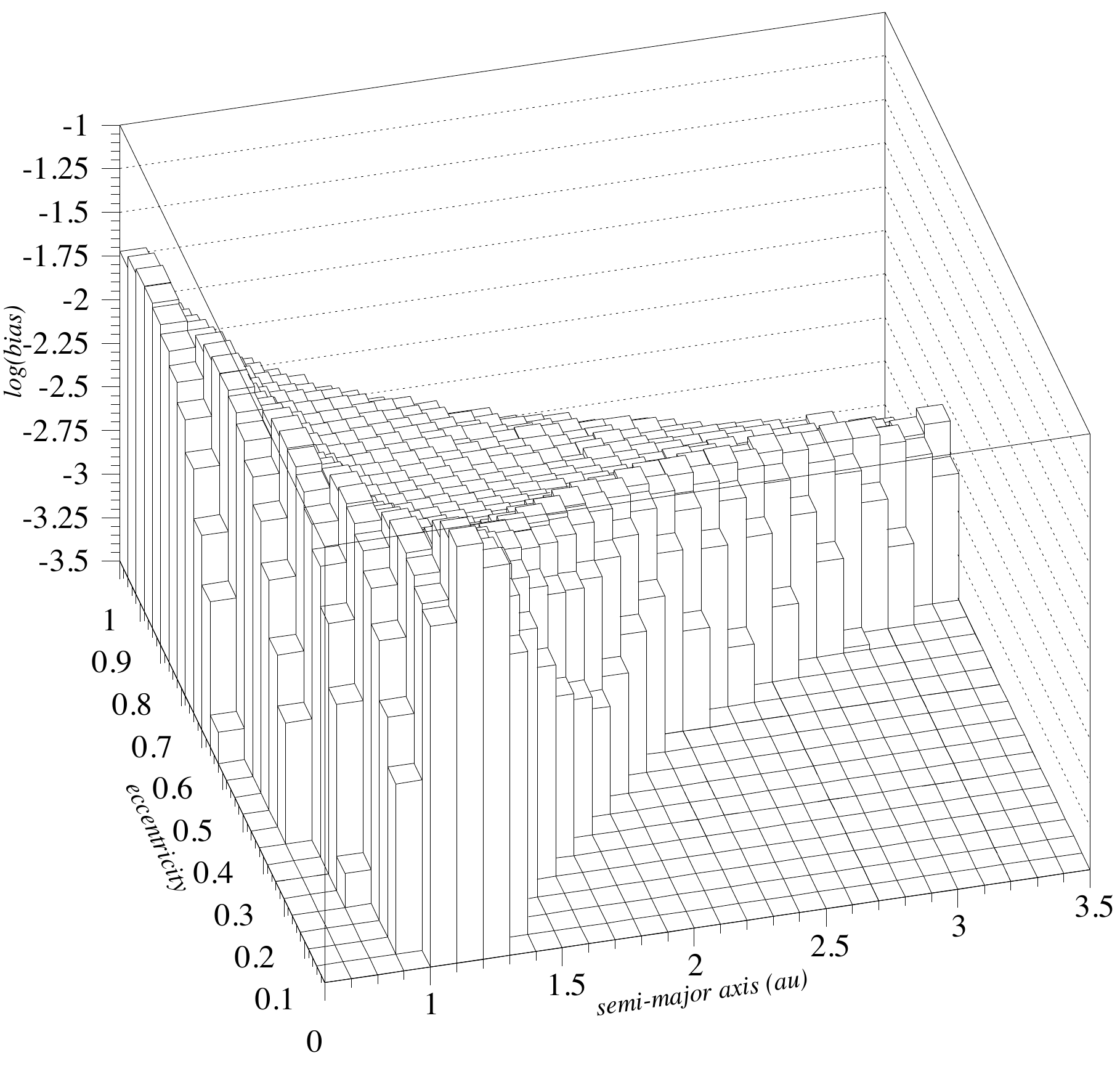}
  \caption{A 2-d slice through the 4-d detection efficiency,
    $\epsilon(a,e,i,H)$, with $i=2\deg$ and $H=22.875$ for the G96
    survey ($H=22.875$ corresponds to objects of about 100$\meter$
    diameter).  To enhance the regions with small efficiency the fig
    shows $\log\epsilon(a,e,i,H)$ as a function of semi-major axis and
    eccentricity.}
  \label{fig:G96-ae-bias}
\end{figure}

The observed four-dimensional $(a,e,i,H)$ distributions in
Fig.\ \ref{fig:obsaeiH} are the convolution of the actual distribution
of NEOs with the observational selection effects ($\epsilon(a,e,i,H)$)
as described in Section~\ref{sec:biasmethods}.  The calculation of the
four-dimensional $\epsilon(a,e,i,H)$ is non-trivial but was performed for
the Spacewatch survey \citep{bot2002a}, for the Catalina Sky Survey
G96 and 703 sites employed herein \citep{2016Icar..266..173J}, and
most recently for a combination of many NEO surveys
\citep{Tricarico2016,Tricarico2017}.  It is impossible to directly
compare the calculated detection efficiency in these publications
because they refer to different asteroid surveys for different periods
of time.  The fact that \citet{Tricarico2017}'s final cumulative $H$
distribution is in excellent agreement with the $H$ distribution found
in this work suggests that both bias calculations must be accurate to
within the available statistics.

The $\epsilon(a,e,i,H)$ slice in Fig.\ \ref{fig:G96-ae-bias}
illustrates some of the features of the selection effects that are
manifested in the observations shown in Fig.\ \ref{fig:obsaeiH}.  The
`flat' region with no values in the lower-right region represents bins
that do not contain NEO orbits.  The flat region in the lower-left
corresponds to orbits that can not be detected by CSS because they are
usually too close to the Sun.  There is a `ridge' of relatively high
detection efficiency along the $q=1\au$ line that corresponds to the
enhanced detection of objects in the $e$ vs. $a$ panels in
Fig.\ \ref{fig:obsaeiH}. That is, higher detection efficiency along
the ridge means that more objects are detected. Perhaps
counter-intuitively, the peak efficiency for this $(i,H)$ combination
occurs for objects with $(a\sim1.15\au,e\sim0.05)$ while objects on
orbits with $1.0\au \le a \le 1.1\au$ are less efficiently detected.
This is because the synodic period between Earth and an asteroid with
$a=1.1\au$ is 11 years but only about 7.5 years for objects with
$a=1.15\au$, close to the 7-year survey time period considered here.
Thus, asteroids with $a=1.1\au$ are not detectable as frequently as
those with $a=1.15\au$. Furthermore, those with smaller semi-major
axis have faster apparent rates of motion when they are detectable.
Interestingly, the detection efficiency is relatively high for
Aten-class objects on orbits with high eccentricity because they are
at aphelion and moving relatively slowly when they are detectable in
the night sky from Earth. 

The bias against detecting NEOs rapidly becomes severe for smaller
objects (Supplementary Animation 1 and Supplementary Fig.~1), and 
only those 
that have close approaches to the Earth on low-inclination orbits are 
even remotely detectable. For more details on the bias calculation and 
a discussion of selection effects in general and for the CSS we refer 
the reader to \citet{jed2002a} and \citet{2016Icar..266..173J}.

\section{NEO orbit distributions}
\label{sec:integrations}

\subsection{Identification of ERs in the MAB}

In order to find an exhaustive set of ERs in the MAB \citet{gra2017a}
used the largest MBOs with $H$ magnitudes below the assumed
completeness limit and integrated them for 100~Myr or until they
entered the NEO region.

\citet{gra2017a} started from the orbital elements and $H$ magnitudes
of the 587,129 known asteroids as listed on July 21, 2012 in the MPC's
\verb|MPCORB.DAT| file. For MBOs interior to the 3:1 MMR with Jupiter
(centered at $a\sim2.5\au$) they selected all non-NEOs ($q > 1.3\au$)
that have $H \leq 15.9$. Exterior to the 3:1 MMR they selected all
non-NEOs that have $H \leq 14.4$ and $a < 4.1\au$. To ensure that the
sample is complete they iteratively adjusted these criteria to result
in a set of objects that had been discovered prior to Jan 1,
2012---that is, no objects fulfilling the above criteria had been
discovered in the $\sim7$-month period leading to the extraction date.

\begin{figure}
  \centering
  \includegraphics[width=\columnwidth]{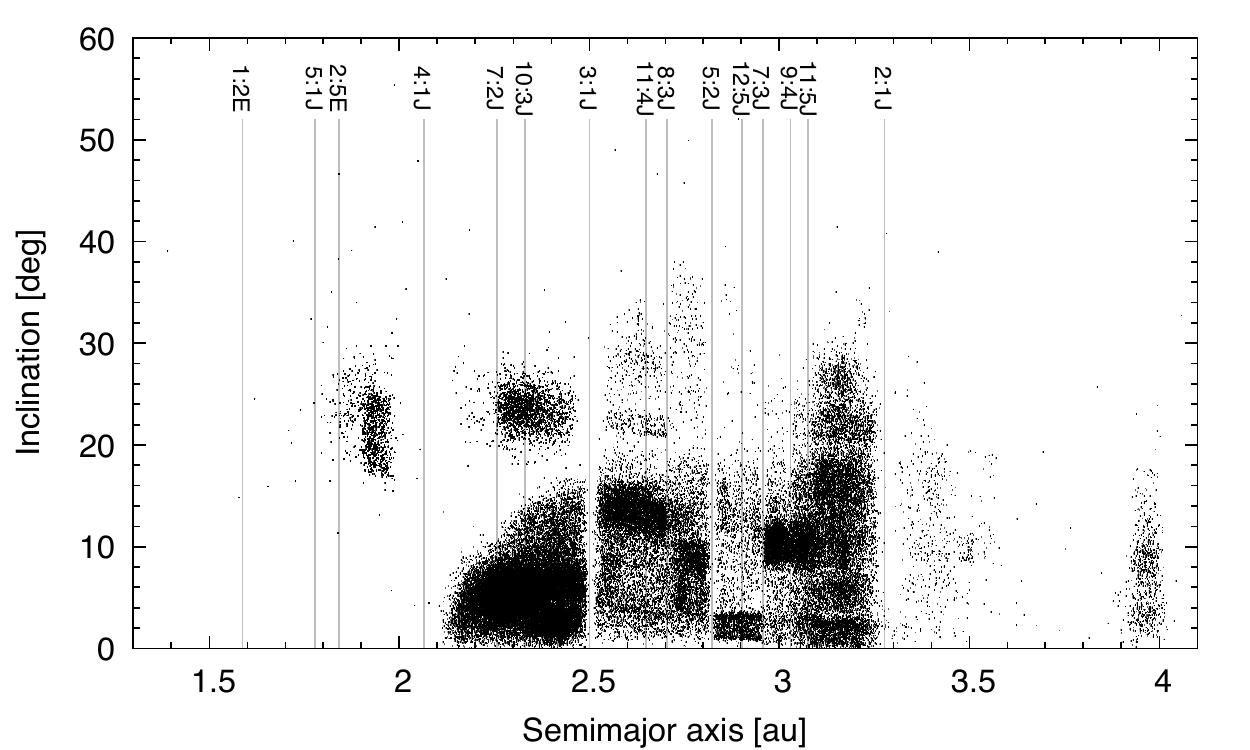}
  \includegraphics[width=\columnwidth]{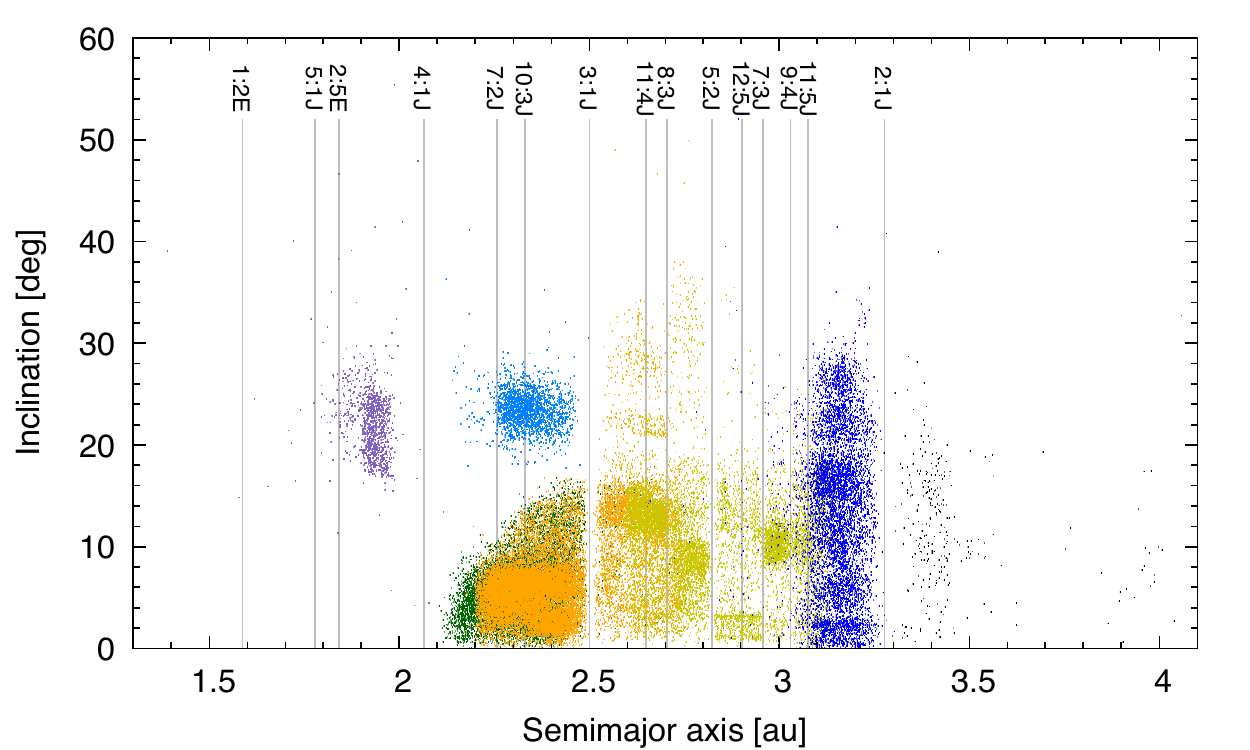}
  \includegraphics[width=\columnwidth]{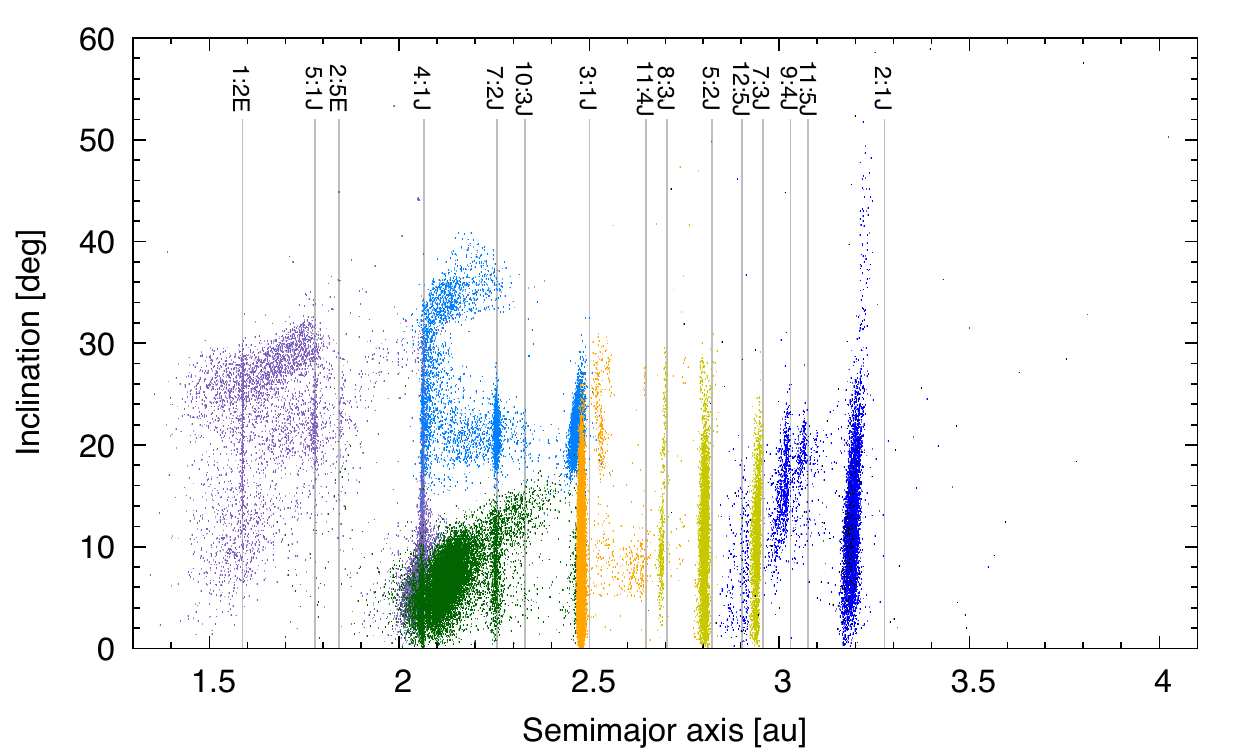}
  \caption{(Top) Initial ($a$,$i$) distribution of test
    asteroids. (Middle) Initial ($a$,$i$) distribution for test
    asteroids that entered the NEO region ($q=1.3\au$) during the
    100-Myr integration. (Bottom) ($a$,$i$) distribution of test
    asteroids at the time they entered the NEO region
    ($q=1.3\au$). The color coding in the middle and bottom plots
    correspond to the nominal set of ERs defined in
    Sec.~\ref{sec:sourceselection}. The ERs were defined based on
    initial orbital elements (Hungarias and Phocaeas) or on orbital
    elements at the epoch when the test asteroids enter the NEO region
    (the $\nu_6$, 3:1J, 5:2J, and 2:1J complexes).}
  \label{fig:escaperoutes}
\end{figure}

To guarantee a reasonable accuracy for the orbital elements and $H$
magnitudes \citet{gra2017a} also required that the selected objects
have been observed for at least 30 days, which translates to a
relative uncertainty of about 1\% for semimajor axis, eccentricity and
inclination for MBOs \citep{mui2006a}. It is well known that the $H$
magnitudes may have errors of some tenths of a magnitude. However,
what is important for the present study is that any systematic effects
affect the entire sample in the same way, so that the $H$ cut is done
in a similar fashion throughout the MAB. In the end \citet{gra2017a}
were left with a sample of 92,449 MBO orbits where Hungaria and
Phocaea test asteroids were cloned 7 and 3 times, respectively
(Fig.~\ref{fig:escaperoutes}, top and middle). They then assigned a
diameter of 100~m and a random spin obliquity of $\pm90\deg$ to each
test asteroid. The test asteroids were integrated with a 1-day
timestep for 100 Myr under the influence of a Yarkovsky-driven
semimajor-axis drift and accounting for gravitational perturbations by
all planets (Mercury through Neptune). During the course of the
integrations 70,708 test asteroids entered the NEO region ($q <
1.3\au$) and their orbital elements were recorded with a time
resolution of 10~kyr.

The orbital elements ($a,e,i,\Omega,\omega,M_0$) at the MBO-NEO
boundary ($q=1.3\au$) define the locations of the escape routes from
the MB and form the initial conditions for the NEO residence-time
integrations (Fig.~\ref{fig:escaperoutes}, bottom). We cloned the test
asteroids associated with the $\nu_{6,o}$ SR, and the 7:2J and 8:3J
MMRs 5 times to increase the sample in these otherwise undersampled
ERs. The total number of test asteroids was thus increased to 80,918.

Our approach to limit ourselves to only 100-m-diameter test asteroids
could be problematic because Yarkovsky drift in the MAB may affect the
resultant NEO steady-state orbit distribution.  The effect would arise
because different drift rates imply that asteroids drifting into
resonances will spend a different amount of time in or close to the
resonances. In cases where the bodies are drifting slowly, they could
become trapped in tiny resonances and pushed out of the MAB prior to
when our model results predict
\citep[e.g.,][]{1998AJ....116.3029N,bot2002b}.  In other cases, SRs
such as the $\nu_6$ with the adjacent $\nu_{16}$ can change the
inclination of the asteroid
\citep{1986A&A...166..326F,1986A&A...170..138S} and the amount of
change depends on the time it takes for the asteroid to evolve to the
NEO region.  Unfortunately, we are not yet at the point where full-up
models including accurate representations of the Yarkovsky and YORP
effects can be included for tens of thousands of asteroids across the
MAB. Our work in this paper represents a compromise between getting
the dynamics as correct as possible and ensuring computational
expediency. Our main concern here is that using a drift rate that
varies with size could lead to steady-state orbit distributions that
are correlated with asteroid diameter.

To test this scenario we selected the test asteroids in set 'B' in
\citet{gra2017a} that entered the NEO region and produced steady-state
distributions for $D=0.1\km$ and $D=3\km$ NEOs that escape through the
3:1J MMR and the $\nu_6$ SR. There was thus a factor of 30 difference
in semimajor-axis drift rate. To save time we decided to continue the
integrations for only up to 10 Myr instead of integrating all test
asteroids until they reach their respective sinks. Since the average
lifetime of all NEOs is $\lesssim10$~Myr this choice nevertheless
allowed most test asteroids to reach their sinks. We then discretized
and normalized the distributions (Eq.~\ref{eq:resnorm}), and computed
the difference between the distributions for small and large test
asteroids.

In the case of the 3:1J MMR we found no statistically significant
differences in NEO steady-state orbit distributions between large and
small test asteroids when compared to the noise (Supplementary
Fig.~2). For $\nu_6$ we found that while the ``signal'' is stronger
compared to 3:1J so is the noise for the $D=3\km$ case. A priori we
would expect a stronger effect for the $\nu_6$ resonance because the
semi-major axis drift induced by the Yarkovsky effect can change the
position of the asteroid relative to an SR, but not relative to an
MMR, which reacts adiabatically. However, based on our numerical
simulations we concluded that the Yarkovsky drift in the MAB results
in changes in the NEO steady-state orbit distributions that are
negligible for the purposes of our work. Therefore we decided to base
the orbital integrations, that are required for constructing the
steady-state NEO orbit distributions, on the test asteroids with
$D=0.1\km$ that escape the MAB \citep{gra2017a}.

\subsection{Orbital evolution of NEOs originating in the MAB}

Next we continued the forward integration of the orbits of the 80,918
100-meter-diameter test asteroids that entered the NEO region with a
slightly different configuration as compared to the MBO integrations
described in the previous subsection. To build smooth orbit
distributions we recorded the elements of all test asteroids with a
time resolution of 250~yr. The average change in the orbital elements
over 250~yr based on our integrations is $\Delta a = 0.004\au$,
$\Delta e = 0.006$, and $\Delta i = 0.728\deg$. The average change
sets a limit on the resolution of the discretized orbit distribution
in order to avoid artefacts, although the statistical nature of the
steady-state orbit distribution softens discontinuities in the orbital
tracks of individual test asteroids.

A non-zero Yarkovsky drift in semimajor axis has been measured for
tens of known NEOs
\citep{2003Sci...302.1739C,2012AJ....144...60N,2013Icar..224....1F,2015aste.book..509V},
but the common assumption is that over the long term the Yarkovsky
effect on NEO orbits is dwarfed by the strong orbital perturbations
caused by their frequent and close encounters with terrestrial
planets. For a km-scale asteroid the typical measured drift in
semimajor axis caused by the Yarkovsky effect is
$\sim2\times10^{-4}\au\Myr^{-1}$ or
$\sim5\times10^{-8}\au\,(250\yr)^{-1}$ whereas the average change of
semimajor axis for NEOs from (size-independent) gravitational
perturbations is $\sim4\times10^{-3}\au\,(250\yr)^{-1}$. The rate of
change of the semimajor axis caused by Yarkovsky is thus several
orders of magnitude smaller than that caused by gravitational
perturbations. We concluded that the effect of the Yarkovsky drift on
the NEO steady-state orbit distribution is negligible compared to the
gravitational perturbations caused by planetary encounters. Hence we
omitted Yarkovsky modeling when integrating test asteroids in the NEO
region.

Integrations using a 1-day timestep did not correctly resolve close
solar encounters. This results in the steady-state orbital
distribution at low $a$ and large $e$ to be very "unstable", as one
can see by comparing the distributions in the top left corners of the
($a,e$) plots in Supplementary Fig.~1, obtained by selecting
alternatively particles with even or odd identification numbers. So we
reduced the nominal integration timestep to 12 hours and restarted the
integrations. These NEO integrations required on the order of 2
million CPU hours and solved the problem (see top left corners of
left-hand-side ($a,e$) plots in
Figs.~\ref{fig:6astsourcedefresolutiona} and
\ref{fig:6astsourcedefresolutionb}). We note that \citet{gre2012a}
used a timestep of only 4 hours to ensure that encounters with Venus
and Earth are correctly resolved even in the fastest
encounters. Considering that the required computing time would have
tripled if we had used a four-hour integration step and considering
that all the evidence we have suggests that the effect is negligible,
we saw no obvious reason to reduce the timestep by an additional
factor of 3. We stress, however, that the integration step is
automatically substantially reduced when the integrator detects a
planetary encounter.

The orbital integrations continued until every test asteroid had
collided with the Sun or a planet (Mercury through Neptune), escaped
the solar system or reached a heliocentric distance in excess of
$100\au$. For the last possibility we assume that the likelihood of
the test asteroid re-entering the NEO region ($a<4.2\au$) is
negligible as it would have to cross the outer planet region without
being ejected from the solar system or colliding with a planet. The
longest lifetimes among the integrated test asteroids are several Gyr.

Out of the 80,918 test asteroids about to enter the NEO region we
followed 79,804 (98.6\%) to their respective sink. The remaining 1.4\%
of the test asteroids did not reach a sink for reasons such as ending
up in a stable orbit with $q>1.3\au$. As the Yarkovsky drift was
turned off these orbits were found to remain virtually stable over
many Gyr and thus the test asteroids were unable to drift into
resonances that would have brought them back into the NEO region. In
addition, the output data files of some test asteroids were corrupted,
and in order not to skew the results we omitted the problematic test
asteroids when constructing the NEO steady-state orbit distributions.

\subsubsection{NEO lifetimes and sinks}

As expected, the most important sinks are i)
a collision with the Sun
and ii) an escape from the inner solar system after a close encounter
with, primarily, Jupiter (Table \ref{table:sinksmeanlife}).

\begin{sidewaystable}[h]
  \small
\caption{Fraction of test asteroids ending up in various sinks
  (collisions with the Sun or planets, or an escape from the inner
  solar system, $a>100\au$) and mean lifetime of test asteroids in the
  NEO region ($q<1.3\au$, $a<4.2\au$, $e<1$, $i<180\deg$) as a
  function of the ER that is provided in the left column. We provide
  four different scenarios for each ER-sink pair: no thermally-driven
  destruction at small $q$, and thermally-driven destruction when
  $q<0.058\au$, $q<0.076\au$, $q<0.184\au$. The three perihelion
  distances correspond to km-scale, average, and dm-scale NEOs
  \protect \citep{2016Natur.530..303G}. The horizontal lines divides
  the ERs into the six asteroidal groups that are used for the nominal
  model in Sec.~\ref{sec:nominal7smodel}. Note that the relatively
  high fraction of Earth-impacts from 11:5J is due to small number
  statistics.}\label{table:sinksmeanlife}
\begin{tabular}{cccccccccc}
\hline
ER & \# of test & Sun            & Mercury       & Venus         & Earth         & Mars          & Jupiter & Escape         &  $\langle L \rangle$ \\
   & asteroids  & [\%]           & [\%]          & [\%]          & [\%]          & [\%]          & [\%]    & [\%]           &          [Myr]  \\ 
\hline
\multicolumn{10}{c}{\bf Hungaria} \\
Hung. & 8128   &  77.5/0.0/0.0/0.0  &  1.1/0.7/0.6/0.3  &  6.2/4.9/4.6/3.4  &  6.4/5.4/5.2/4.3  &  2.2/2.1/2.0/2.0  &  0.1/0.1/0.1/0.1  &  6.9/5.8/5.6/4.1      &  43.3/37.2/36.0/32.5 \\
\hline
\multicolumn{10}{c}{\bf $\nu_6$ complex} \\
$\nu_6$ & 13029  &  81.2/0.0/0.0/0.0  &  0.5/0.4/0.3/0.1  &  2.5/2.0/1.9/1.4  &  2.3/1.9/1.8/1.5  &  0.4/0.4/0.4/0.4  &  0.2/0.2/0.2/0.1  &  12.9/11.8/11.3/8.4   &  9.4/7.2/6.8/5.5 \\
4:1     & 1899   &  75.8/0.0/0.0/0.0  &  0.7/0.5/0.5/0.2  &  4.4/3.7/3.6/2.5  &  3.7/3.2/3.1/2.8  &  0.2/0.2/0.2/0.2  &  0.1/0.1/0.1/0.1  &  15.2/14.0/13.5/9.7   &  10.2/8.5/8.2/6.6 \\
7:2     & 4969   &  83.2/0.0/0.0/0.0  &  0.1/0.1/0.1/0.1  &  1.2/0.9/0.8/0.5  &  1.3/1.1/1.1/0.9  &  0.3/0.3/0.3/0.2  &  0.3/0.3/0.3/0.2  &  13.7/12.5/12.1/8.9   &  6.7/5.5/5.3/4.2 \\
\hline
\multicolumn{10}{c}{\bf Phocaea} \\
Phoc. & 8309   &  89.5/0.0/0.0/0.0  &  0.2/0.1/0.1/0.0  &  1.3/1.0/0.9/0.5  &  1.0/0.6/0.6/0.4  &  0.4/0.3/0.3/0.3  &  0.1/0.0/0.0/0.0  &  7.9/5.0/4.7/2.9      &  14.3/11.2/10.7/8.9 \\
\hline
\multicolumn{10}{c}{\bf 3:1J complex} \\
3:1J & 7753   &  76.1/0.0/0.0/0.0  &  0.1/0.0/0.0/0.0  &  0.4/0.4/0.3/0.2  &  0.2/0.1/0.1/0.1  &  0.0/0.0/0.0/0.0  &  0.6/0.5/0.5/0.4  &  22.6/20.4/19.7/14.6  &  2.2/1.5/1.3/0.9 \\
$\nu_{6,o}$ & 3792   &  70.9/0.0/0.0/0.0  &  0.0/0.0/0.0/0.0  &  0.2/0.1/0.1/0.1  &  0.3/0.2/0.2/0.2  &  0.2/0.2/0.2/0.2  &  0.3/0.3/0.3/0.3  &  28.1/26.1/25.3/22.4  &  3.5/2.5/2.3/1.9 \\
\hline
\multicolumn{10}{c}{\bf 5:2J complex} \\
8:3J & 3080   &  45.1/0.0/0.0/0.0  &  0.0/0.0/0.0/0.0  &  0.0/0.0/0.0/0.0  &  0.2/0.2/0.2/0.2  &  0.0/0.0/0.0/0.0  &  1.0/1.0/1.0/1.0  &  53.6/50.6/49.9/46.1  &  2.2/1.7/1.6/1.4 \\
5:2J & 5955   &  16.9/0.1/0.1/0.1  &  0.0/0.0/0.0/0.0  &  0.1/0.1/0.1/0.1  &  0.1/0.1/0.1/0.1  &  0.0/0.0/0.0/0.0  &  1.2/1.1/1.1/1.1  &  81.8/79.4/78.8/75.0  &  0.5/0.4/0.4/0.3 \\
7:3J & 2948   &  7.4/0.0/0.0/0.0   &  0.0/0.0/0.0/0.0  &  0.0/0.0/0.0/0.0  &  0.0/0.0/0.0/0.0  &  0.0/0.0/0.0/0.0  &  1.1/1.1/1.1/1.1  &  91.5/90.2/89.7/87.7  &  0.2/0.2/0.2/0.2 \\
\hline
\multicolumn{10}{c}{\bf 2:1J complex} \\
9:4J & 443    &  25.3/0.0/0.0/0.0  &  0.0/0.0/0.0/0.0  &  0.0/0.0/0.0/0.0  &  0.0/0.0/0.0/0.0  &  0.0/0.0/0.0/0.0  &  0.9/0.9/0.9/0.9  &  73.8/69.3/68.2/63.0  &  0.5/0.4/0.4/0.3 \\
$z_2$ & 929    &  14.3/0.1/0.1/0.1  &  0.0/0.0/0.0/0.0  &  0.0/0.0/0.0/0.0  &  0.0/0.0/0.0/0.0  &  0.0/0.0/0.0/0.0  &  1.3/1.2/1.2/1.0  &  84.4/82.1/81.5/77.7  &  0.6/0.3/0.3/0.3 \\
11:5J & 133    &  43.6/0.0/0.0/0.0  &  0.0/0.0/0.0/0.0  &  0.0/0.0/0.0/0.0  &  0.8/0.8/0.8/0.8  &  0.0/0.0/0.0/0.0  &  0.8/0.8/0.0/0.0  &  54.9/47.4/45.9/39.8  &  0.5/0.4/0.4/0.4 \\
2:1J & 5321   &  22.1/0.3/0.2/0.1  &  0.0/0.0/0.0/0.0  &  0.0/0.0/0.0/0.0  &  0.1/0.1/0.1/0.1  &  0.0/0.0/0.0/0.0  &  1.0/0.9/0.9/0.8  &  76.8/70.8/69.2/61.5  &  0.5/0.4/0.4/0.3 \\
\hline
\end{tabular}
\end{sidewaystable}

The estimation of NEO lifetimes, that is, the time asteroids and
comets spend in the NEO region before reaching a sink when starting
from the instant when they enter the NEO region (that is,
$q\leq1.3\au$ for the first time), is complicated by the fact that
NEOs are also destroyed by thermal effects
\citep{2016Natur.530..303G}. The typical heliocentric distance for a
thermal disruption depends on the size of the asteroid. For large
asteroids with $D\gtrsim1\km$ the typical perihelion distance at which
the disruption happens is $q\sim0.058\au$. We see a 10--50\%
difference in NEO lifetimes when comparing the results computed with
and without thermal disruption (Table \ref{table:sinksmeanlife}). For
smaller asteroids the difference is even greater because the
disruption distance is larger.

We define the mean lifetime of NEOs to be the average time it takes
for test asteroids from a given ER to reach a sink when starting from
the time that they enter the NEO region. Our estimate for the mean
lifetime of NEOs originating in the 3:1J MMR is comparable to that
provided by \citet{bot2002a}, but for $\nu_6$ and the resonances in
the outer MAB our mean lifetimes are about 50\% and 200\% longer,
respectively (Table \ref{table:sinksmeanlife}).  These differences
can, potentially, be explained by different initial conditions and the
longer timestep used for the integrations carried out by
\citet{bot2002a}. A longer timestep would have made the orbits more
unstable than they really are and close encounters with terrestrial
planets would not have been resolved. An accurate treatment of close
encounters would pull out test asteroids from resonances whereas an
inability to do this would lead to test asteroids rapidly ending up in
the Sun and thereby also to a shorter average lifetime.

\subsubsection{Steady-state orbit distributions and their uncertainties}

\begin{figure*}
  \centering
  {\bf Hungarias} \vspace{1mm} \\
  \includegraphics[width=\columnwidth]{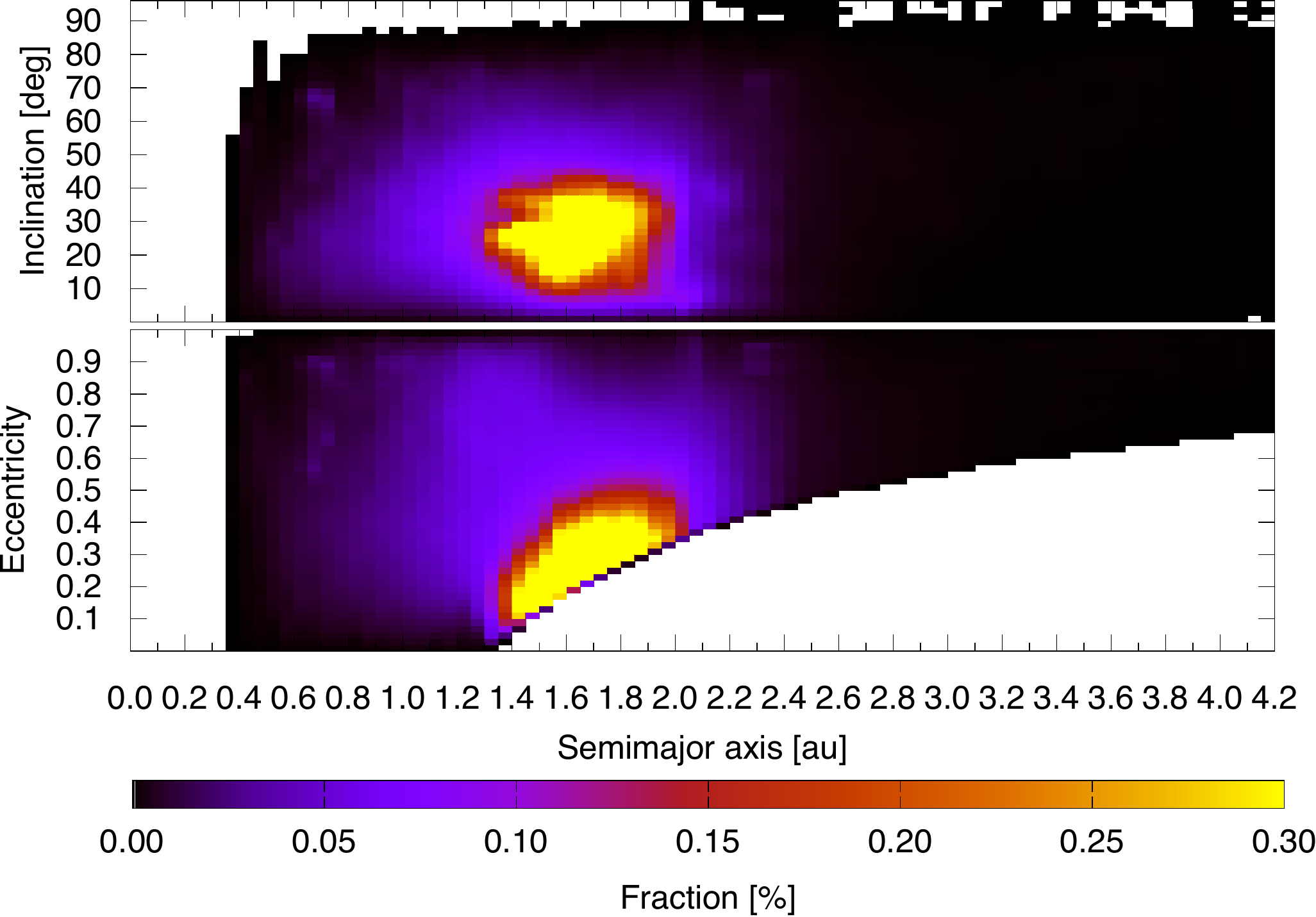} \vspace{2mm}
  \includegraphics[width=\columnwidth]{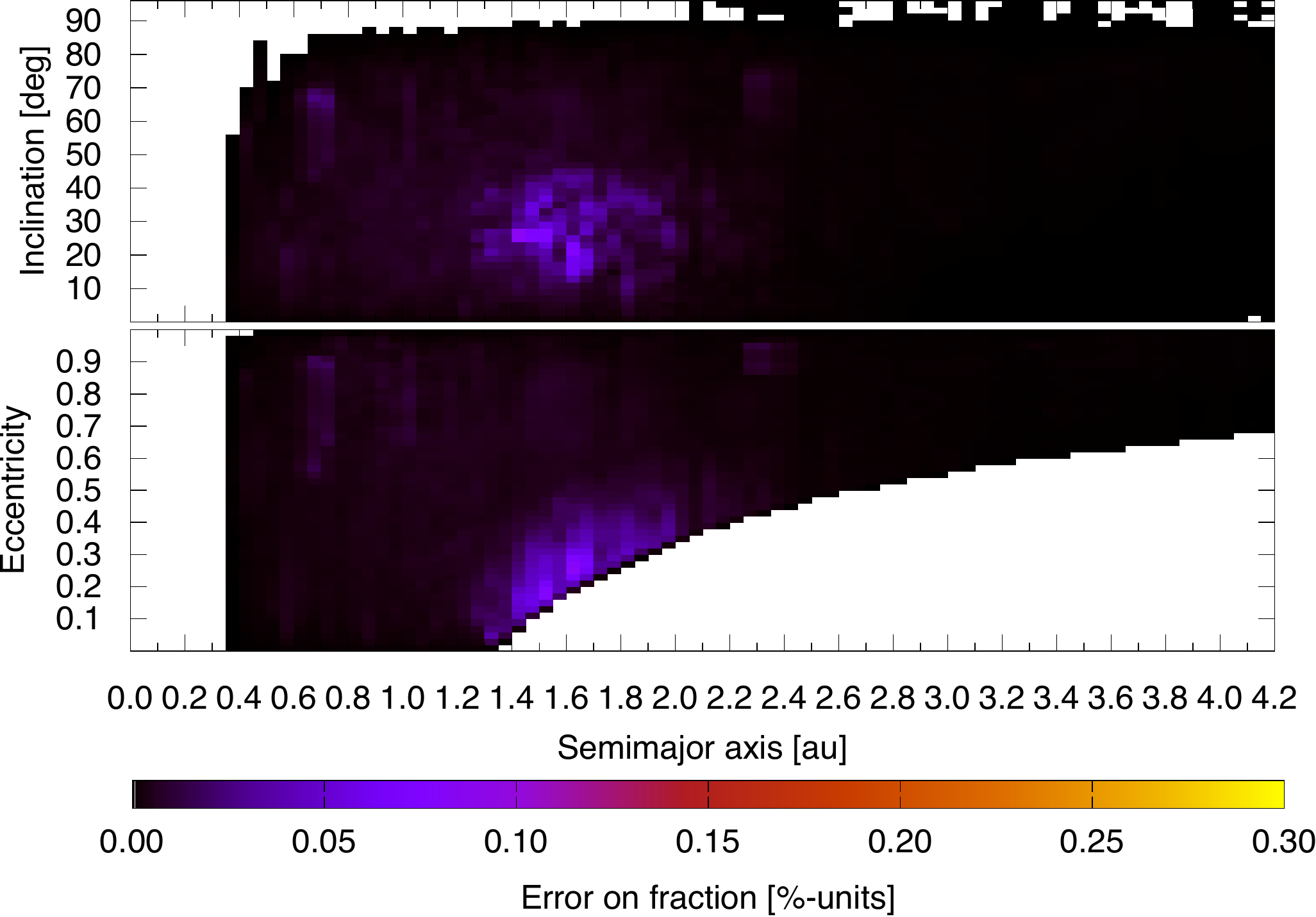} \\
  {\bf $\nu_6$ complex} \vspace{1mm} \\
  \includegraphics[width=\columnwidth]{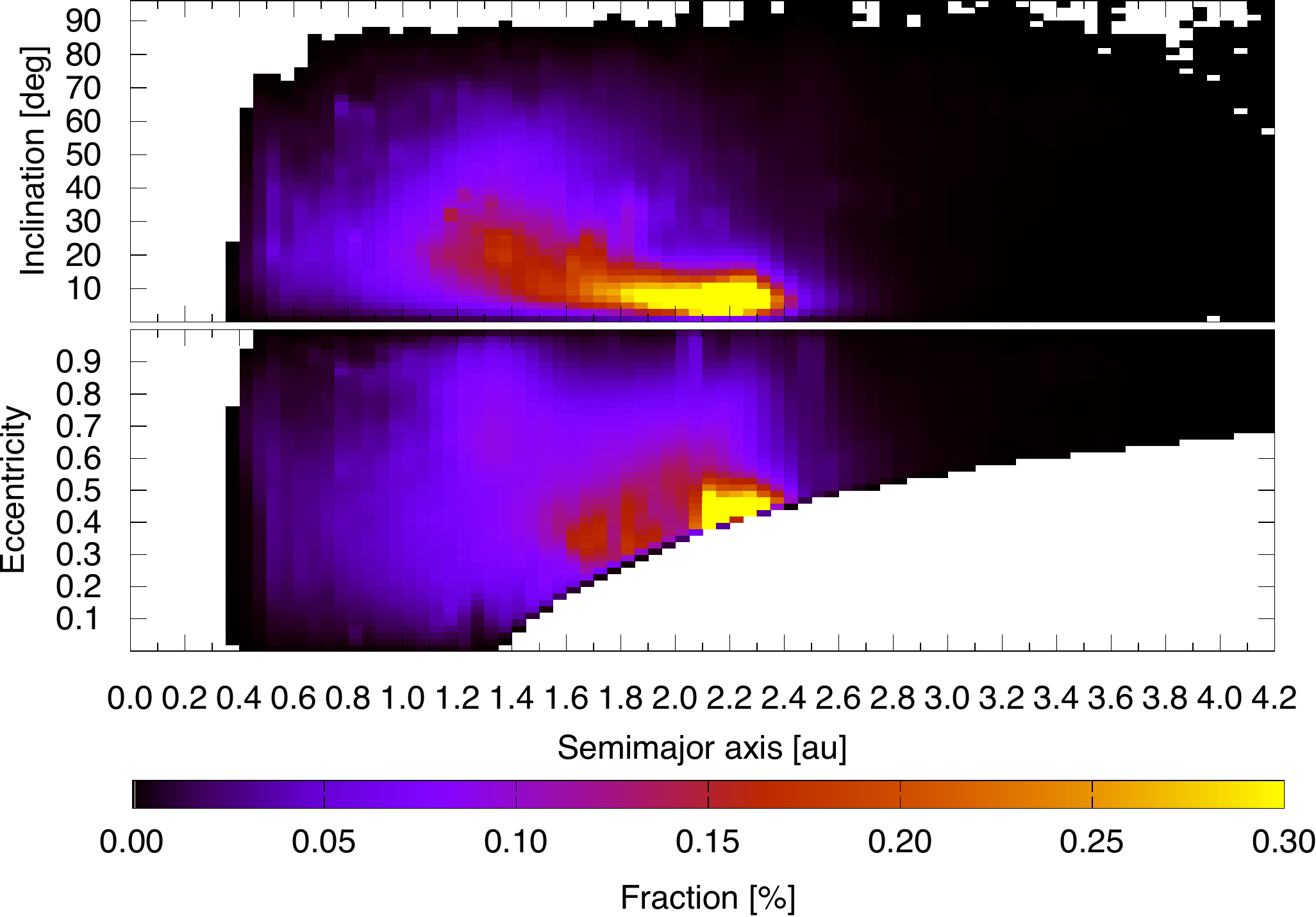} \vspace{3mm}
  \includegraphics[width=\columnwidth]{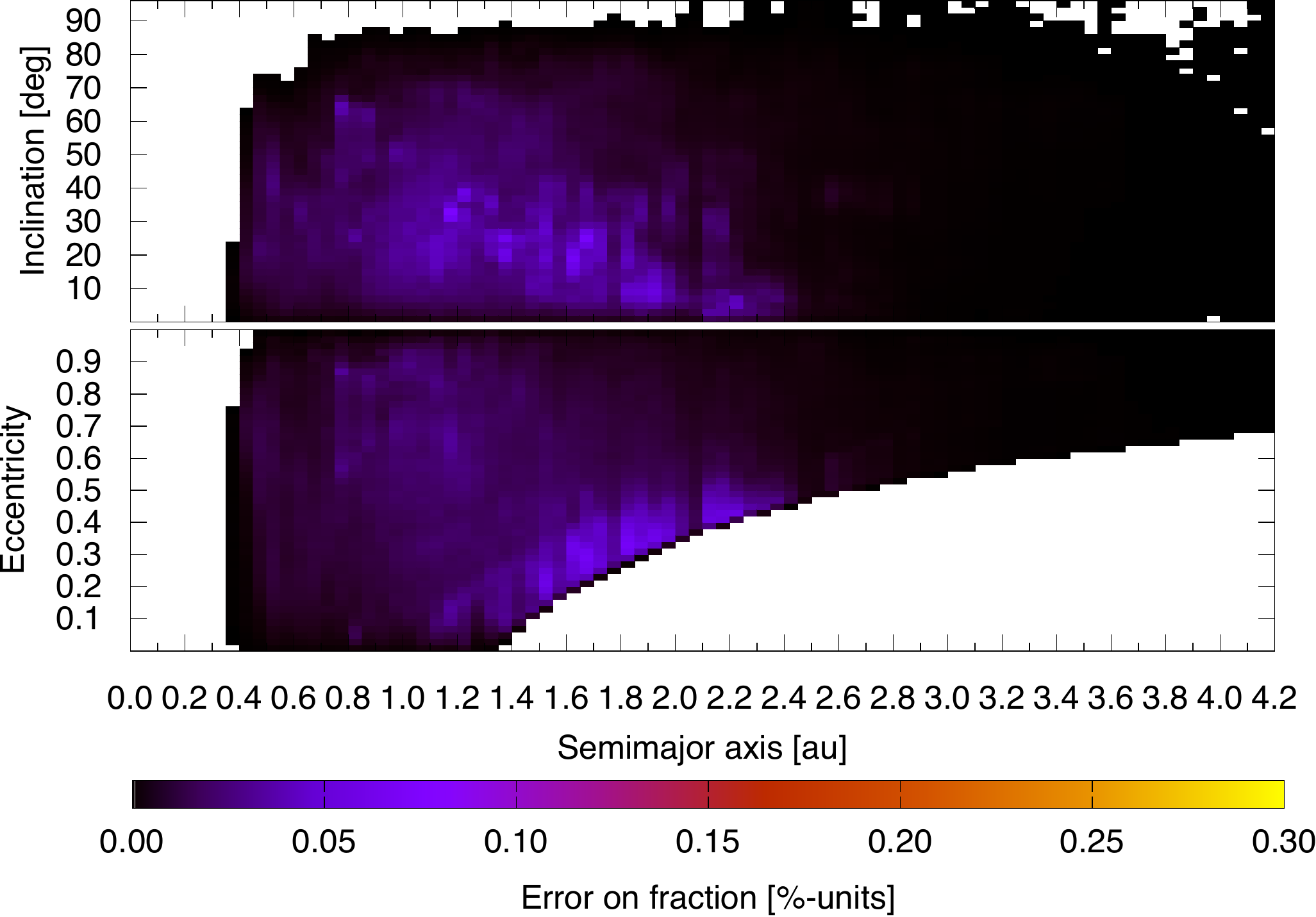} \\
  {\bf Phocaeas} \vspace{1mm} \\
  \includegraphics[width=\columnwidth]{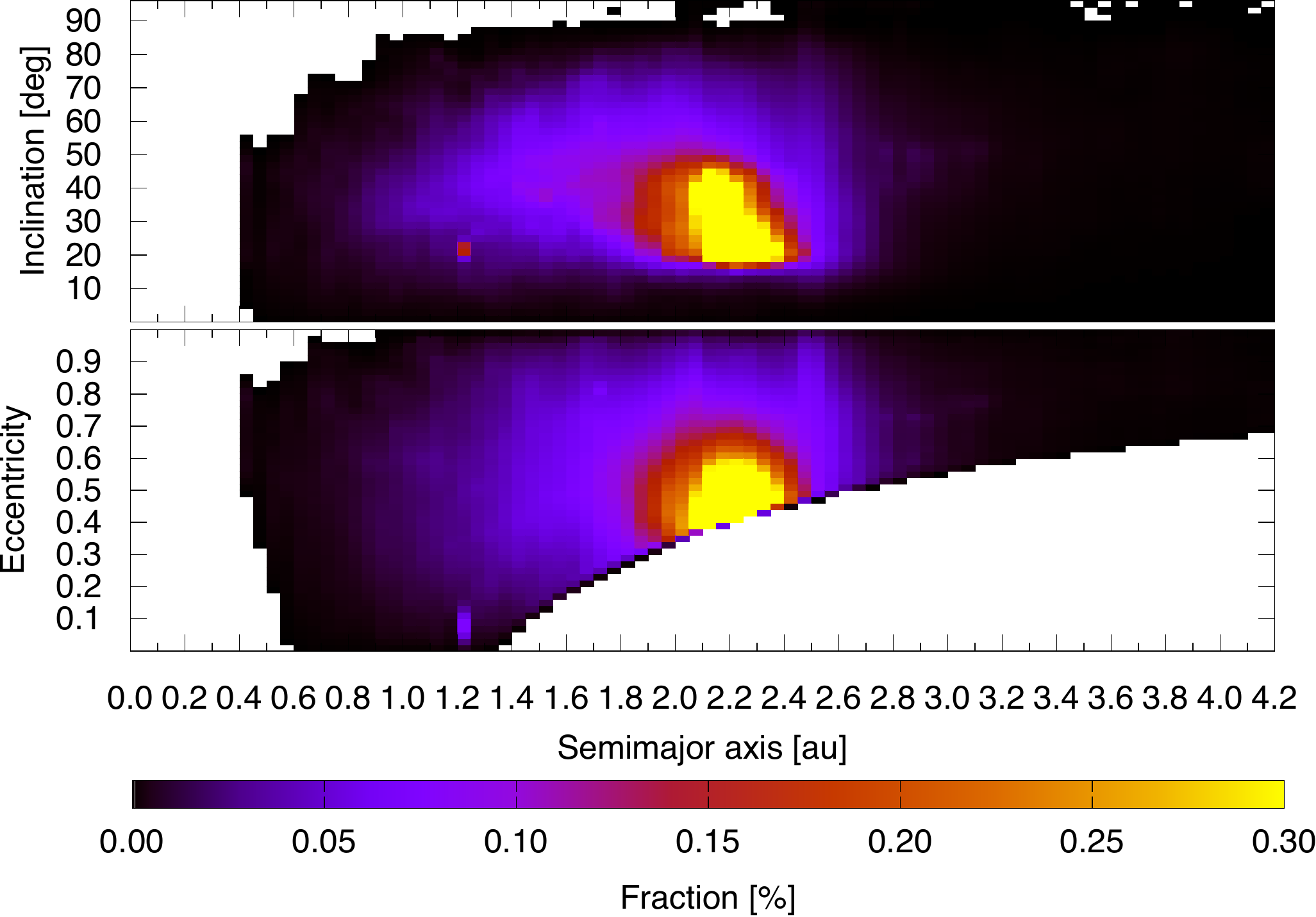}
  \includegraphics[width=\columnwidth]{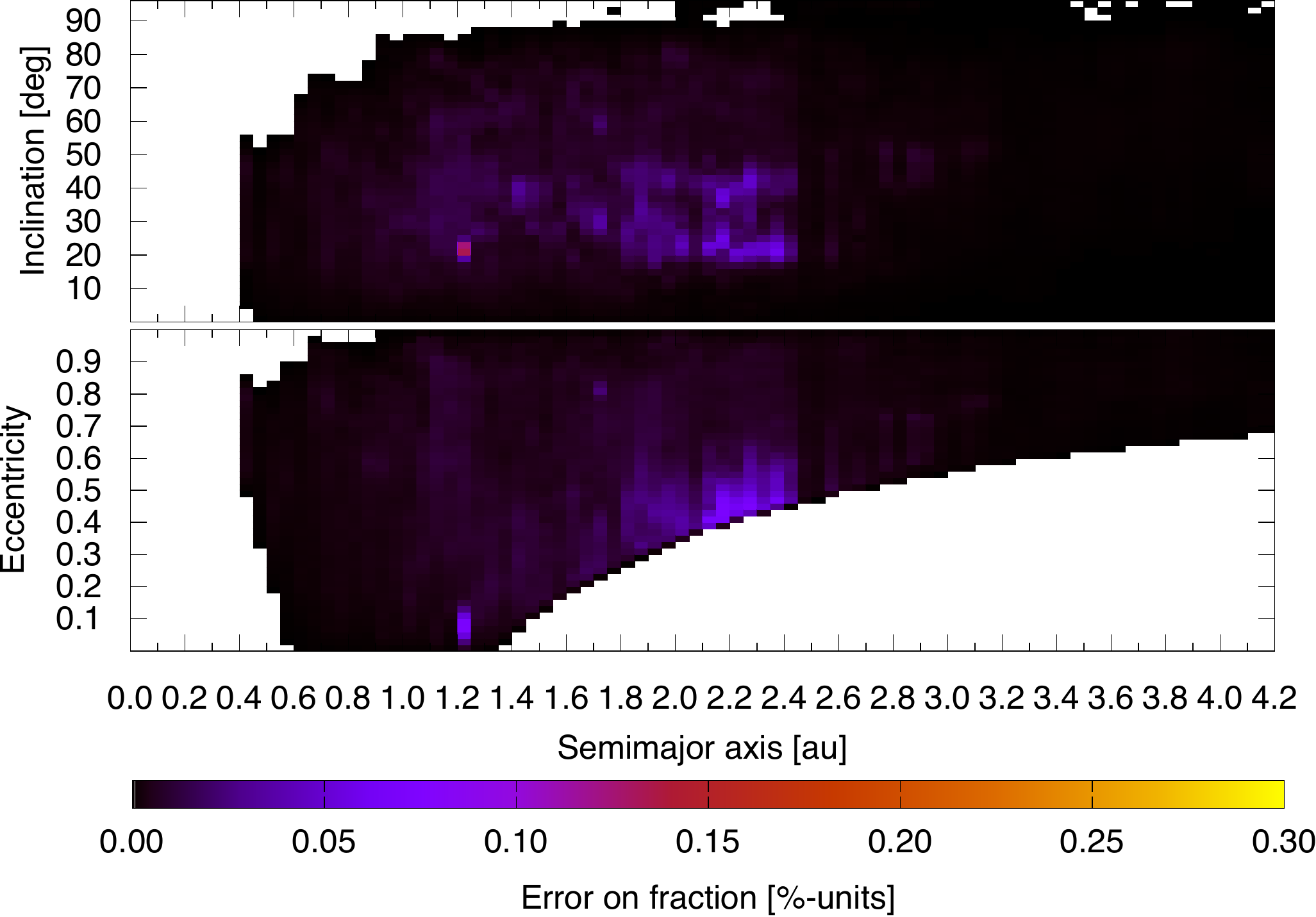}
  \caption{Steady-state orbit distributions (left) and the
    corresponding uncertainty distributions (right) for NEOs
    originating in asteroidal ERs: Hungarias (top panel), $\nu_6$
    complex (middle panel), and Phocaeas (bottom panel).}
  \label{fig:6astsourcedefresolutiona}
\end{figure*}

\begin{figure*}
  \centering
  {\bf 3:1J complex} \vspace{1mm} \\
  \includegraphics[width=\columnwidth]{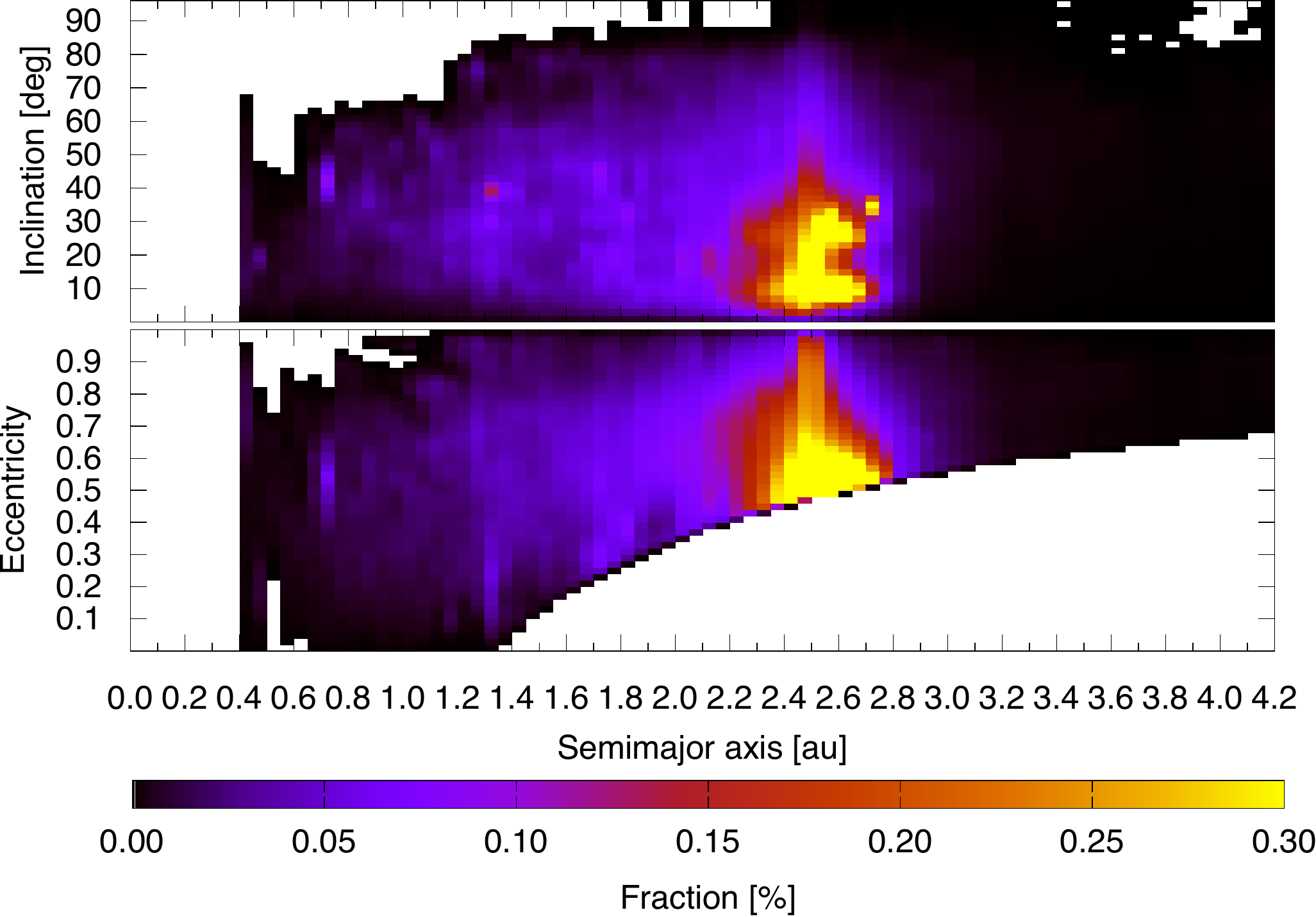} \vspace{2mm}
  \includegraphics[width=\columnwidth]{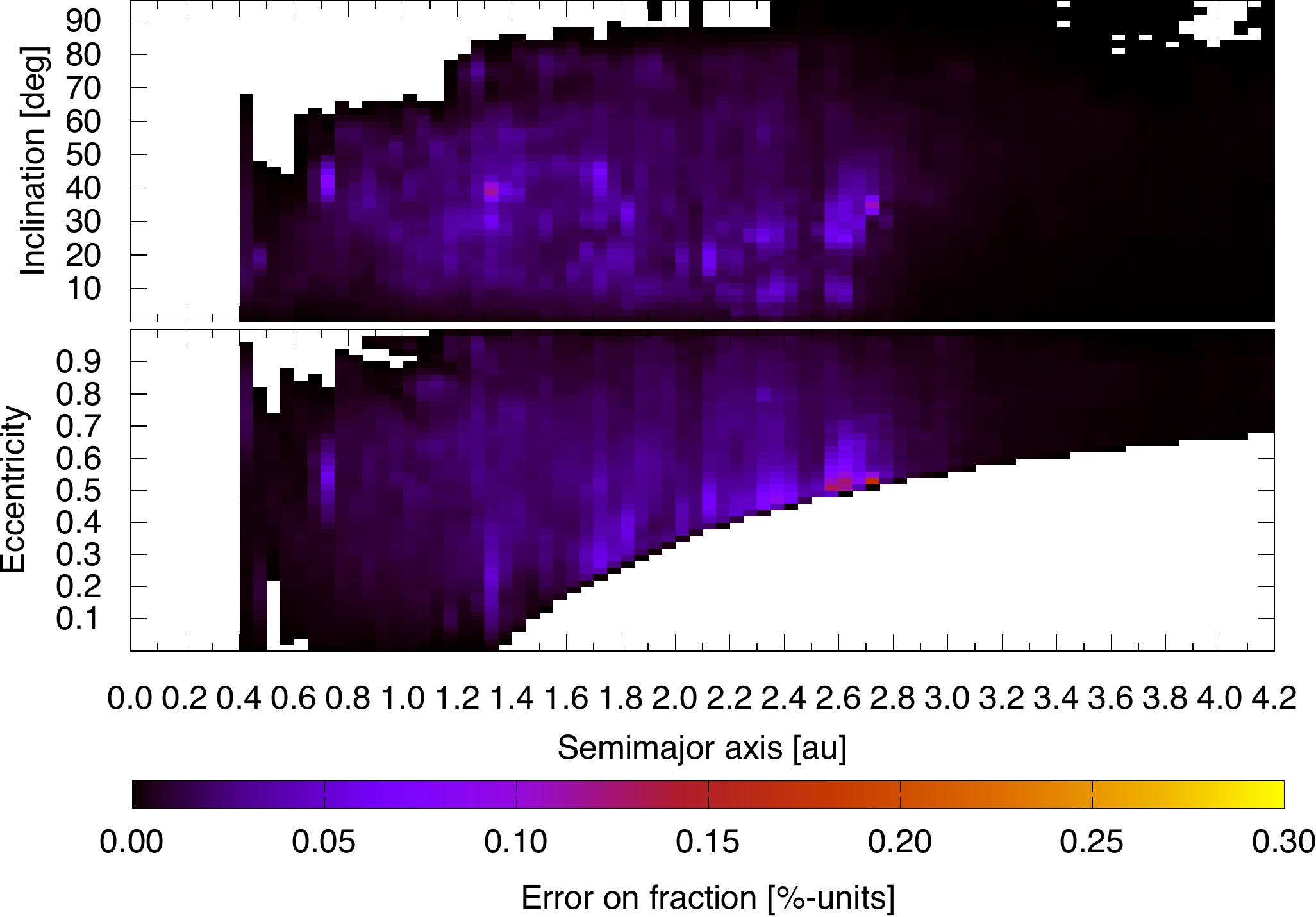}
  {\bf 5:2J complex} \vspace{1mm} \\
  \includegraphics[width=\columnwidth]{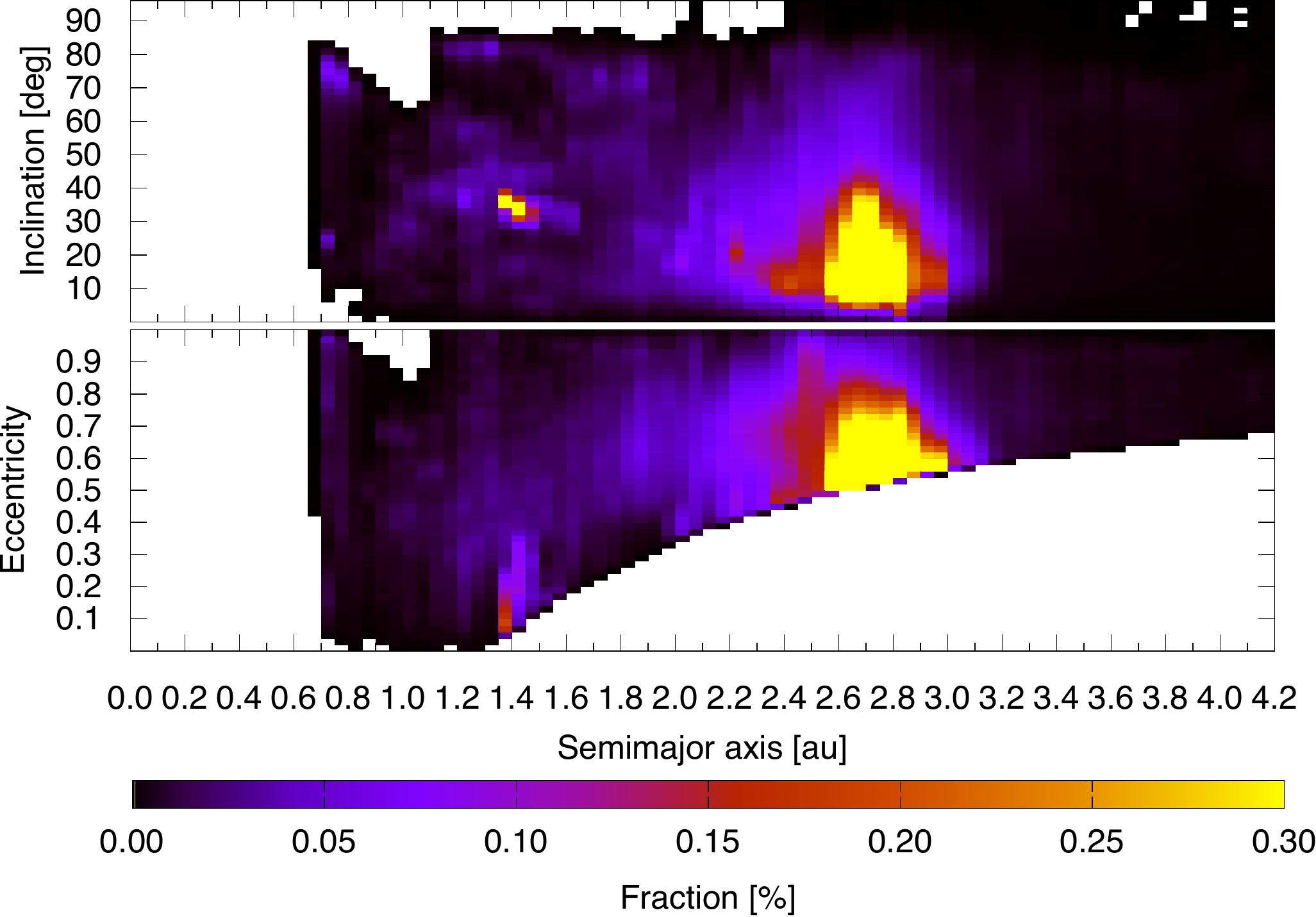} \vspace{2mm}
  \includegraphics[width=\columnwidth]{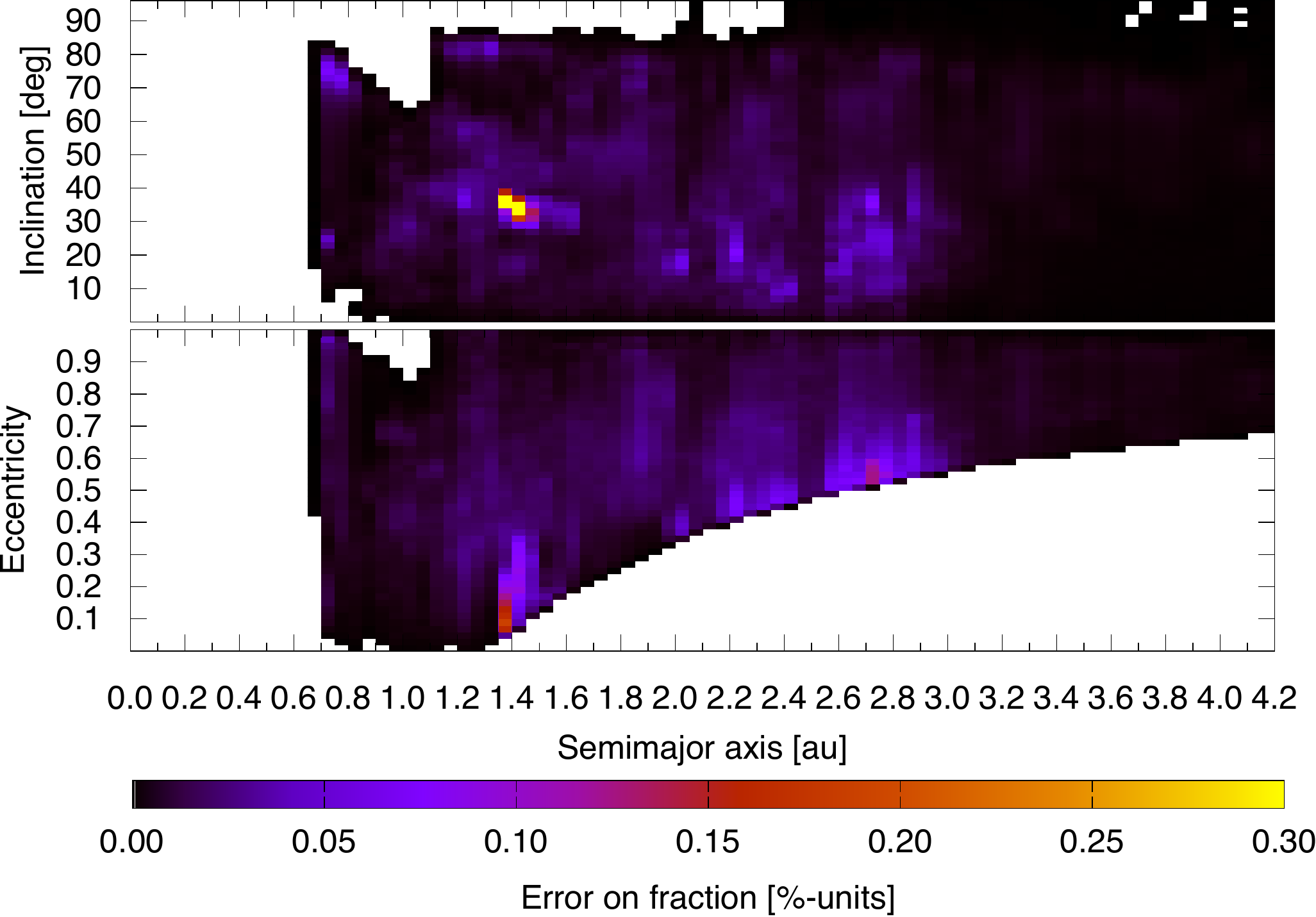}
  {\bf 2:1J complex} \vspace{1mm} \\  \includegraphics[width=\columnwidth]{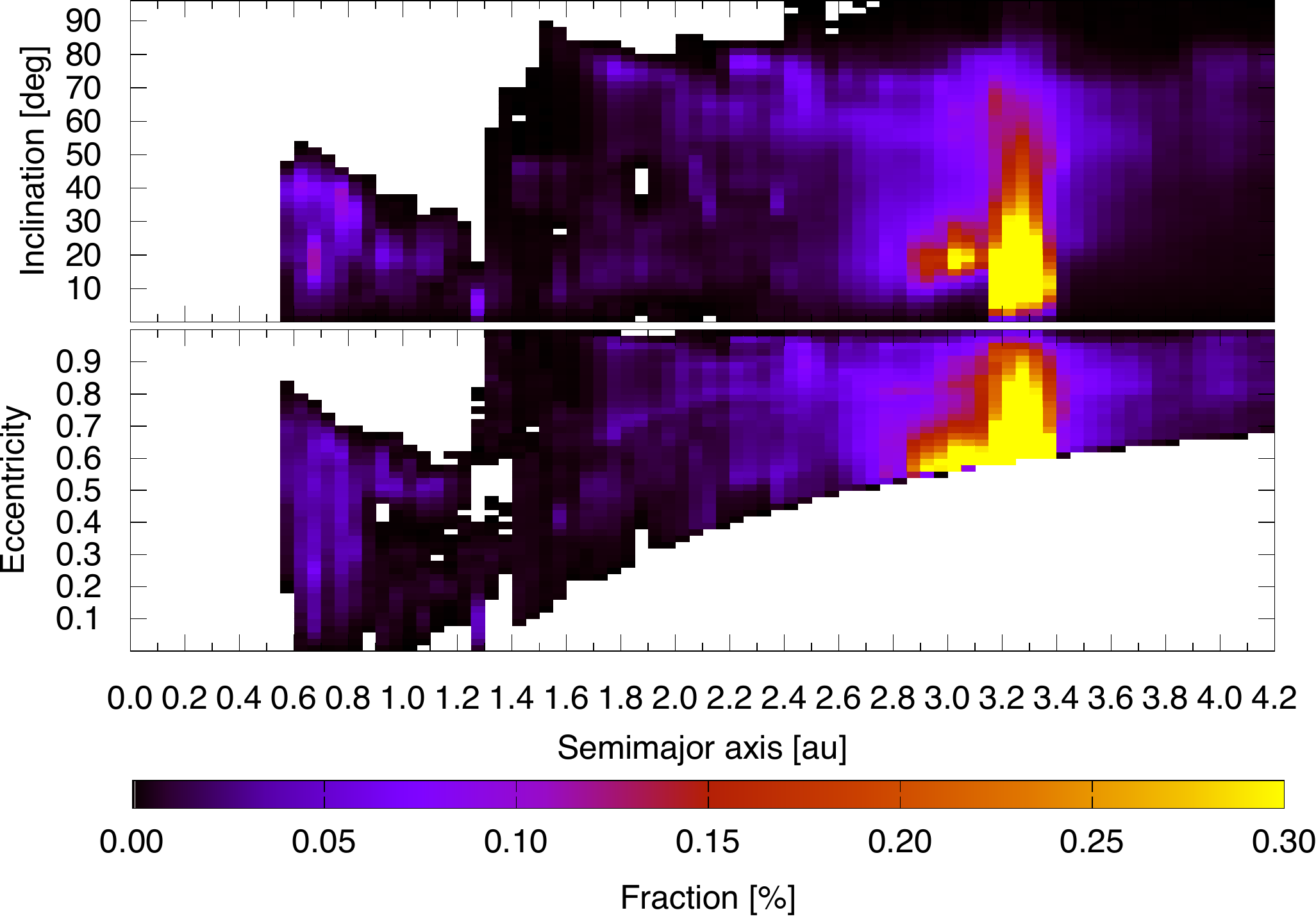}
  \includegraphics[width=\columnwidth]{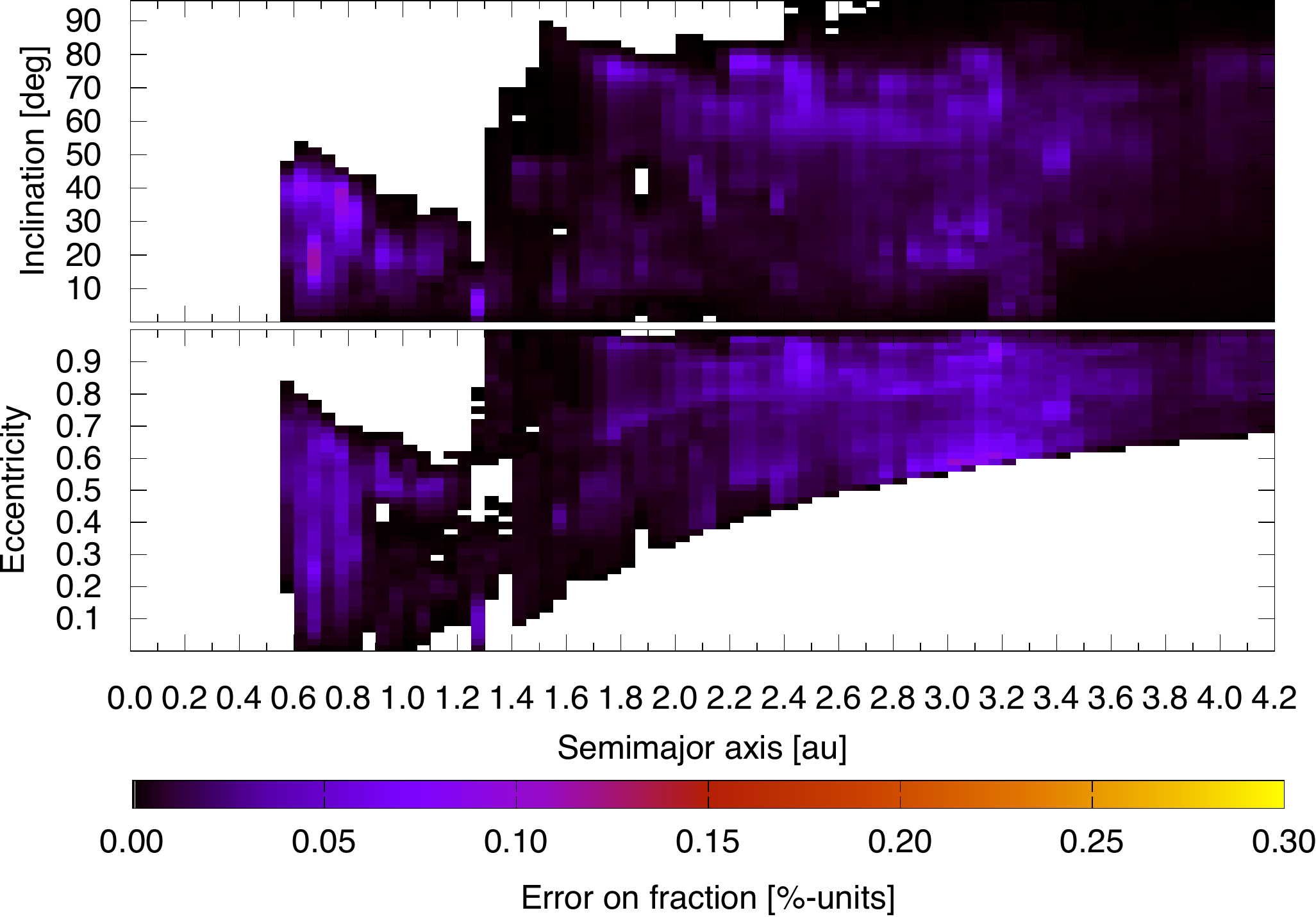}
  \caption{Steady-state orbit distributions (left) and the
    corresponding uncertainty distributions (right) for NEOs
    originating in asteroidal ERs: 3:1J (top panel), 5:2J complex
    (middle panel), and 2:1J complex (bottom panel).}
  \label{fig:6astsourcedefresolutionb}
\end{figure*}

We combined the evolutionary tracks of test asteroids that enter the
NEO region through 12 ERs and from 2 additional SRs into 6
steady-state orbit distributions by summing up the time that the test
asteroids spend in various parts of the binned ($a,e,i$) space (left
column in Figs.~\ref{fig:6astsourcedefresolutiona} and
\ref{fig:6astsourcedefresolutionb}).

To understand the statistical uncertainty of the orbit distributions,
we divided the test asteroids for each orbit distribution into
even-numbered and odd-numbered groups and estimated the uncertainty of
the overall orbit distribution by computing the difference between the
orbit distributions composed of even-numbered and odd-numbered test
asteroids. The difference distribution was then normalized by using
the combined orbit distribution so as to result in a distribution with
the same units as the combined distribution (right column in
Figs.~\ref{fig:6astsourcedefresolutiona} and
\ref{fig:6astsourcedefresolutionb}).

\subsubsection{JFC steady-state orbit distribution}

\begin{figure}[h]
  \centering
  \includegraphics[width=\columnwidth]{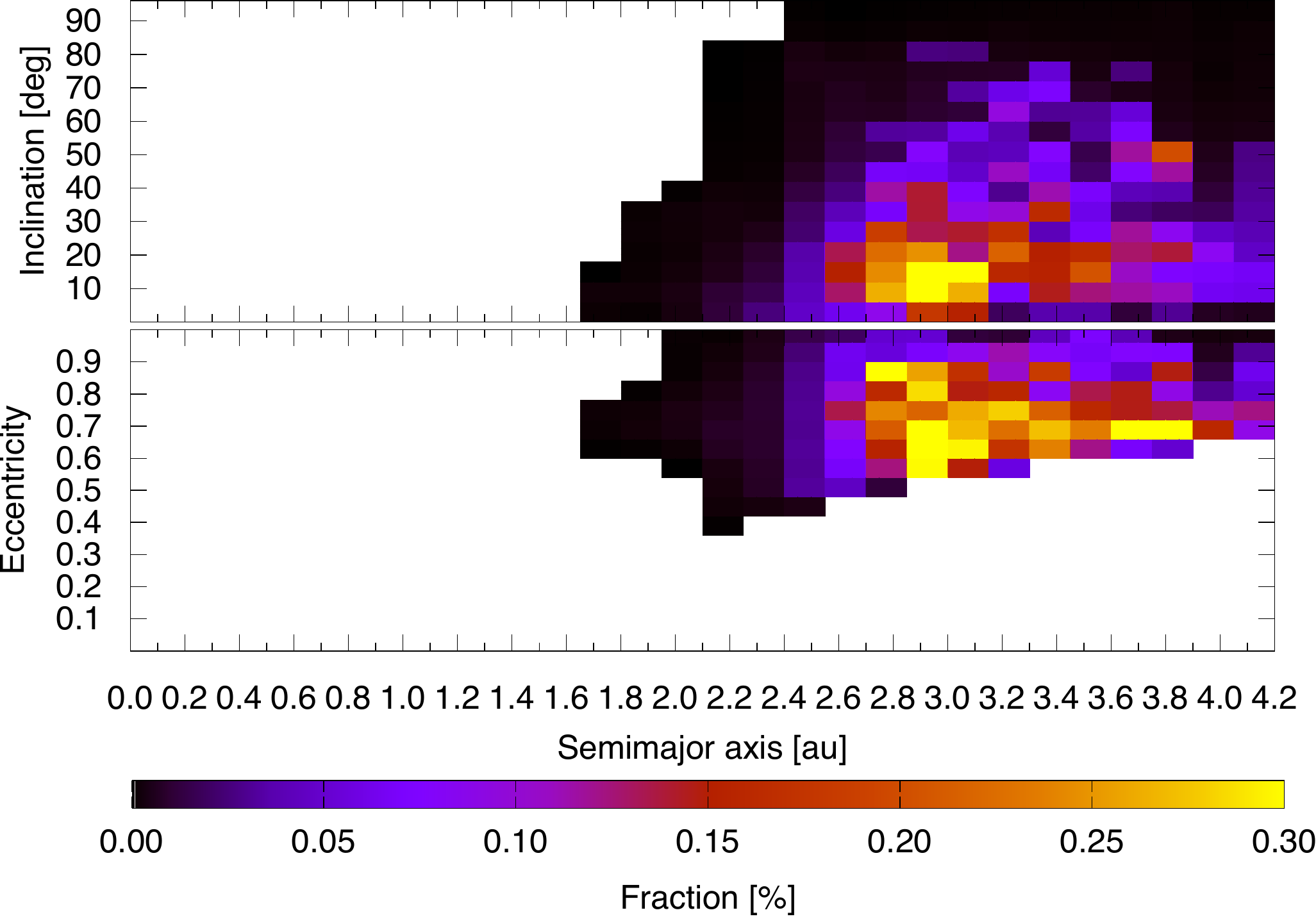}
  \caption{JFC steady-state orbit distribution.}
  \label{fig:JFC-encke}
\end{figure}

There are several works in the published literature which computed the
steady-state orbital distribution of Jupiter-family comets by
integrating particles coming from the trans-Neptunian region up to
their ultimate dynamical removal. The pioneering work was that of
\citet{1997Icar..127...13L}, followed by \citet{2006Icar..182..161L},
\citet{2009Icar..203..140D}, and \citet{2013Icar..225...40B}. The
resulting JFC orbital distributions have been kindly provided to us by
the respective authors. We have compared them and selected the one
from the \citet{2006Icar..182..161L} work because it is the only one
constructed using simulations that accounted for the gravitational
perturbations by the terrestrial planets. Thus, unlike the other
distributions, this one includes "comets" on orbits decoupled from the
orbit of Jupiter (i.e., not undergoing close encounters with the giant
planet at their aphelion) such as comet Encke.  We think that this
feature is important to model NEOs of trans-Neptunian origin. We
remind the reader that \citet{bot2002a} used the JFC distribution from
\citet{1997Icar..127...13L}, given that the results of
\citet{2006Icar..182..161L} were not yet available. Thus, this is
another improvement of this work over \citet{bot2002a}. The JFC
orbital distribution we adopted is shown in Fig.~\ref{fig:JFC-encke}.

There is an important difference between what we have done in this
work and what was done in the earlier models of the orbital
distribution of {\it active} JFCs because they included a JFC fading
parameter. In essence, a comet is considered to become active when its
perihelion distance decreases below some threshold (typically 2.5~au)
for the first time. That event starts the "activity clock". Particles
are assumed to contribute to the distribution of JFCs only up to a
time $T_{\rm active}$ of the activity clock. Limiting the physical
lifetime is essential to reproduce the observed inclination
distribution of active JFCs, as first shown in
\citet{1997Icar..127...13L}. What happens after $T_{\rm active}$ is
not clear. The JFCs might disintegrate or they may become
dormant. Only in the second case, of course, can the comet contribute
to the NEO population with an asteroidal appearance.  We believe the
second case is much more likely because JFCs are rarely observed to
disrupt, unlike long period comets. Besides, several studies argued
for the existence of dormant JFCs
\citep[e.g.,][]{2005AJ....130..308F,2006Icar..185..211F}. Thus, in
order to build the distribution shown in Fig.~\ref{fig:JFC-encke} we
have used the original numerical simulations of
\citet{2006Icar..182..161L} but suppressed any limitation on a
particle's age.

\section{Debiased NEO orbit and absolute-magnitude distributions}
\label{sec:results}

\subsection{Selecting the preferred combination of steady-state orbit distributions}
\label{sec:sourceselection}

We first needed to find the optimum combination of ERs. To strike a
quantitative balance between the goodness of fit and the number of
parameters we used the AICc metric defined by Eq.\ (\ref{eq:aicc}). We
tested 9 different ER models out of which all but one are based on
different combinations of the steady-state orbit distributions
described in Sect.\ \ref{sec:integrations}. The one additional model
is the integration of Bottke-like initial conditions by
\citet{gre2012a}. The combination of different ERs was done by summing
up the residence-time distributions of the different ERs, that is,
prior to normalizing the orbit distributions. Hence the initial orbit
distribution and the direct integrations provided the relative shares
between the different orbit distributions that were combined.

\begin{figure}[h]
  \centering
  \includegraphics[width=\columnwidth]{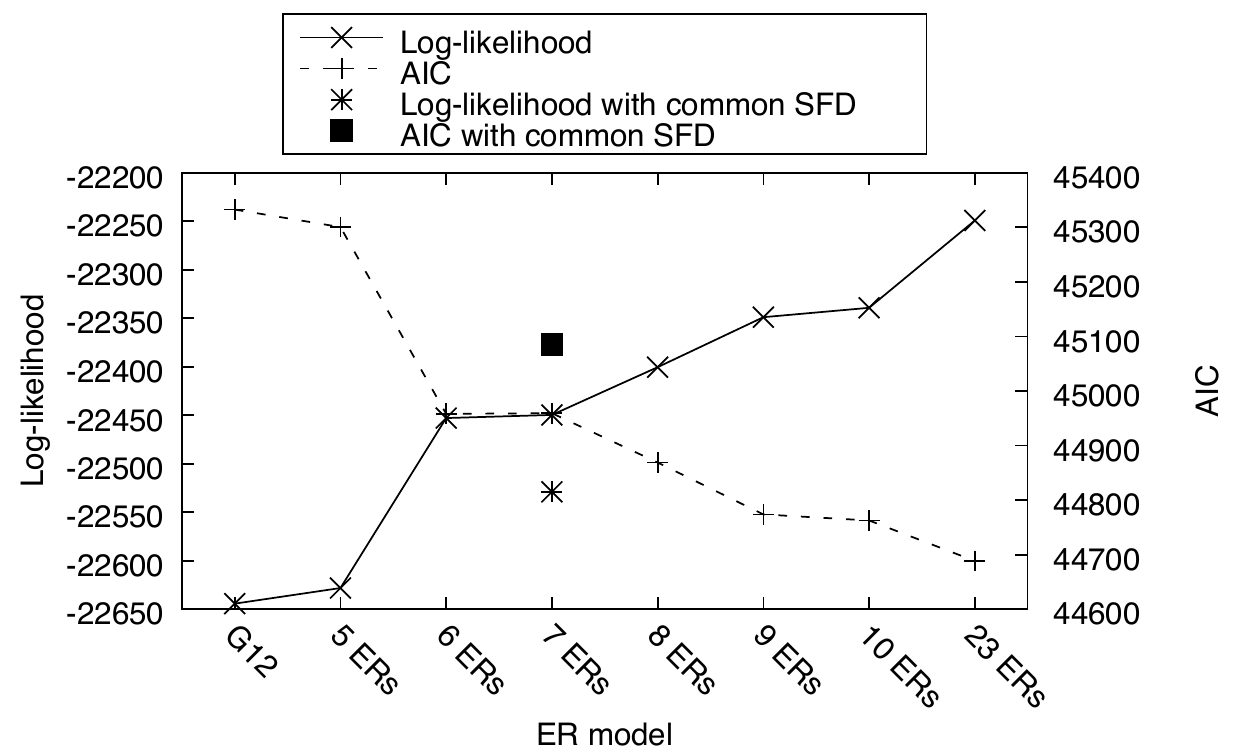}
  \caption{The log-likelihood of the best-fit solution (left axis and
    solid line) and the corresponding AICc metric (right axis and
    dashed line) as a function of the number of ERs included in the
    model. G12 stands for the five-component orbit model by \protect
    \citet{gre2012a}.}
  \label{fig:diffcombsources}
\end{figure}

As expected, the maximum likelihood (ML) constantly improves as the
steady-state orbit distribution is divided into a larger number of
subcomponents (Fig.~\ref{fig:diffcombsources}). The ML explicitly
shows that our steady-state orbit distributions lead to better fits
compared to the Bottke-like orbit distributions by \citet{gre2012a},
even when using the same number of ERs, that is, four asteroidal and
one cometary ER. The somewhat unexpected outcome of the analysis is
that we do not find a minimum for the AICc metric, which would have
signaled an optimum number of model parameters
(Fig.~\ref{fig:diffcombsources}). Instead the AICc metric improves all
the way to the most complex model tested which contains 23 different
ERs and hence 94 free parameters!

The largest drop in AICc per additional source takes place when we go
from a five-component model to a six-component model. The difference
between the two being that the former divides the outer MAB into two
components and lack Hungarias and Phocaeas whereas the latter has a
single-component outer-MAB ER and includes Hungarias and
Phocaeas. Continuining to the seven-component model we again split up
the outer-MAB ER into two components (the 5:2J and 2:1J
complexes). The difference between the five-component model and the
seven-component model is thus the inclusion of Hungarias and Phocaeas
in the latter. The dramatic improvement in AICc shows that the
Hungarias and Phocaeas are relevant components of an NEO orbit model.

Although a model with 94 free parameters is formally acceptable, we
had some concern that it would lead to degenerate sets of model
parameters. It might also (partly) hide real phenomena that are
currently not accounted for and thus should show up as a disagreement
between observations and our model's predictions. Therefore we took a
heuristic approach and compared the steady-state orbit distributions
to identify those that are more or less overlapping and can thus be
combined. After a qualitative evaluation of the orbit distributions we
concluded that it is sensible to combine the asteroidal ERs into six
groups (Figs.~\ref{fig:6astsourcedefresolutiona} and
\ref{fig:6astsourcedefresolutionb}). Of these six groups the Hungaria
and Phocaea orbit distributions are uniqely defined based on the
initial orbits of the test asteroids whereas the four remaining groups
are composed of complexes of escape routes ($\nu_6$, 3:1J, 5:2J, and
2:1J) that produce overlapping steady-state orbit distributions.

\begin{figure}[h]
  \centering
  \includegraphics[width=\columnwidth]{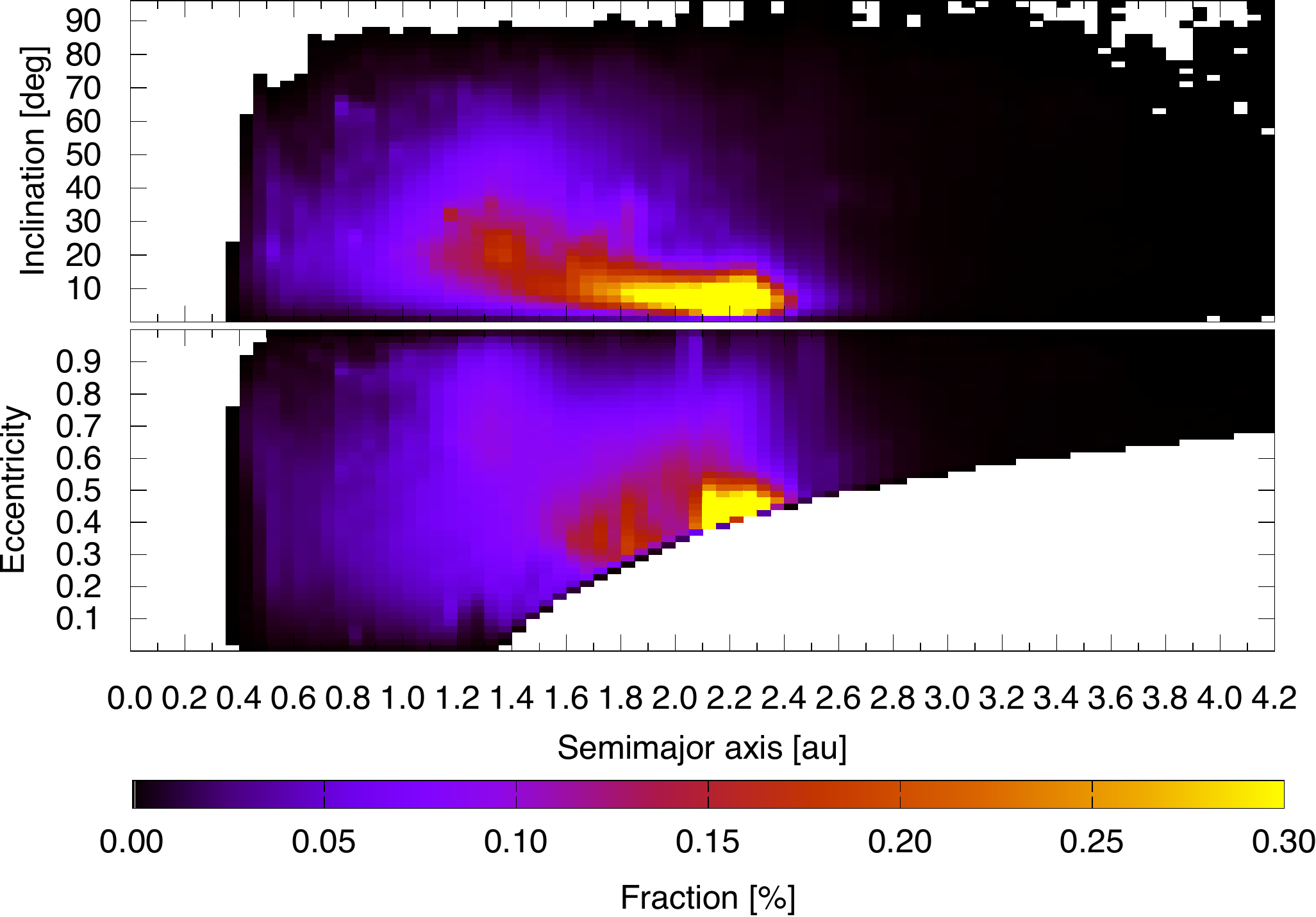}
  \includegraphics[width=\columnwidth]{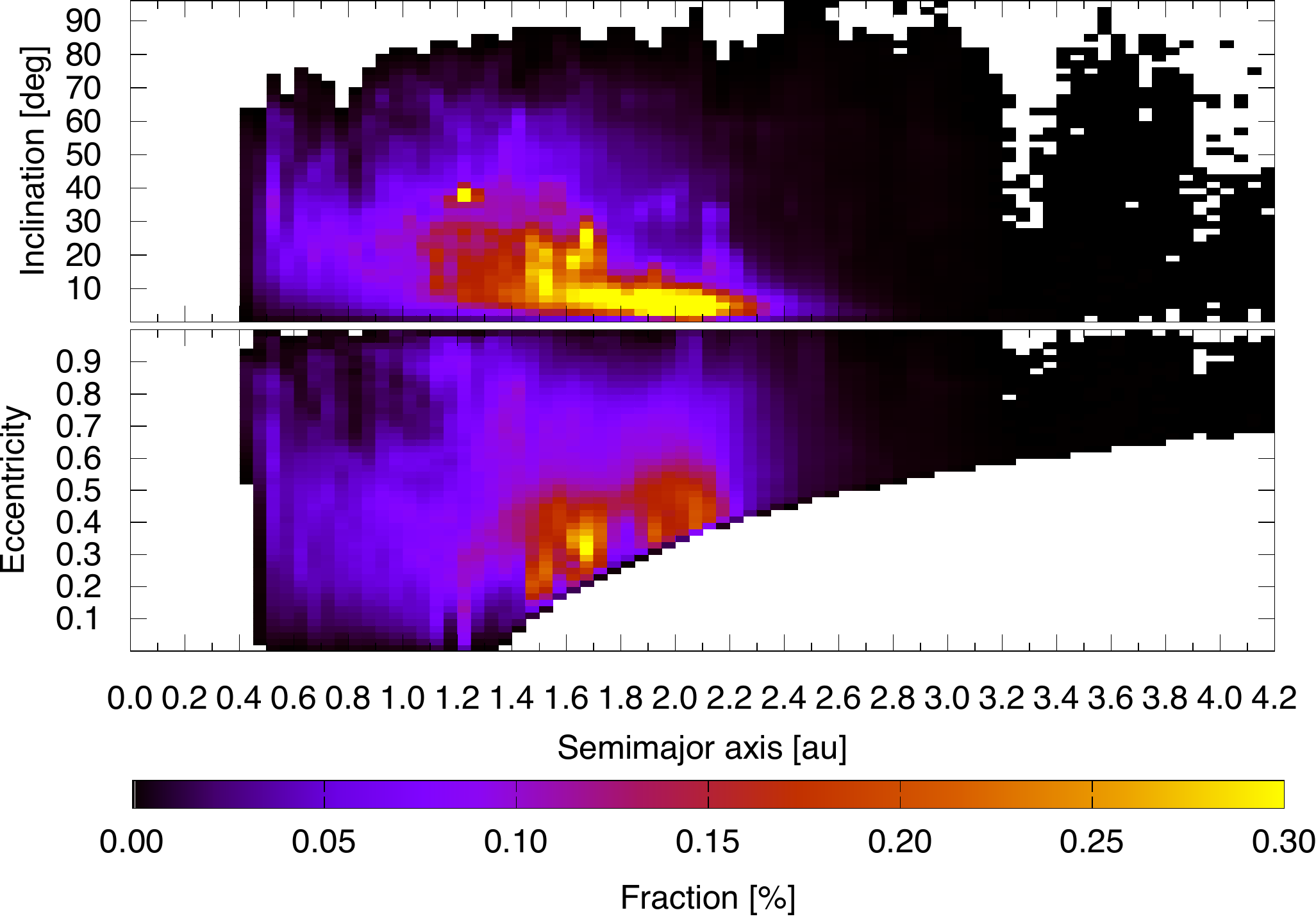}
  \caption{Steady-state orbit distributions for $\nu_6$ and 7:2J (top)
    and 4:1J (bottom).}
  \label{fig:nu6complex}
\end{figure}

\begin{figure}[h]
  \centering
  \includegraphics[width=\columnwidth]{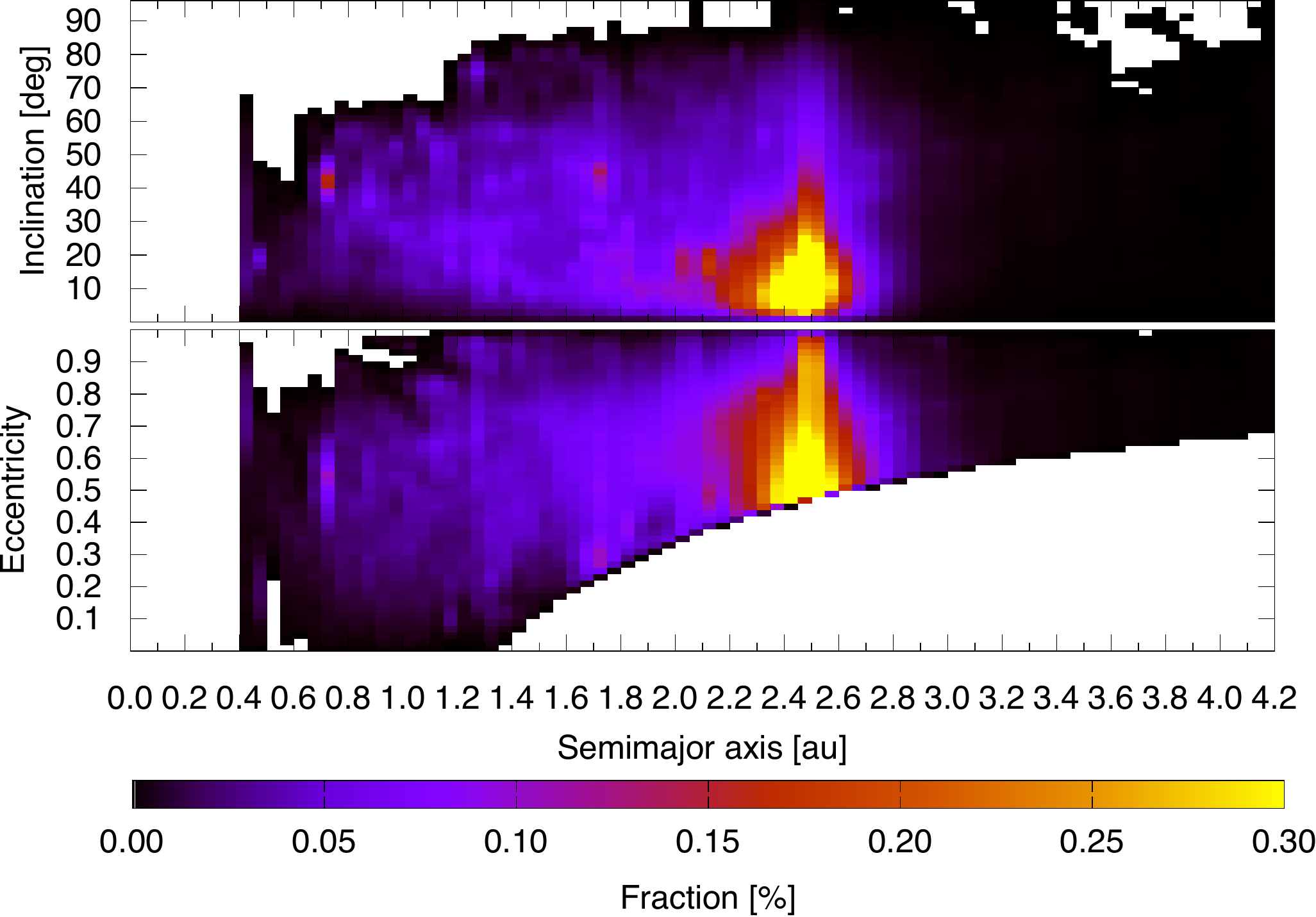}
  \includegraphics[width=\columnwidth]{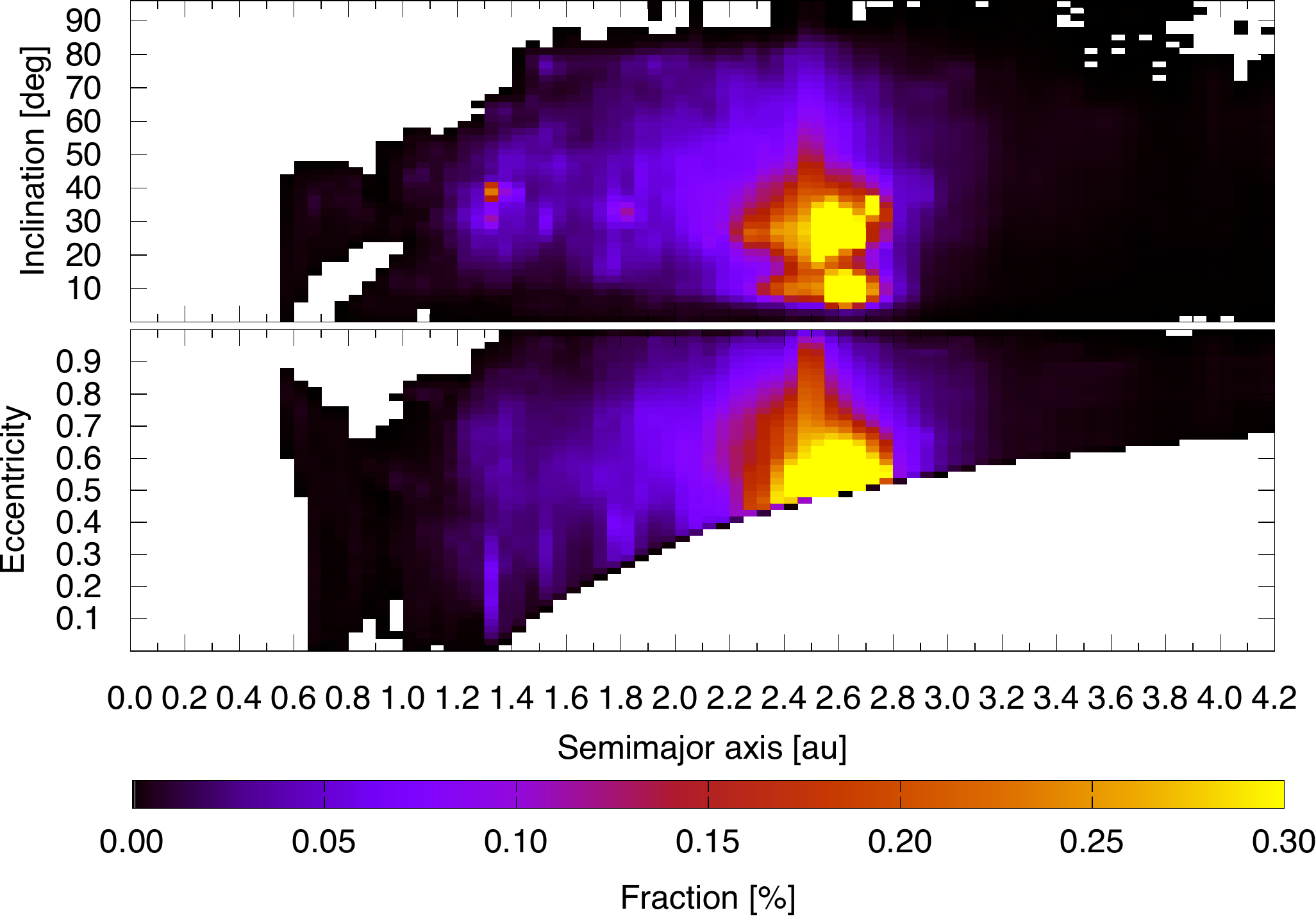}
  \caption{Steady-state orbit distributions for 3:1J (top) and the
    low-$i$ and high-$i$ components of the outer $\nu_6$ (bottom).}
  \label{fig:3:1complex}
\end{figure}

In principle one could argue, based on Fig.~\ref{fig:diffcombsources},
that it would make sense to use 9 ERs because then the largest drop in
the AICc metric would have been accounted for. The difference between
7 and 9 ERs is that the 4:1J has been separated from the $\nu_6$
complex (Fig.~\ref{fig:nu6complex}) and the $\nu_6$ component external
to the 3:1J has been separated from the 3:1J
(Fig.~\ref{fig:3:1complex}). However, the differences between the
$\nu_6$ complex and the 4:1J orbit distributions are small with the
most notable difference being that the $\nu_6$ distribution extends to
larger $a$. Similarly, the $\nu_6$ component external to the 3:1J has
some clear structure compared to the 3:1J component but this structure
is also clearly visible in the combined 3:1J orbit distribution (top
panel in Fig.~\ref{fig:6astsourcedefresolutionb}). There is thus a
substantial overlap in the orbit distributions and we therefore
decided to include one cometary and six asteroidal ERs in the model.

\subsection{The best-fit model with 7 ERs}
\label{sec:nominal7smodel}

Having settled on using 7 steady-state orbit distributions for the
nominal model we then turned to analyzing the selected model in
greater detail. As described in the Introduction, it is impossible to
reach an acceptable agreement between the observed and predicted orbit
distribution unless the disruption of NEOs at small $q$ is accounted
for \citep{2016Natur.530..303G}. Here we solved the discrepancy by
fitting for the two parameters that describe a penalty function
against NEOs with small $q$ (see Sect.~\ref{sec:theory}), in addition
to the parameters describing the $H$ distributions. The best-fit
parameters for the penalty function are $k=1.40\pm0.07\au^{-1}$ and
$q_0=0.69\pm0.02\au$. Although a direct comparison of the best-fit
penalty function and the physical model by \citet{2016Natur.530..303G}
is impossible we find that the penalty function $p=0.86\pm0.05$ at
$q=0.076\au$ and, of course, even higher for $q<0.076\au$.
Considering that the 3$\sigma$ value at $q=0.076\au$ reaches unity we
find the agreement satisfactory.

\begin{figure}[h]
  \centering
  \includegraphics[width=0.49\columnwidth]{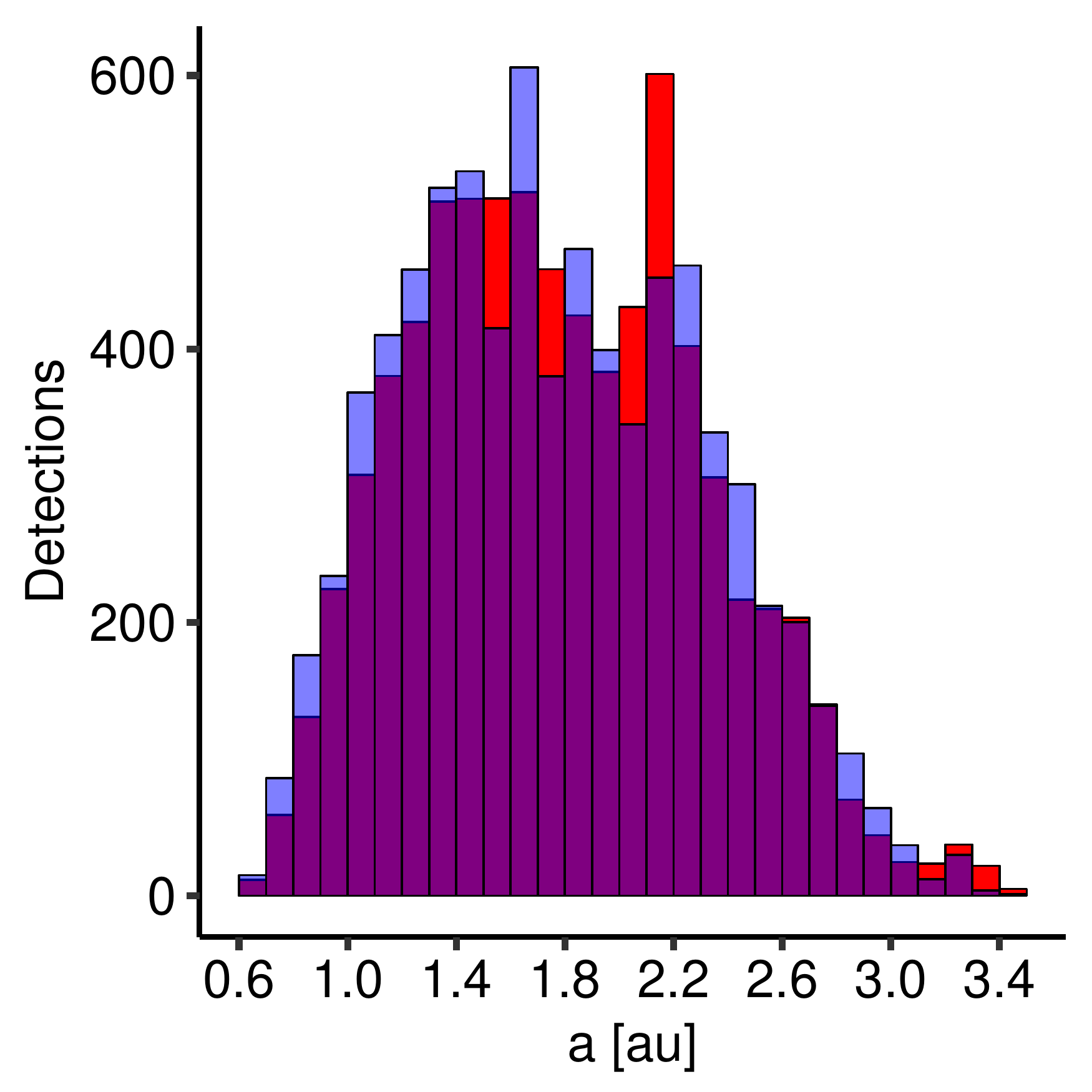}
  \includegraphics[width=0.49\columnwidth]{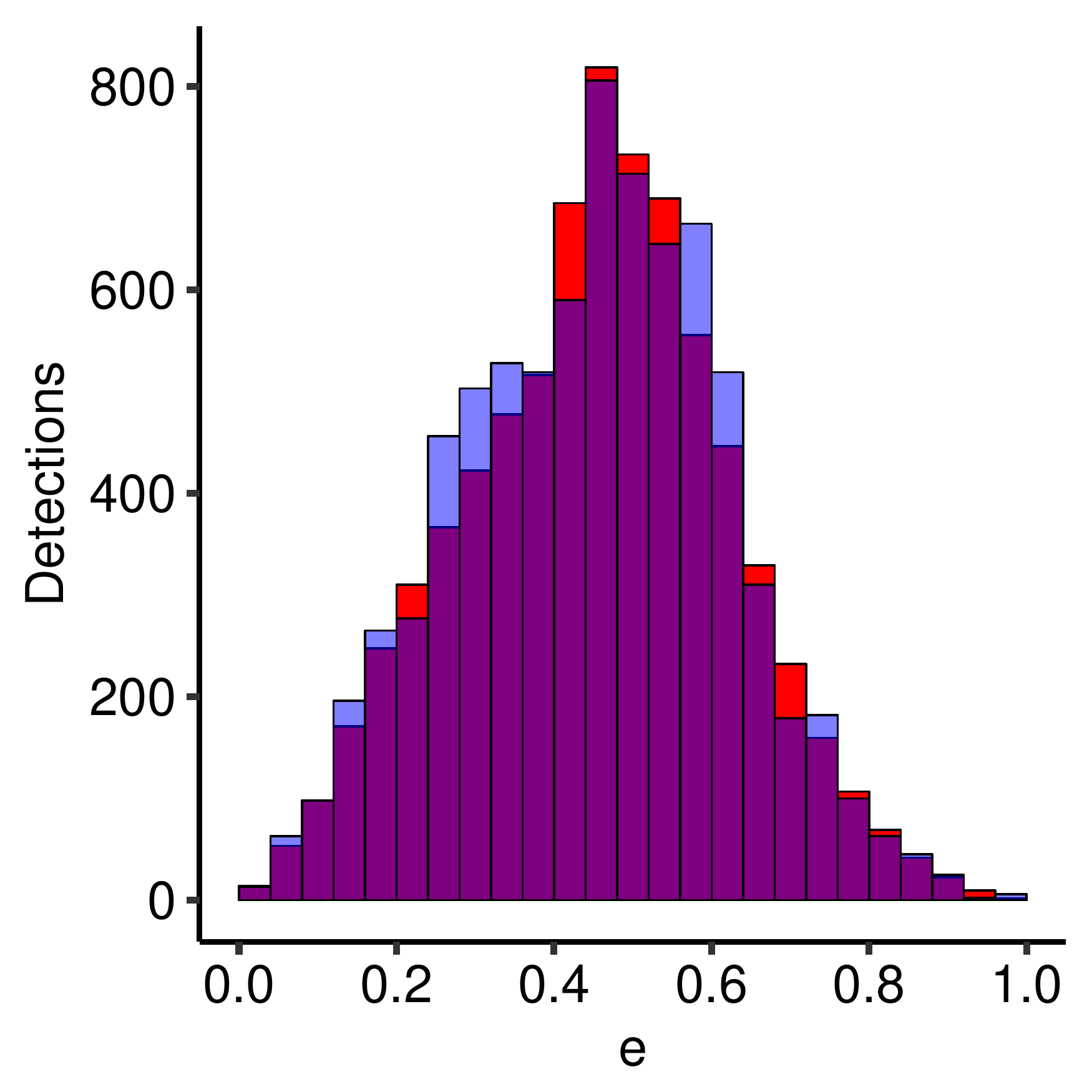}
  \includegraphics[width=0.49\columnwidth]{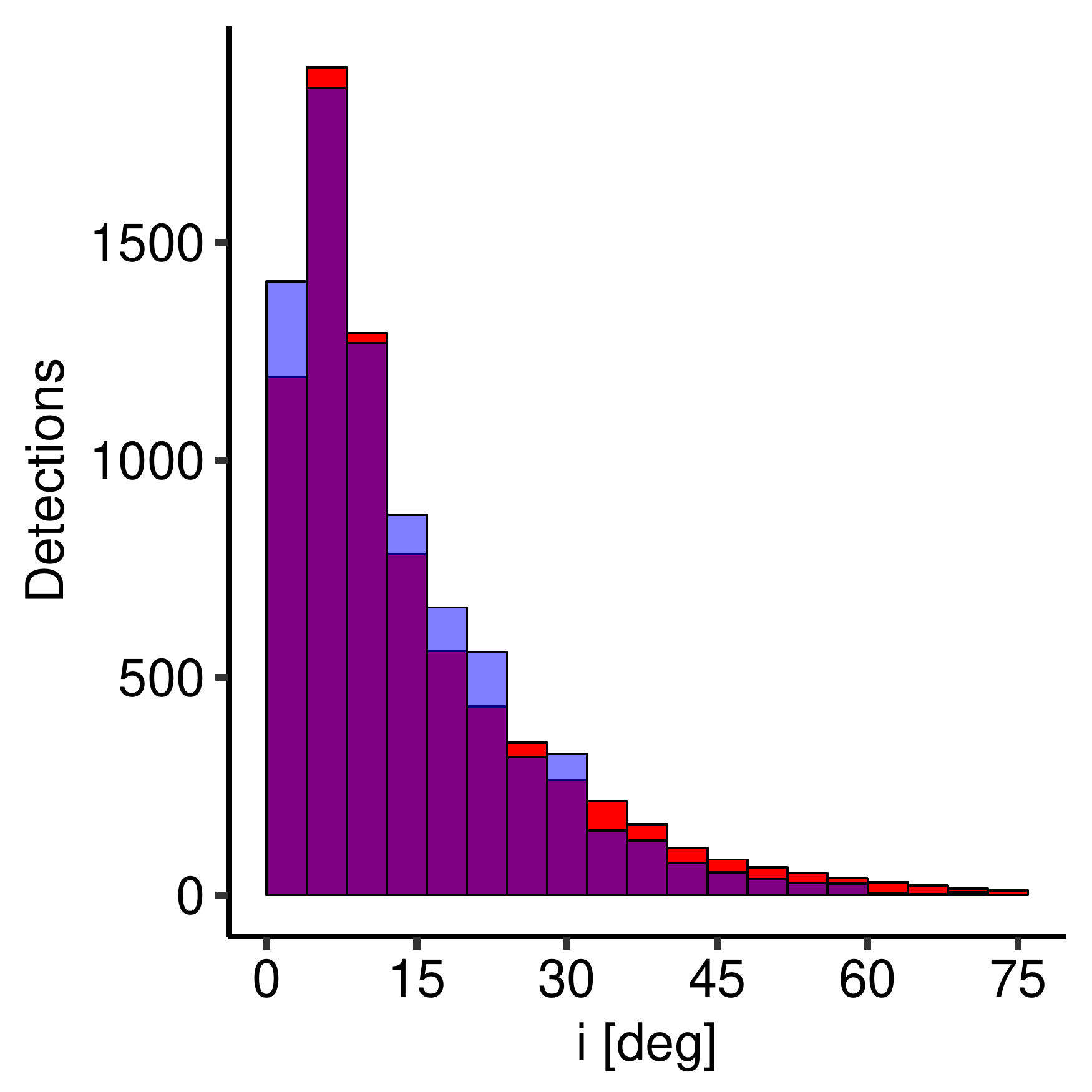}
  \includegraphics[width=0.49\columnwidth]{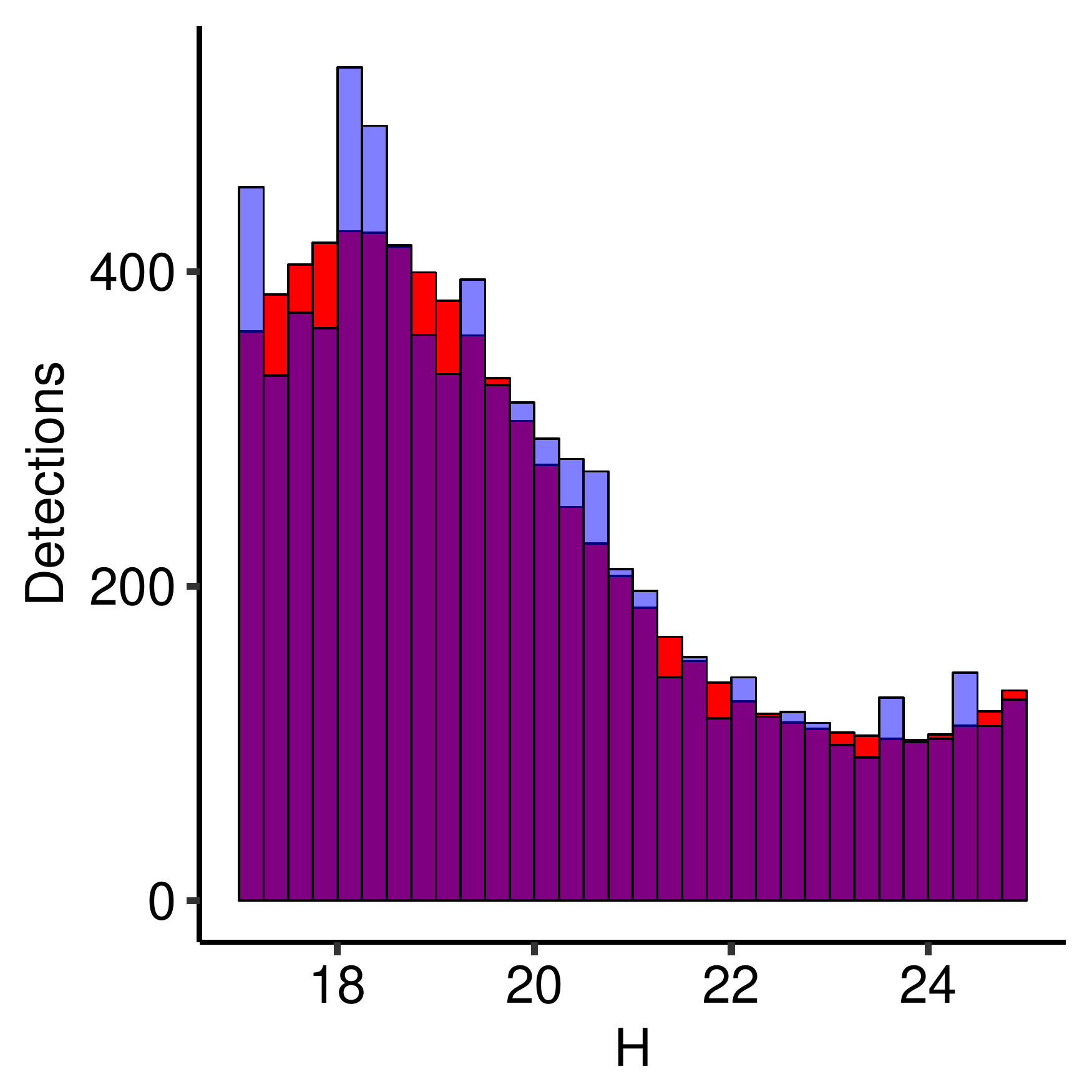}
  \caption{Comparison between G96 and 703 detections (blue;
    $n(a,e,i,H)$) and prediction based on the best-fit model (red;
    $\epsilon(a,e,i,H)\,N(a,e,i,H)$). The purple color indicates
    overlapping distributions.}
  \label{fig:bestfitaeih}
\end{figure}

\begin{figure}[h]
  \centering
  \includegraphics[width=0.49\columnwidth]{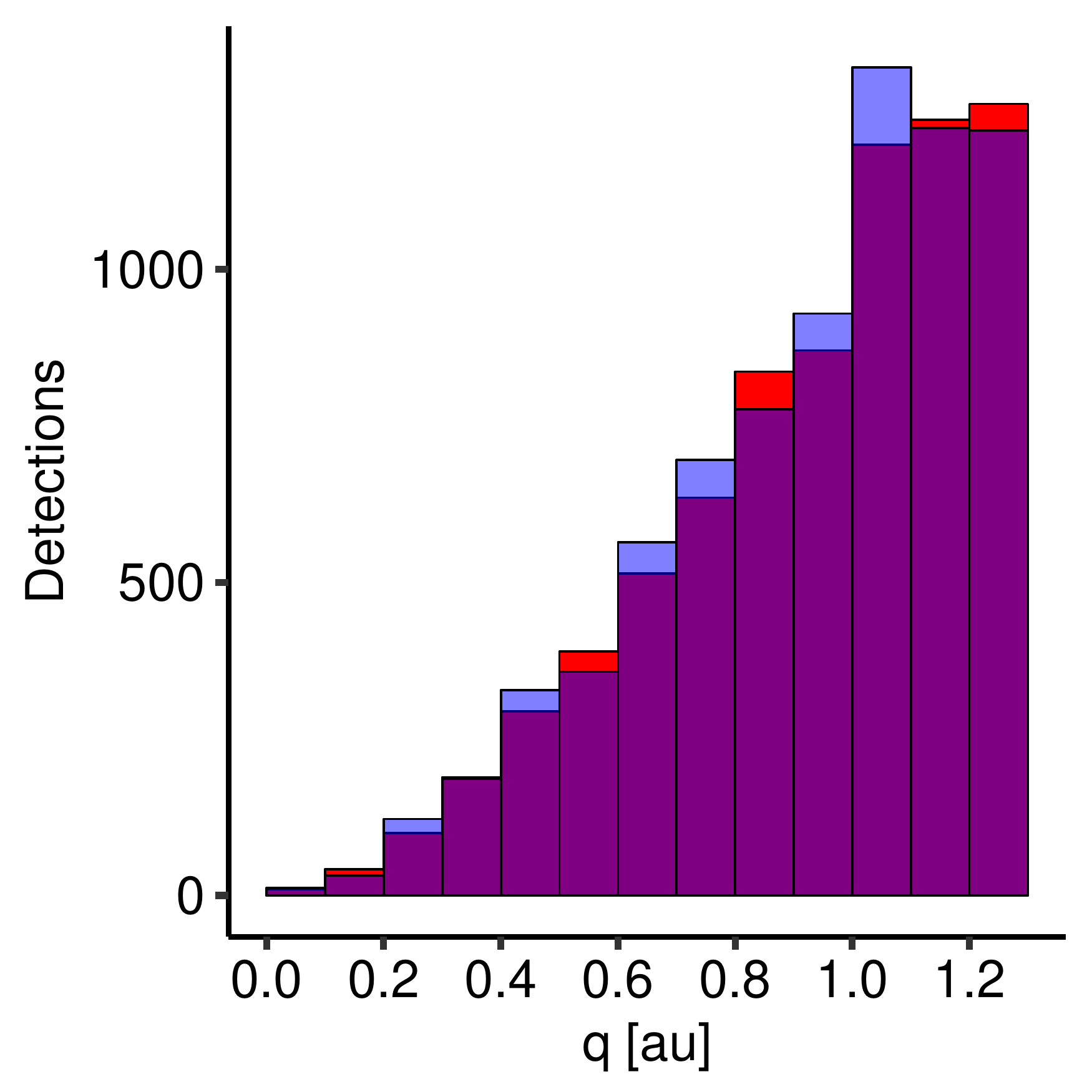}
  \caption{As Fig.~\ref{fig:bestfitaeih} but for perihelion distance
    $q$.}
  \label{fig:bestfitq}
\end{figure}

A comparison between the observed and predicted number (i.e., marginal
distributions of $\epsilon(a,e,i,H)\,N(a,e,i,H)$) of NEO detections
shows that the best-fit model accurately reproduces the observed
($a$,$e$,$i$,$H$) distributions (Fig.~\ref{fig:bestfitaeih}) as well
as the observed $q$ distribution (Fig.~\ref{fig:bestfitq}). Thus the
penalty function is able to mitigate the problem caused by not
including a physical model of NEO disruptions close to the Sun.

We used the same approach as \citet{bot2002a} to evaluate the
statistical uncertainty of the 28 parameters (four parameters for each
of the 7 HFDs and two parameters for the penalty function) that define
our model.  We first generated 100 random realizations of the biased
best-fit four-dimensional model (the marginal distributions of which
are shown in red in Fig.~\ref{fig:bestfitaeih}). Each realization
contains 7,769 virtual detections, that is, the number of detections
reported by the G96 and 703 surveys that we used for the nominal
fit. When re-fitting the model to each of these synthetic data sets we
obtained slightly different values for the best-fit parameters
(Supplementary Figs.~3 and 4), which we interpret as being caused by
the statistical uncertainty.  Note that the parameter distributions
are often non-Gaussian but nevertheless relatively well constrained.

We interpret the RMS with respect to the best-fit parameters obtained
for the real data to be a measure of the statistical uncertainty for
the parameters. The best-fit parameters describing the HFDs as well as
their uncertainties are reported in Table~\ref{table:parameters}. The
best-fit parameters for the penalty function are $k=1.40 \pm 0.07$ and
$q_0 = 0.69 \pm 0.02$. The minimum HFD slope is statistically distinct
from zero only for $\nu_6$ and 3:1J and the slope minimum typically
occurs around $H\sim20$. This implies that the HFD slope should be at
its flattest around $H\sim20$ as seen in the apparent $H$ distribution
observed by G96 (Fig.~\ref{fig:obsaeiH}). A constant slope is
acceptable (curvature within 1$\sigma$ of zero) only for Phocaeas and
5:2J, which shows that the decision to allow for a more complex
functional form for the HFDs law was correct.
\begin{sidewaystable}
\centering
\caption{The best-fit model parameters $N_0(H_0=17)$, $c$, $H_{\rm
    min}$, and $\alpha_{\rm min}$. The average lifetime $\langle L
  \rangle$ in NEO region ($q<1.3\au$ and $a<4.2\au$) as computed from
  the orbital integrations. The model prediction for the debiased
  number of NEOs $N(17<H<22)$ and $N(17<H<25)$; the relative fraction
  of NEOs from each ER $\beta(17<H<22)$ and $\beta(17<H<25)$; the
  absolute flux of NEOs from each ER into the NEO region $F(17<H<22)$
  and $F(17<H<25)$; and the relative flux of NEOs from each ER into
  the NEO region $\gamma(17<H<22)$ and
  $\gamma(17<H<25)$.}\label{table:parameters}
\begin{tabular}{cccccccc}
\hline
ER & Hungaria & $\nu_6$ complex & Phocaea & 3:1J complex & 5:2J complex & 2:1J complex & JFC \\
\hline
$N_0(H_0=17)$ & $23.1 \pm 3.0$ & $34.6 \pm 2.6$ & $10.3 \pm 3.7$ & $25.1 \pm 6.0$ & $19.3 \pm 5.1$ & $3.6 \pm 1.9$ & $6.6 \pm 2.6$ \\
$c$ & $0.013 \pm 0.007$ & $0.037 \pm 0.003$ & $0.008 \pm 0.027$ & $0.009 \pm 0.007$ & $0.008 \pm 0.019$ & $0.062 \pm 0.023$ & $0.024 \pm 0.019$ \\
$H_{\rm min}$ & $15.6 \pm 1.5$ & $20.4 \pm 0.1$ & $23.4 \pm 3.9$ & $18.0 \pm 4.6$ & $25.0 \pm 3.0$ & $20.9 \pm 0.8$ & $18.0 \pm 2.1$ \\
$\alpha_{\rm min}$ & $0.00 \pm 0.06$ & $0.24 \pm 0.01$ & $0.00 \pm 0.09$ & $0.33 \pm 0.10$ & $0.00 \pm 0.10$ & $0.00 \pm 0.04$ & $0.12 \pm 0.08$ \vspace{0.5mm} \\
$\langle L\rangle$ [Myr] & $37.19 \pm 0.23$ & $6.98 \pm 0.50$ & $11.16 \pm 0.37$ & $1.83 \pm 0.03$ & $0.68 \pm 0.11$ & $0.40 \pm 0.05$ & -- \\
$N(17<H<22)$ & $1,400^{+130}_{-160}$ & $11,470^{+280}_{-220}$ & $670^{+190}_{-110}$ & $7,250^{+420}_{-480}$ & $3,020^{+290}_{-290}$ & $890^{+120}_{-70}$ & $470^{+100}_{-80}$ \vspace{1mm} \\ 
$\beta(17<H<22)$ & $0.056^{+0.005}_{-0.007}$ & $0.456^{+0.011}_{-0.009}$ & $0.027^{+0.008}_{-0.004}$ & $0.288^{+0.017}_{-0.019}$ & $0.120^{+0.012}_{-0.012}$ & $0.035^{+0.005}_{-0.003}$ & $0.019^{+0.004}_{-0.003}$ \vspace{1mm} \\
$F(17<H<22)$ [Myr$^{-1}$]  & $38^{+4}_{-4}$          & $1,640^{+130}_{-120}$    & $60^{+17}_{-10}$        & $3,950^{+240}_{-270}$    & $4,420^{+860}_{-860}$    & $2,200^{+410}_{-330}$ & -- \\ 
$\gamma(17<H<22)$ & $0.003^{+0.001}_{-0.001}$ & $0.133^{+0.010}_{-0.010}$ & $0.005^{+0.001}_{-0.001}$ & $0.321^{+0.019}_{-0.022}$ & $0.359^{+0.070}_{-0.070}$ & $0.179^{+0.033}_{-0.027}$ & -- \\
$N(17<H<25)$ & $143,500^{+26,200}_{-26,600}$ & $296,400^{+17,800}_{-16,500}$ & $1,300^{+25,200}_{-200}$ & $286,400^{+25,200}_{-34,000}$ & $7,200^{+15,400}_{-800}$ & $4,400^{+7,900}_{-1,200}$ & $62,700^{+11,600}_{-12,200}$ \vspace{1mm} \\ 
$\beta(17<H<25)$ & $0.179^{+0.033}_{-0.033}$ & $0.370^{+0.022}_{-0.021}$ & $0.002^{+0.031}_{-0.000}$ & $0.357^{+0.031}_{-0.042}$ & $0.009^{+0.019}_{-0.001}$ & $0.005^{+0.010}_{-0.001}$ & $0.078^{+0.014}_{-0.015}$ \vspace{1mm} \\
$F(17<H<25)$ [Myr$^{-1}$] & $3,900^{+700}_{-700}$    & $42,500^{+4000}_{-3900}$  & $100^{+2300}_{-20}$        & $156,100^{+14,000}_{-18,700}$ & $10,600^{+22,600}_{-2,100}$ & $10,900^{+19,600}_{-3,200}$ & -- \\
$\gamma(17<H<25)$ & $0.017^{+0.003}_{-0.003}$ & $0.190^{+0.018}_{-0.017}$ & $0.0005^{+0.0103}_{-0.0001}$ & $0.697^{+0.062}_{-0.083}$    & $0.047^{+0.101}_{-0.009}$   & $0.049^{+0.087}_{-0.014}$ & -- \\
\hline
\end{tabular}
\end{sidewaystable}

\begin{figure}[h]
  \centering
  \includegraphics[width=\columnwidth]{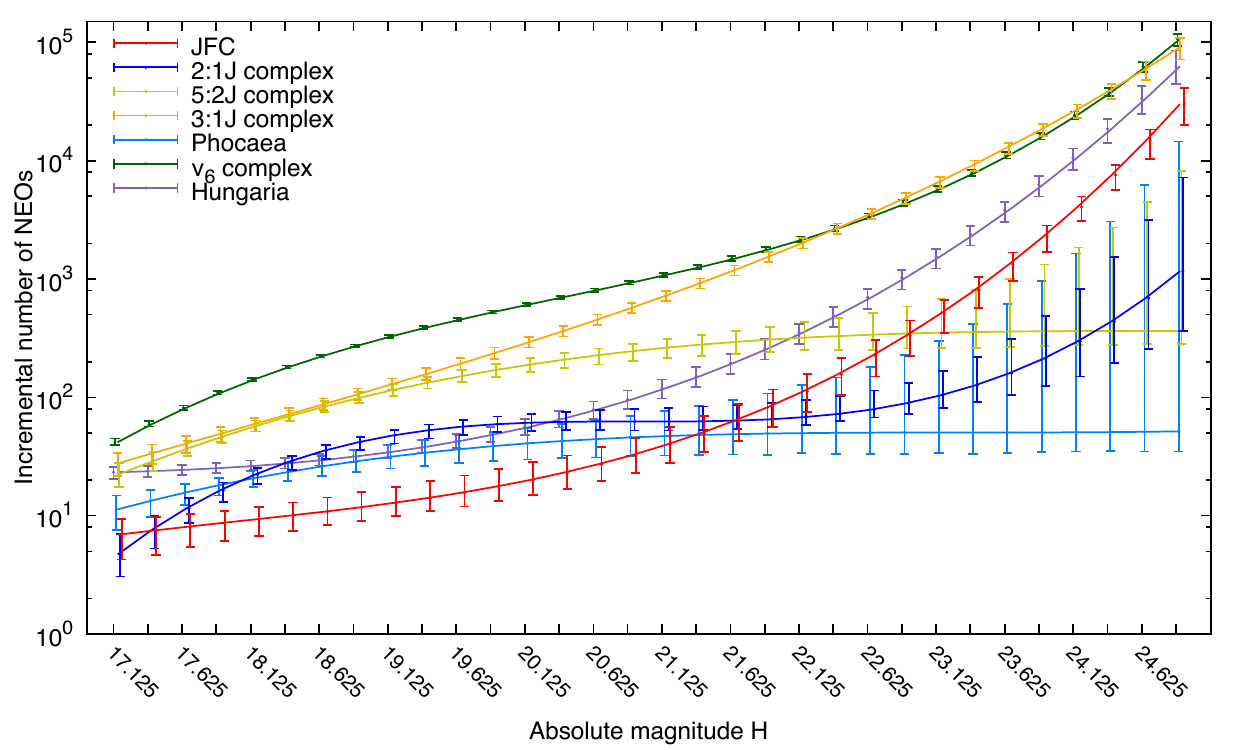}
  \caption{Debiased, incremental absolute-magnitude distributions for
    NEOs from different ERs.}
  \label{fig:neosfds}
\end{figure}

Figure \ref{fig:neosfds} shows a graphical representation of the
parameters describing the HFDs (Table~\ref{table:parameters}). The
statistical uncertainty for the number of Phocaeas, 5:2J, 2:1J and
JFCs is around 1--2 orders of magnitude for $H\gtrsim23$ whereas the
uncertainty for Hungarias, $\nu_6$ and 3:1J is within a factor of a
few throughout the $H$ range. The explanation for the large
uncertainties for the former group is that their absolute observed
numbers are smaller than for the NEOs originating in the latter group
and hence their numbers are more difficult to constrain.

\begin{figure}[h]
  \centering
  \includegraphics[width=\columnwidth]{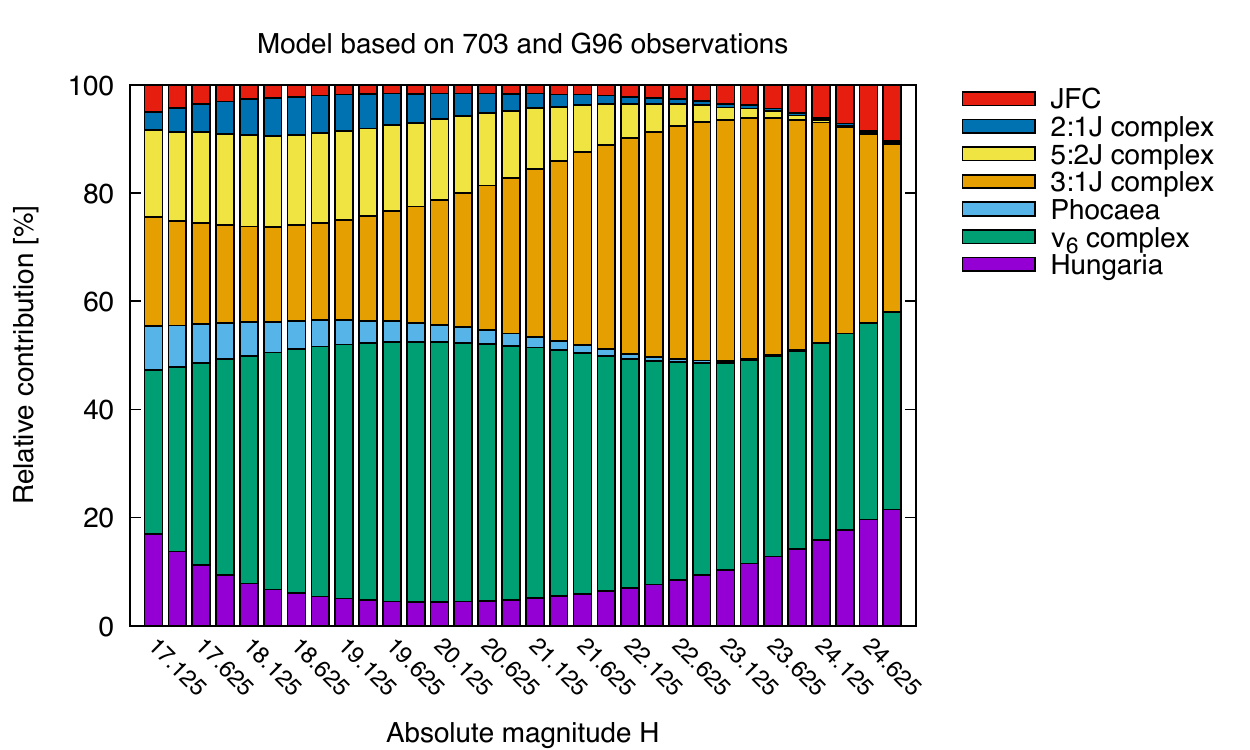}
  \caption{Relative importance of 7 different NEO ERs as a function of
    the absolute magnitude.}
  \label{fig:relsources}
\end{figure}

The ratio between the HFD for each ER and the overall HFD shows that
the contribution from the different ERs varies as a function of $H$
(Fig.~\ref{fig:relsources}). The $\nu_6$ and 3:1J complexes are the
largest contributors to the steady-state NEO population regardless of
$H$ as expected. The $\nu_6$ dominates the large NEOs whereas the 3:1J
is equally important at about $D\sim100\meter$. The Hungarias and JFCs
have a non-negligible contribution throughout the $H$ range whereas
Phocaeas and the 5:2J and 2:1J complexes have a negligible
contribution at small sizes.

The varying contribution from different ERs as a function of $H$ leads
to the overall orbit distribution also varying with $H$ despite the
fact that the ER-specific orbit distributions do not vary (Supplementary
Animation 2 and Supplementary Fig.~5).

Next we propagated the uncertainties of the model parameters into the
number of objects in each cell of the ($a$,$e$,$i$,$H$) space. Even
though the errors in individual cells can be larger than 100\%
(particularly when the expected population in the cell is small) the
overall statistical uncertainty on the number of NEOs with $H<25$ is
$\lesssim5$\% (Table~\ref{table:neoclass}).
\begin{table}
\centering
\caption{Absolute and relative shares belonging to 5 NEO classes based
  on the number-weighted average over $17<H<25$. The uncertainty
  estimates only account for the random
  component.}\label{table:neoclass}
\begin{tabular}{ccccc}
\hline
  NEO class & $N(17<H<25)$ & Relative share [\%] \\
  \hline
Amor   & $316,000^{+19,000}_{-12,000}$ & $39.4^{+2.3}_{-1.5}$ \\
Apollo & $436,000^{+21,000}_{-12,000}$ & $54.4^{+2.6}_{-1.5}$ \\
Aten   & $27,700^{+900}_{-800}$ & $3.46^{+0.12}_{-0.10}$ \\
Atira  & $9,200^{+300}_{-300}$ & $1.15^{+0.04}_{-0.03}$ \\
Vatira & $1,970^{+60}_{-70}$ & $0.25^{+0.01}_{-0.01}$ \\
\hline
\end{tabular}
\end{table}

\begin{figure}[h]
  \centering
  \includegraphics[width=\columnwidth]{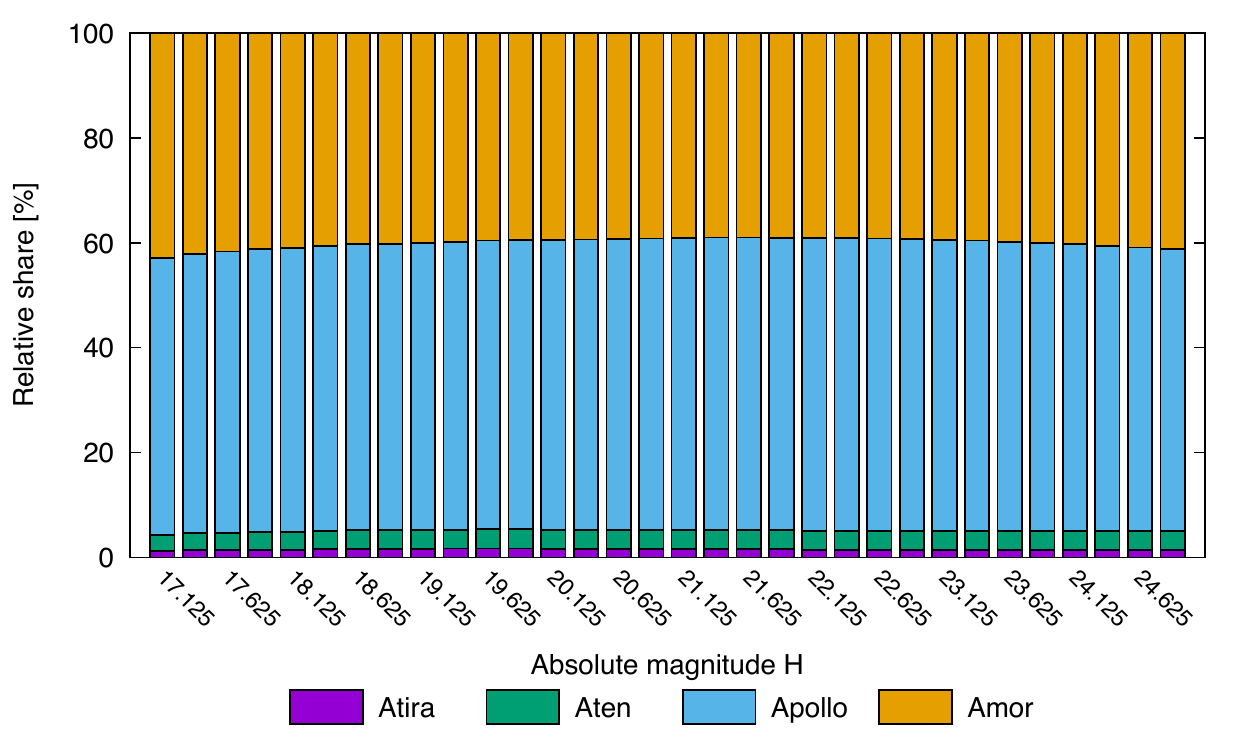}
  \caption{Relative shares between the 4 main NEO classes as a
    function of the absolute magnitude.}
  \label{fig:relclass}
\end{figure}

Given that the relative importance of the 7 ERs vary substantially
with $H$ (Fig.~\ref{fig:relsources}) it is somewhat surprising that
the relative shares of the 4 different NEO classes is hardly changing
with $H$ (Fig.~\ref{fig:relclass}).

\begin{figure}[h]
  \centering
  \includegraphics[width=\columnwidth]{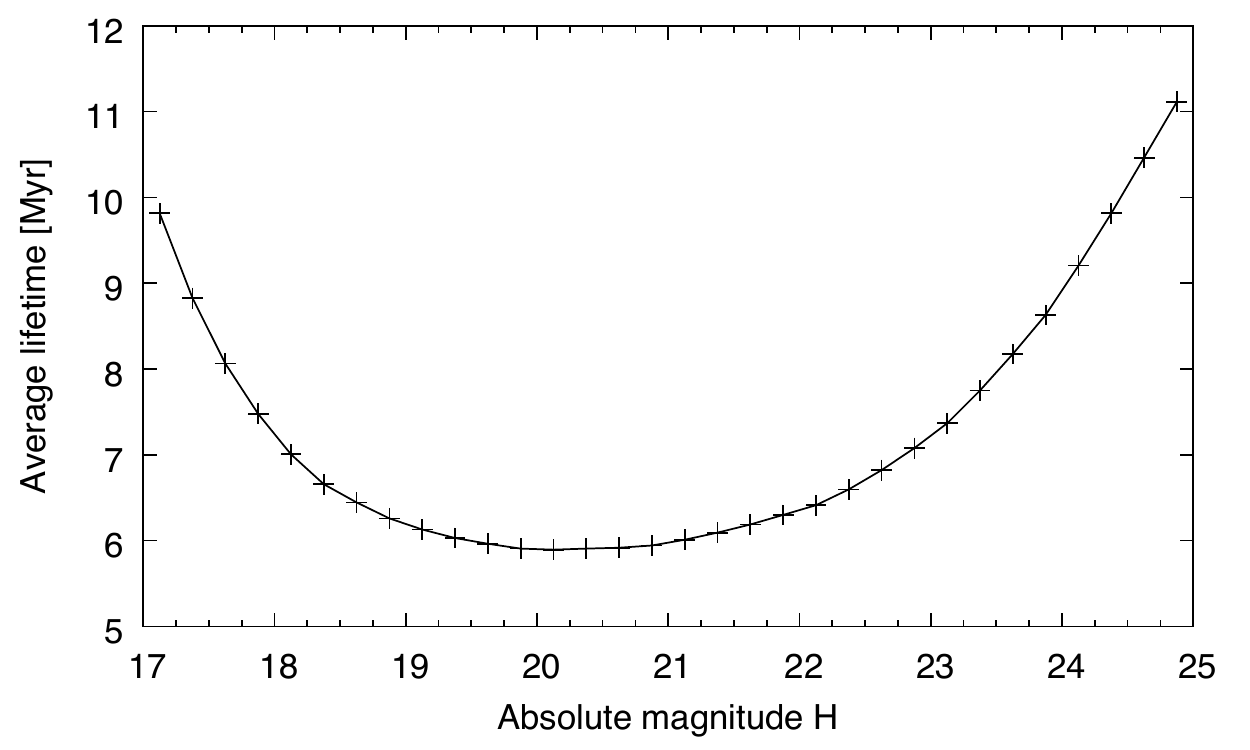}
  \caption{The average lifetime of NEOs as a function of absolute
    magnitude $H$ weighted by the relative contribution from each ER.}
  \label{fig:lifetime}
\end{figure}

Another consequence of the varying contribution from different ERs as
a function of $H$ is that also the average lifetime of NEOs changes
with $H$ when weighted by the relative contribution from each ER
(Fig.~\ref{fig:lifetime}). The average lifetime ranges from about 6 to
11 Myr with the mid-sized NEOs having the shortest lifetimes. The
increased average lifetime for both the largest and the smallest NEOs
considered is driven by the contribution from the Hungaria ER, because
NEOs from that ER have about 4--100 times longer lifetimes than NEOs
from the other ERs considered. Note also the clear correlation between
the average lifetime and the relative contribution of the Hungaria ER
(Fig.~\ref{fig:relsources}).

The analysis above does not account for the systematic uncertainties
arising, for instance, from an imperfect evaluation of the biases or
of the construction of the steady-state orbit distributions of the
NEOs coming from the various ERs. We will next assess systematic
uncertainties quantitatively by i) comparing models constructed using
two independent sets of orbit distributions (Sect.~\ref{sec:sensorb}),
and ii) by comparing models based on two independent surveys
(Sect.~\ref{sec:sensobs}).

\subsection{Sensitivity to variations in orbit distributions}
\label{sec:sensorb}

\begin{figure*}[h!]
  \centering
  \includegraphics[width=0.49\columnwidth]{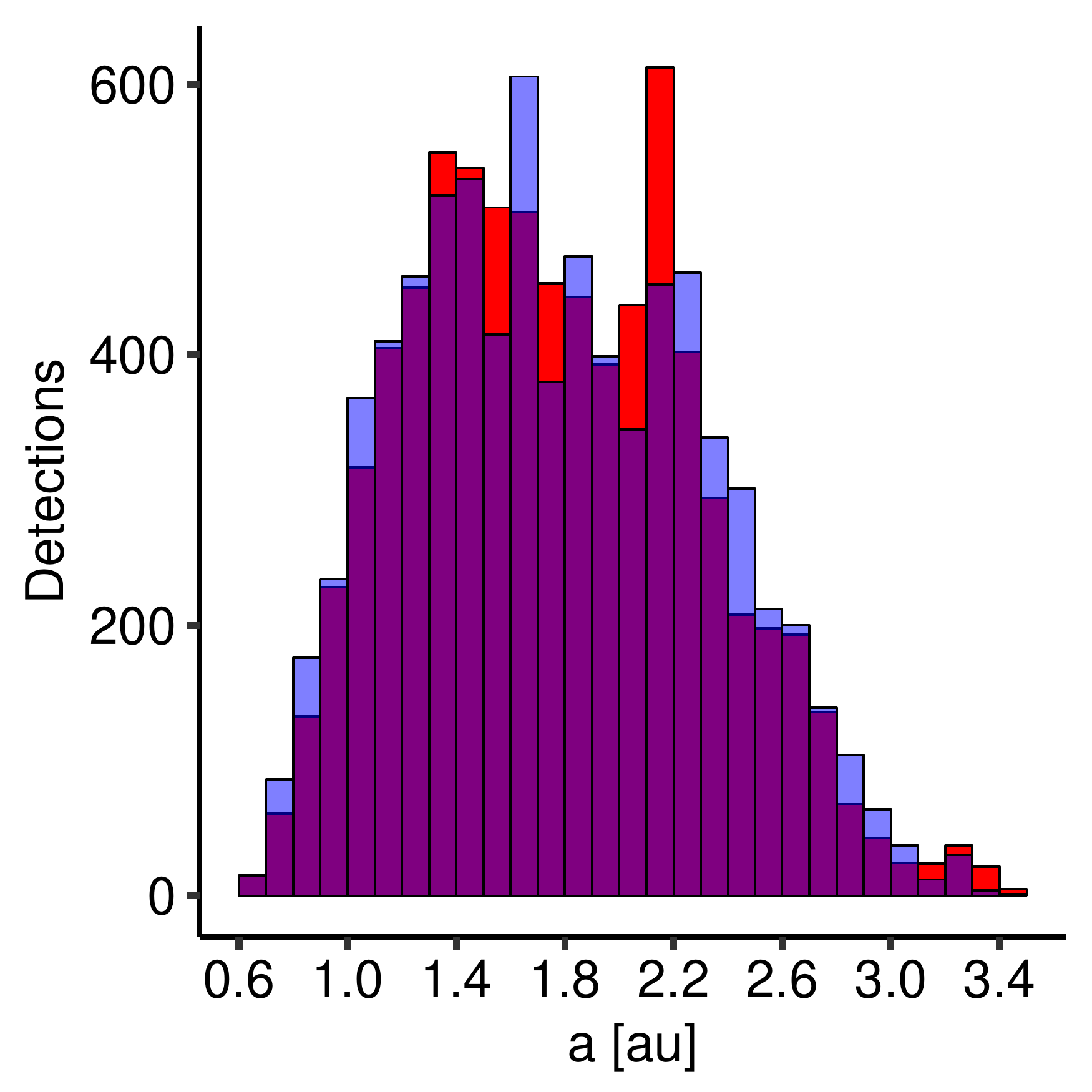}
  \includegraphics[width=0.49\columnwidth]{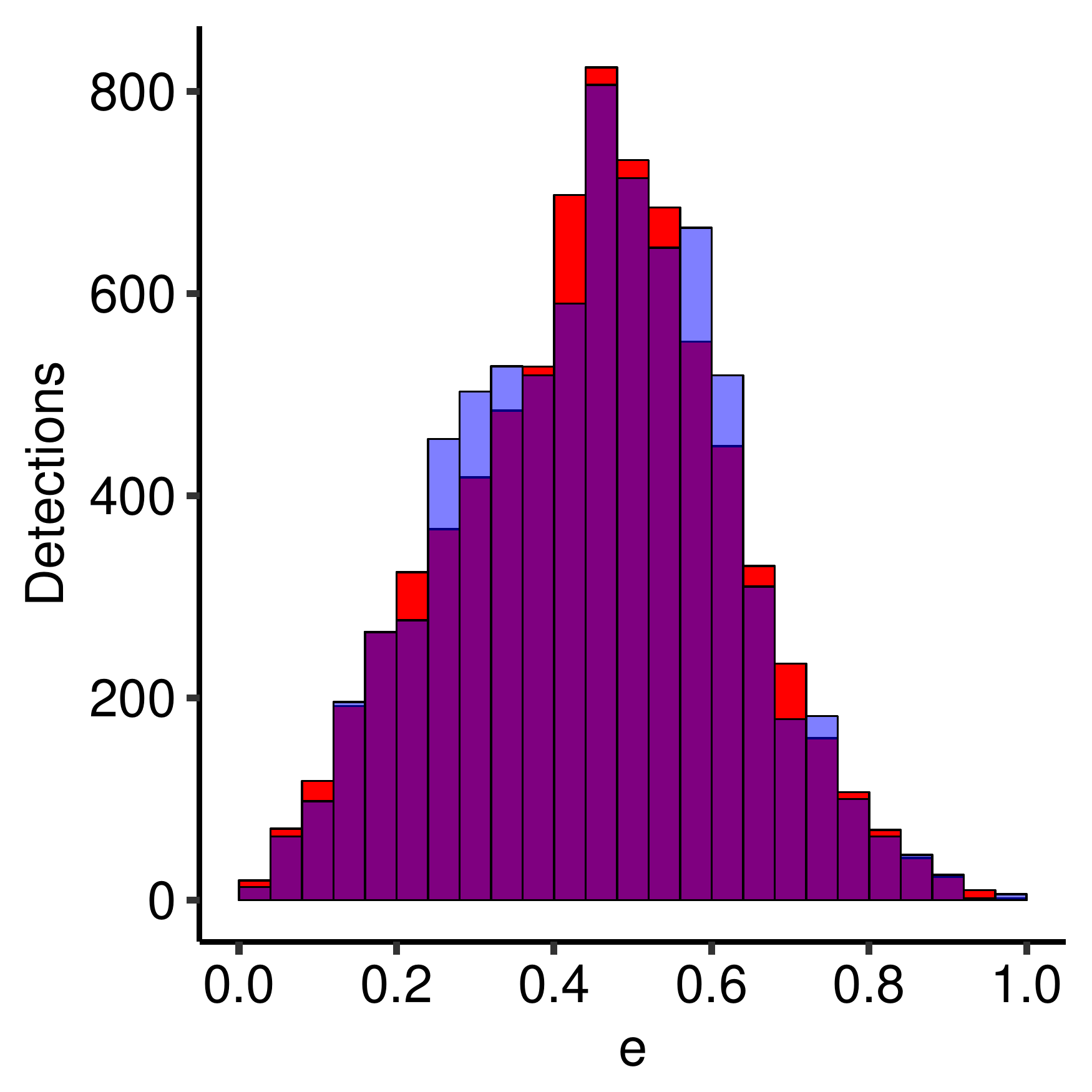}
  \includegraphics[width=0.49\columnwidth]{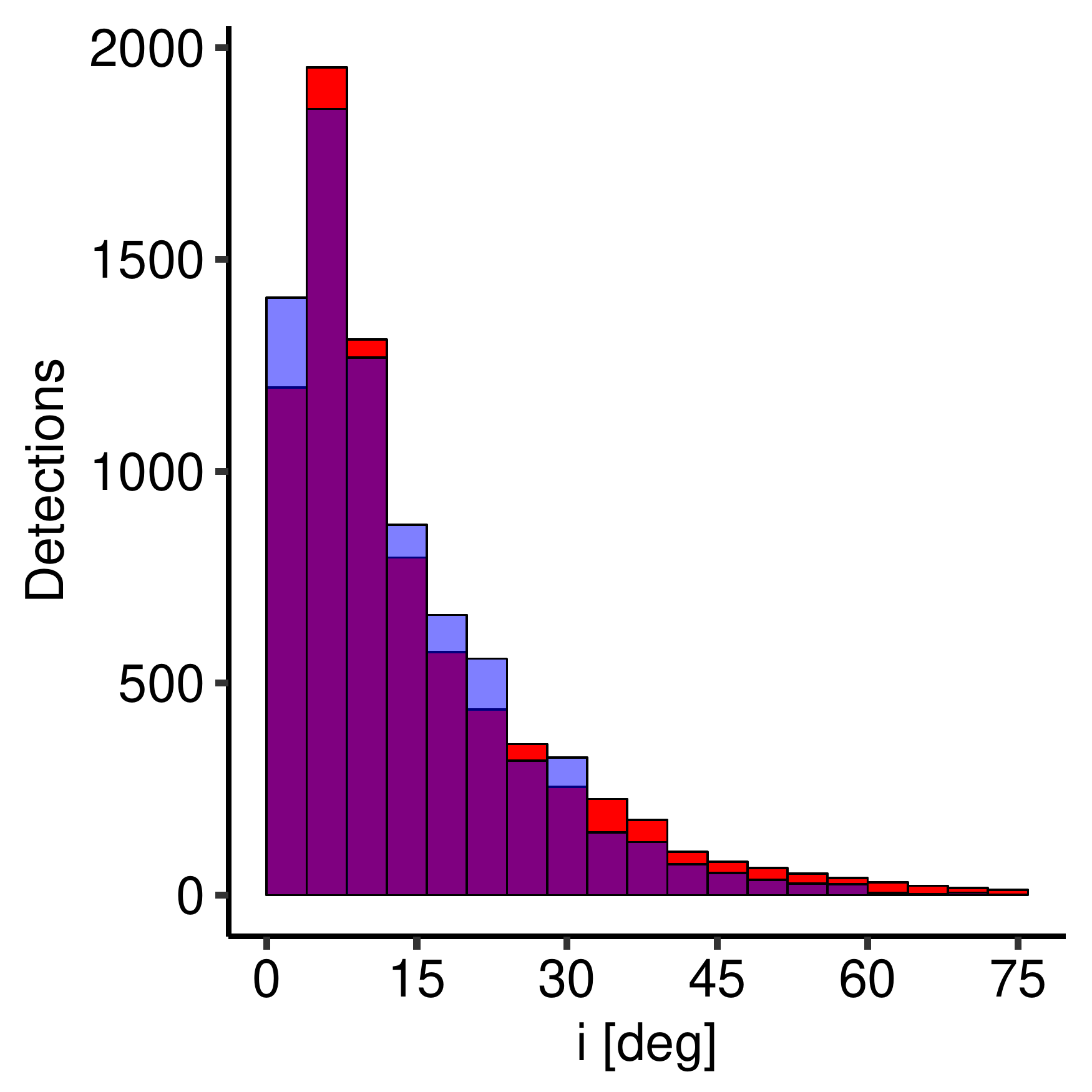}
  \includegraphics[width=0.49\columnwidth]{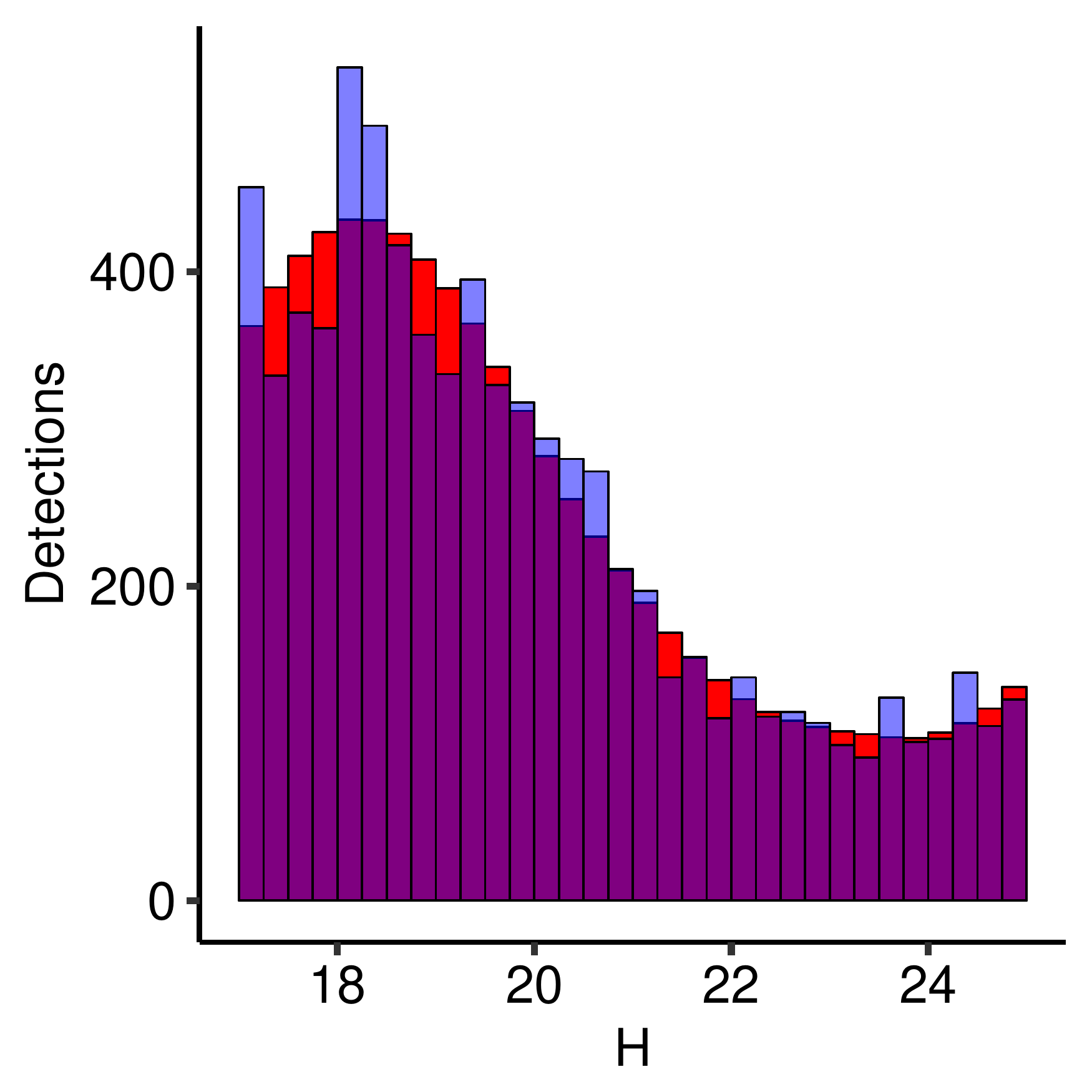}
  \includegraphics[width=0.49\columnwidth]{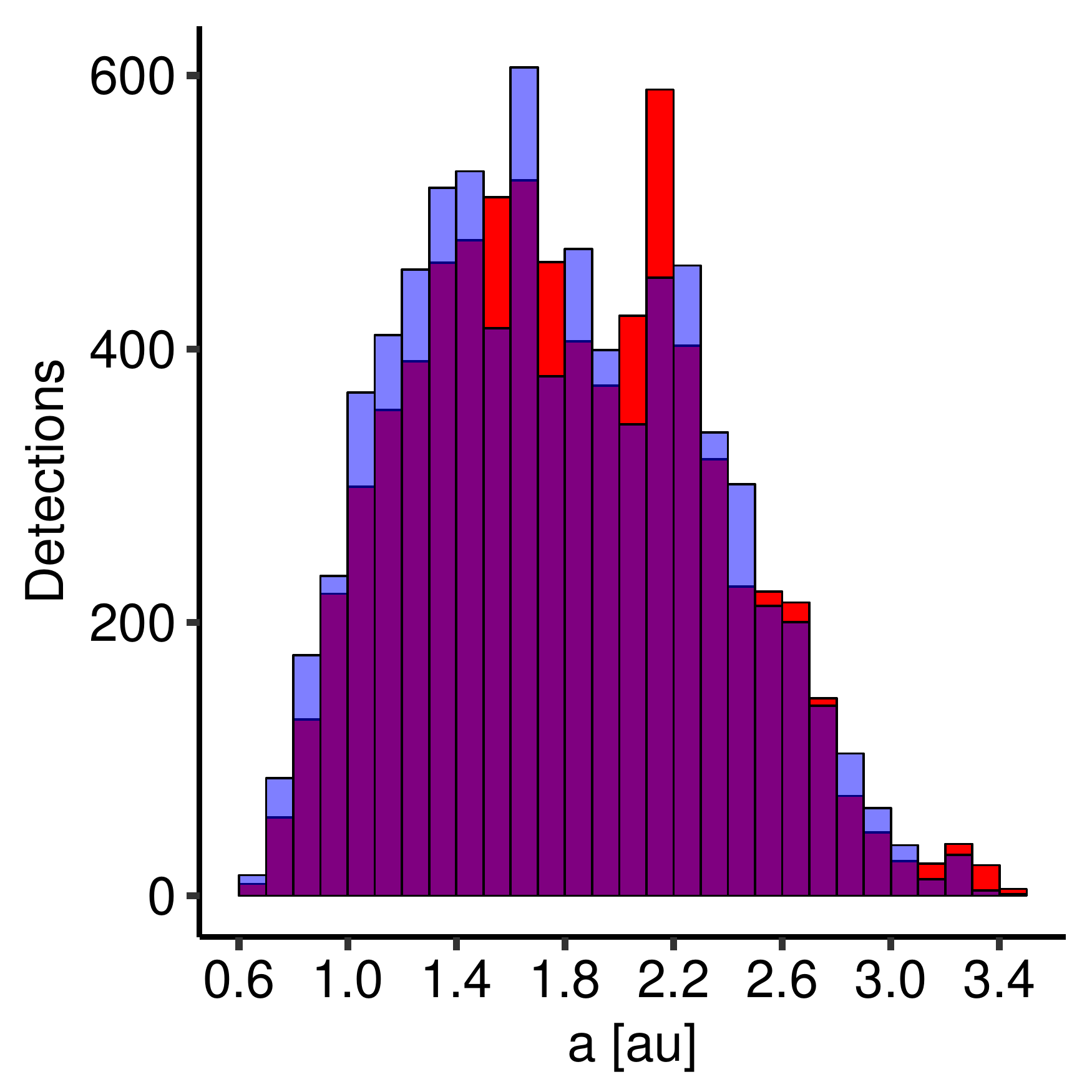}
  \includegraphics[width=0.49\columnwidth]{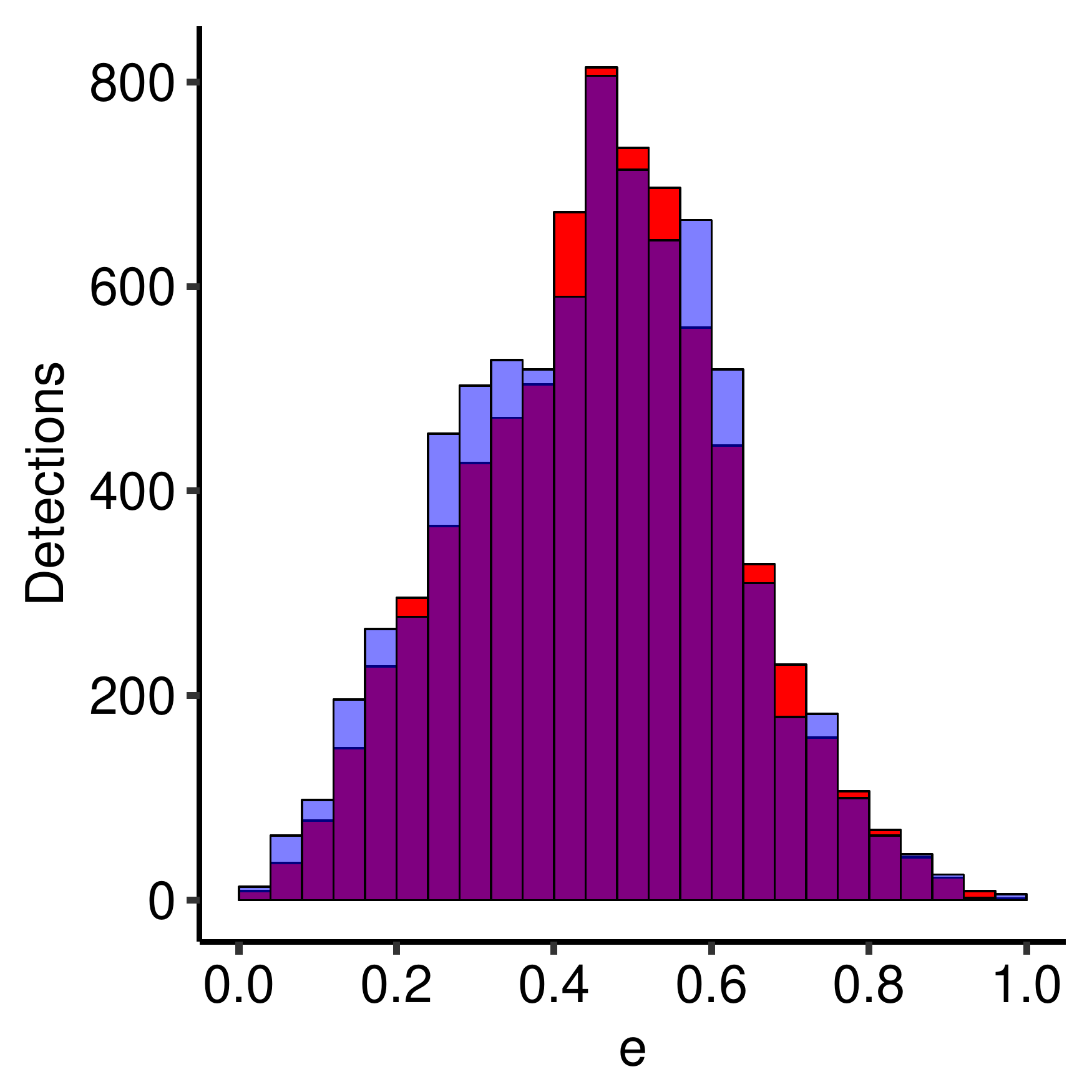}
  \includegraphics[width=0.49\columnwidth]{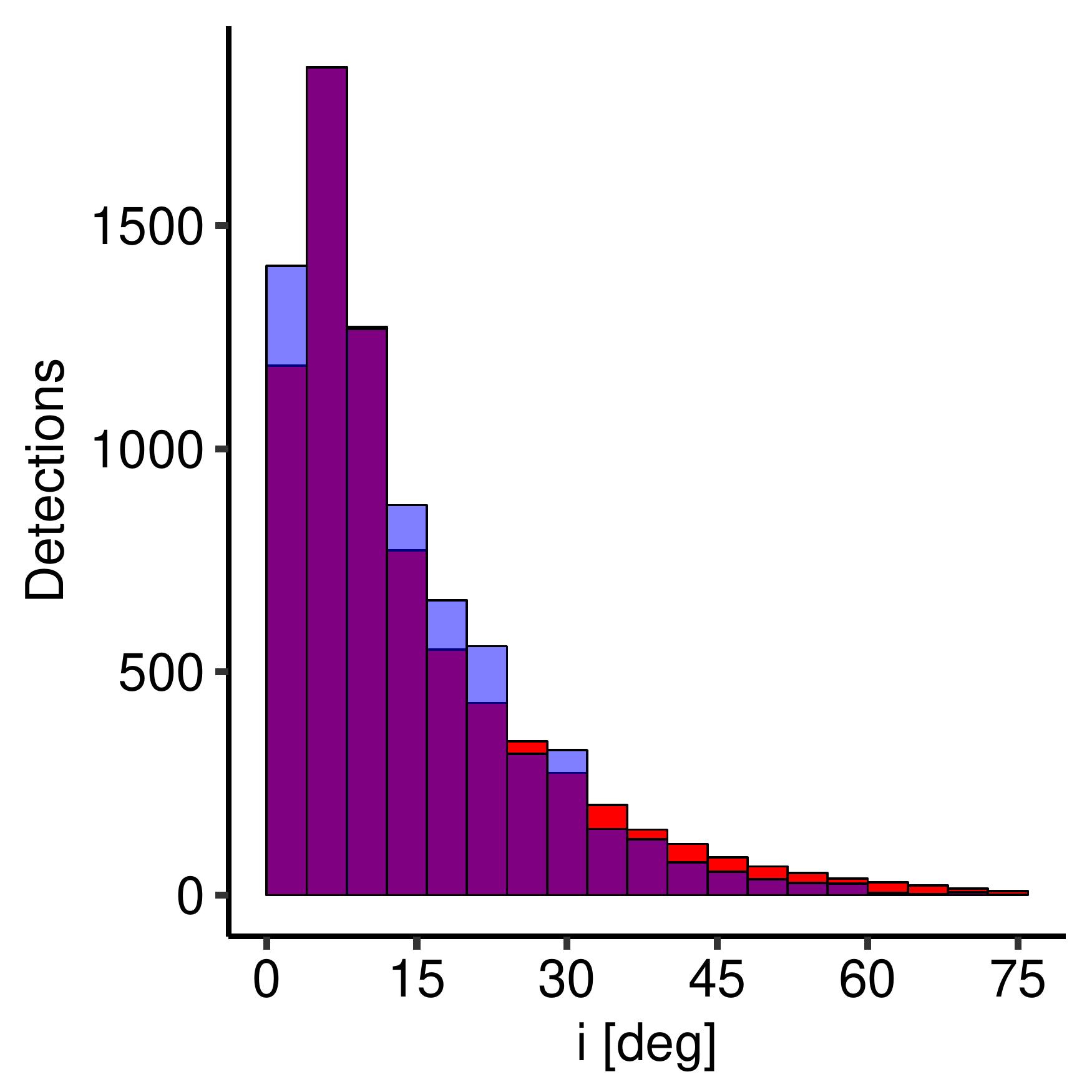}
  \includegraphics[width=0.49\columnwidth]{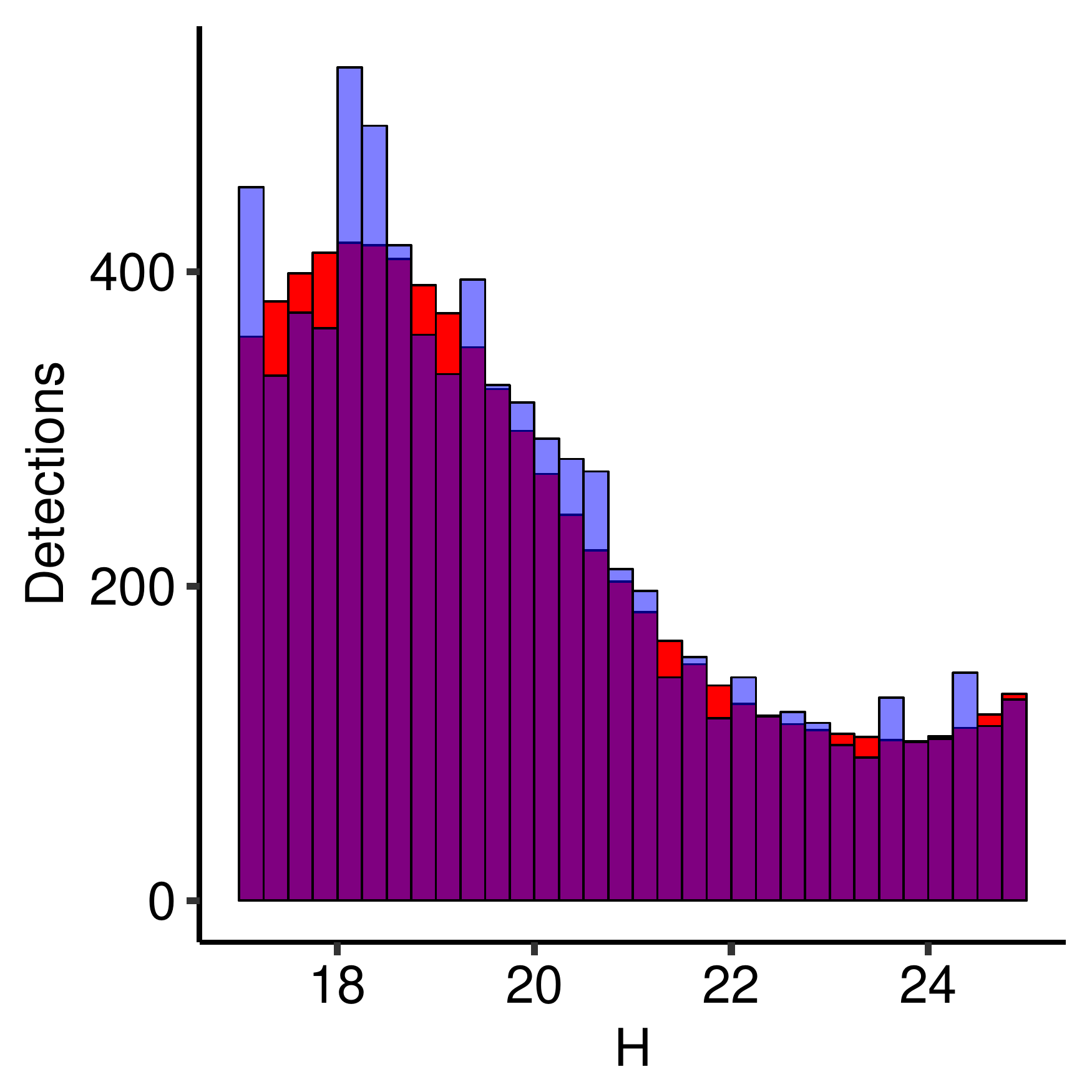}
  \caption{As Fig.~\ref{fig:bestfitaeih} but using steady-state orbit
    distributions based on (top row) even-numbered and (bottom row)
    odd-numbered test asteroids.}
  \label{fig:singlefits_evenodd}
\end{figure*}

To assess the sensitivity of the model on statistical variations in
the 7 included steady-state orbit distributions corresponding to the
ERs we divided the test asteroids into even-numbered and odd-numbered
sets to construct two independent orbit distributions. Then we
constructed the biased marginal $a$, $e$, $i$, and $H$ distributions
by using the two sets of orbit distributions and the best-fit
parameters found for the nominal model
(Table~\ref{table:parameters}). A comparison of the biased
distributions with the observed distributions show that both sets of
orbit distributions lead to an excellent agreement between model and
observations (Fig.~\ref{fig:singlefits_evenodd}). The largest
discrepancy between the biased marginal distributions is found for the
$a$ and $e$ distributions ($a\sim1.3\au$ and $e\sim0.2$). In general,
the variations in the orbit distributions are small and the systematic
uncertainties arising from the orbit distributions are negligible as
far as the nominal model is concerned.

\subsection{Sensitivity to observational data}
\label{sec:sensobs}

\begin{figure*}[h!]
  \centering
  {\bf \small Prediction for 703 detections using model based on 703 detections} \vspace{1mm} \\
  \includegraphics[width=0.45\columnwidth]{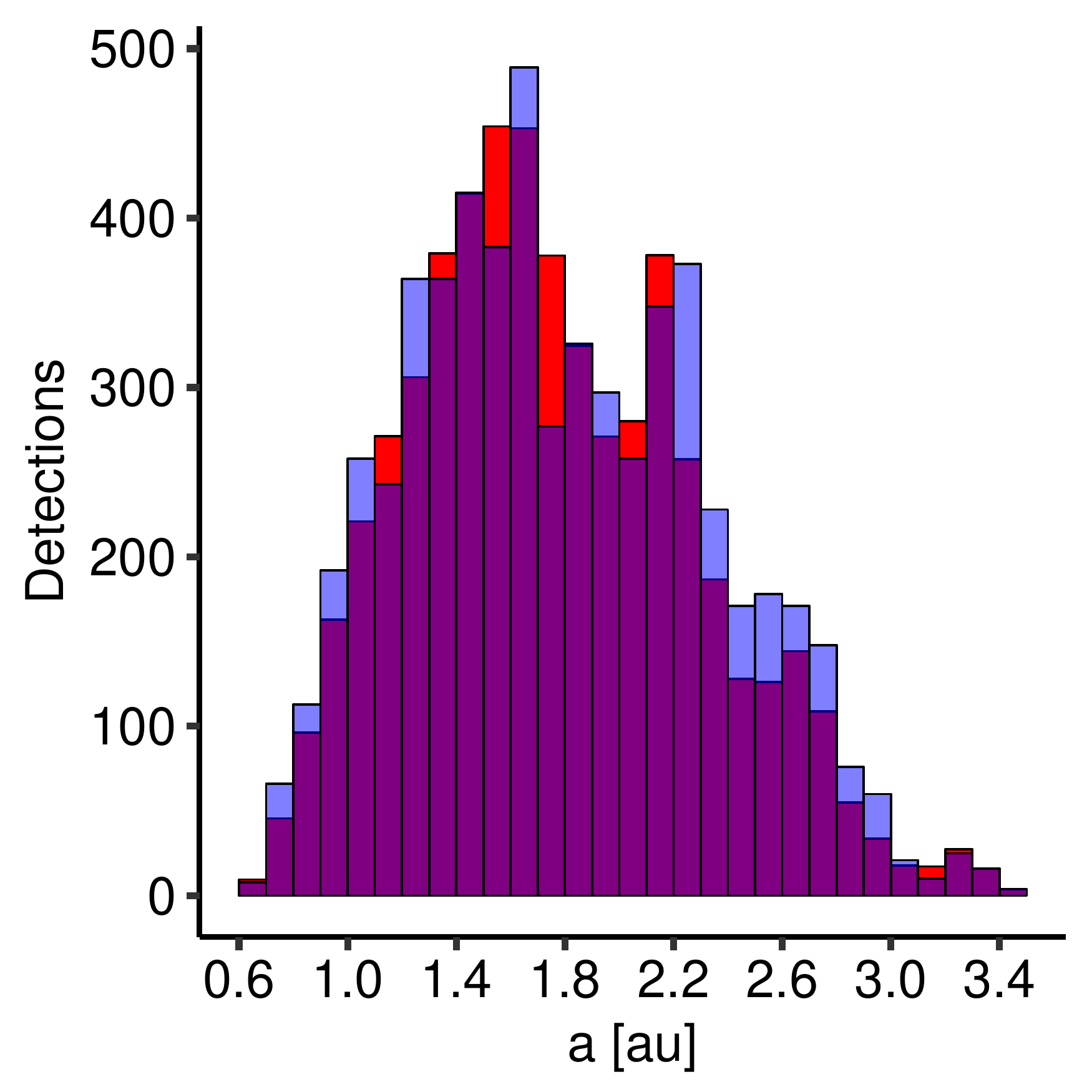} \vspace{2mm}
  \includegraphics[width=0.45\columnwidth]{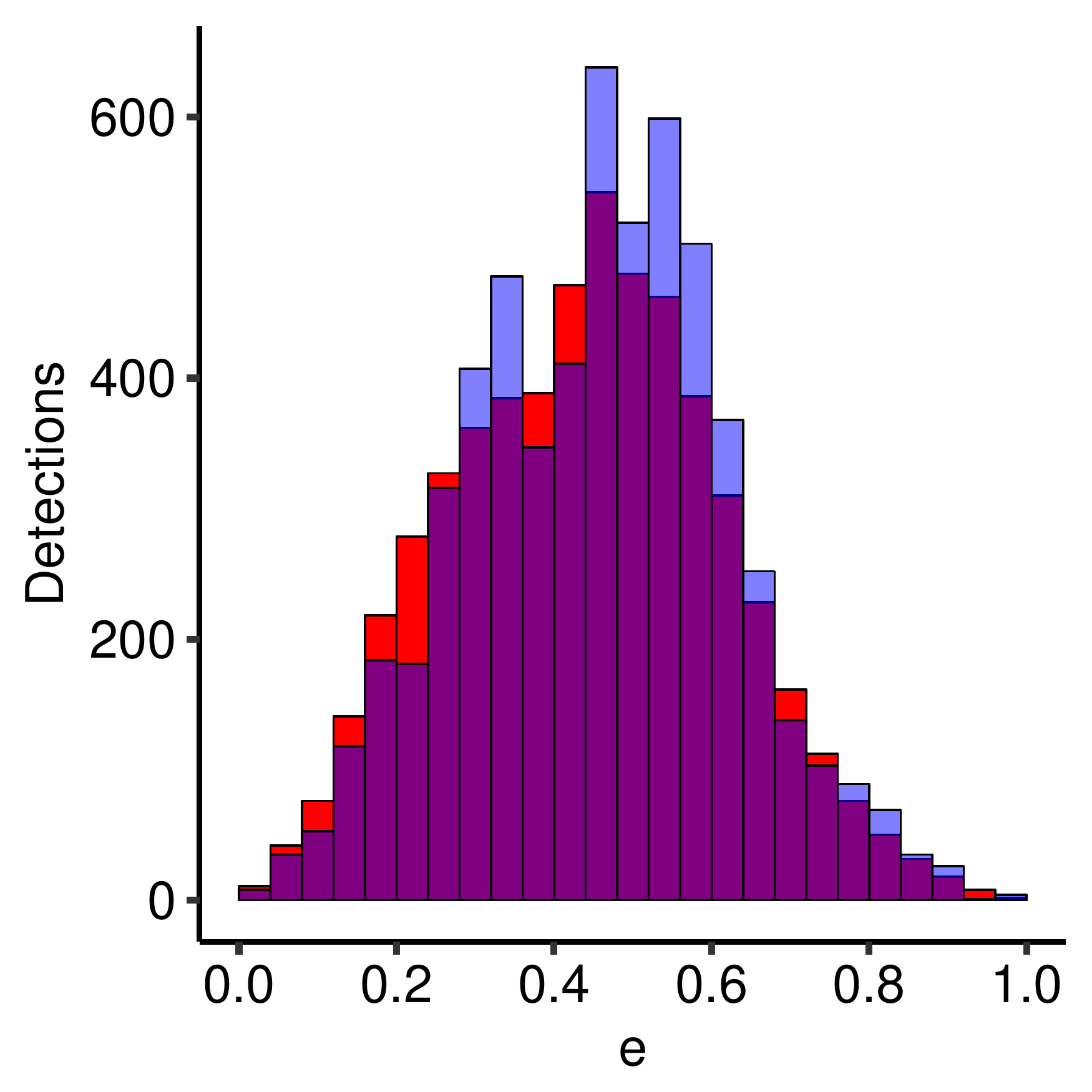}
  \includegraphics[width=0.45\columnwidth]{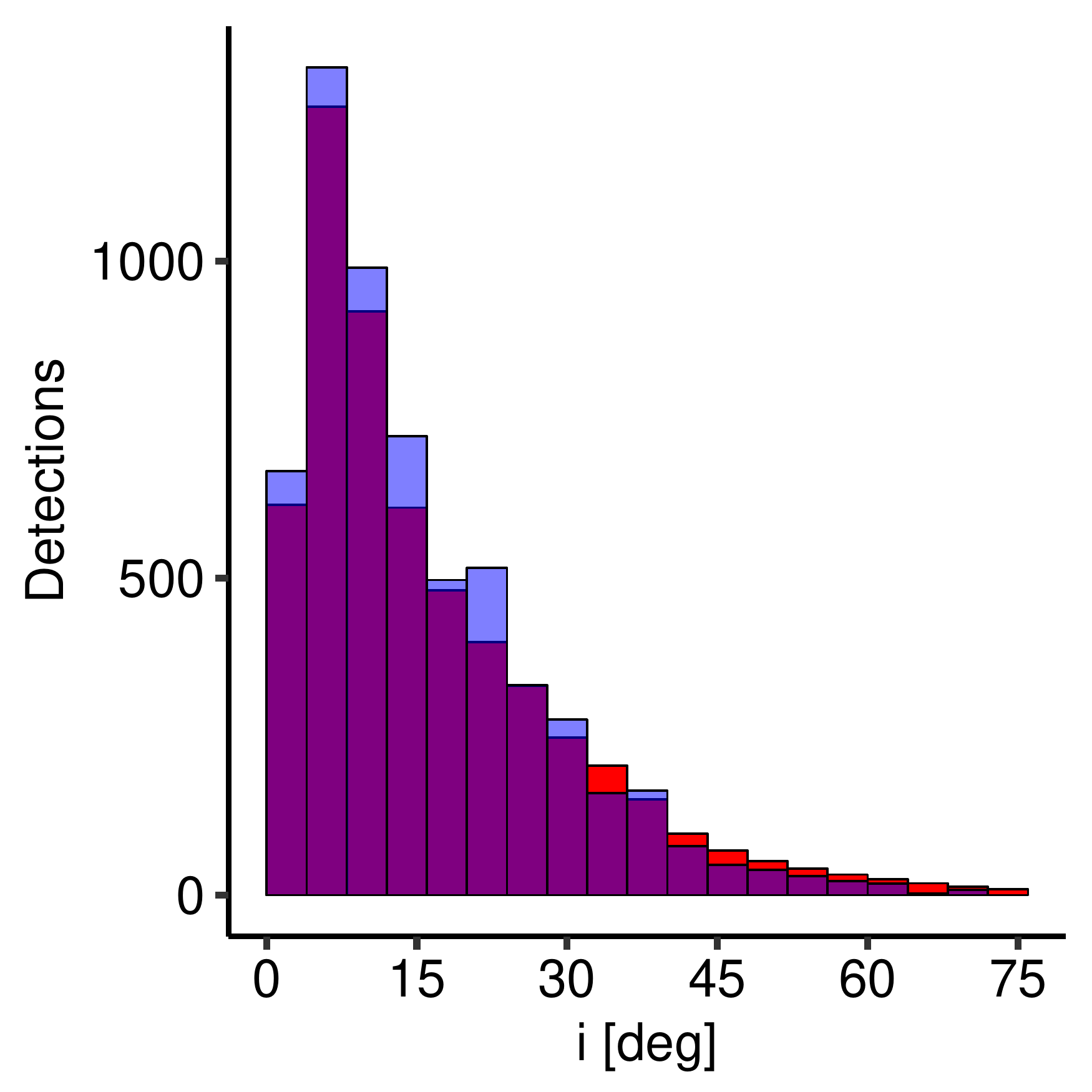}
  \includegraphics[width=0.45\columnwidth]{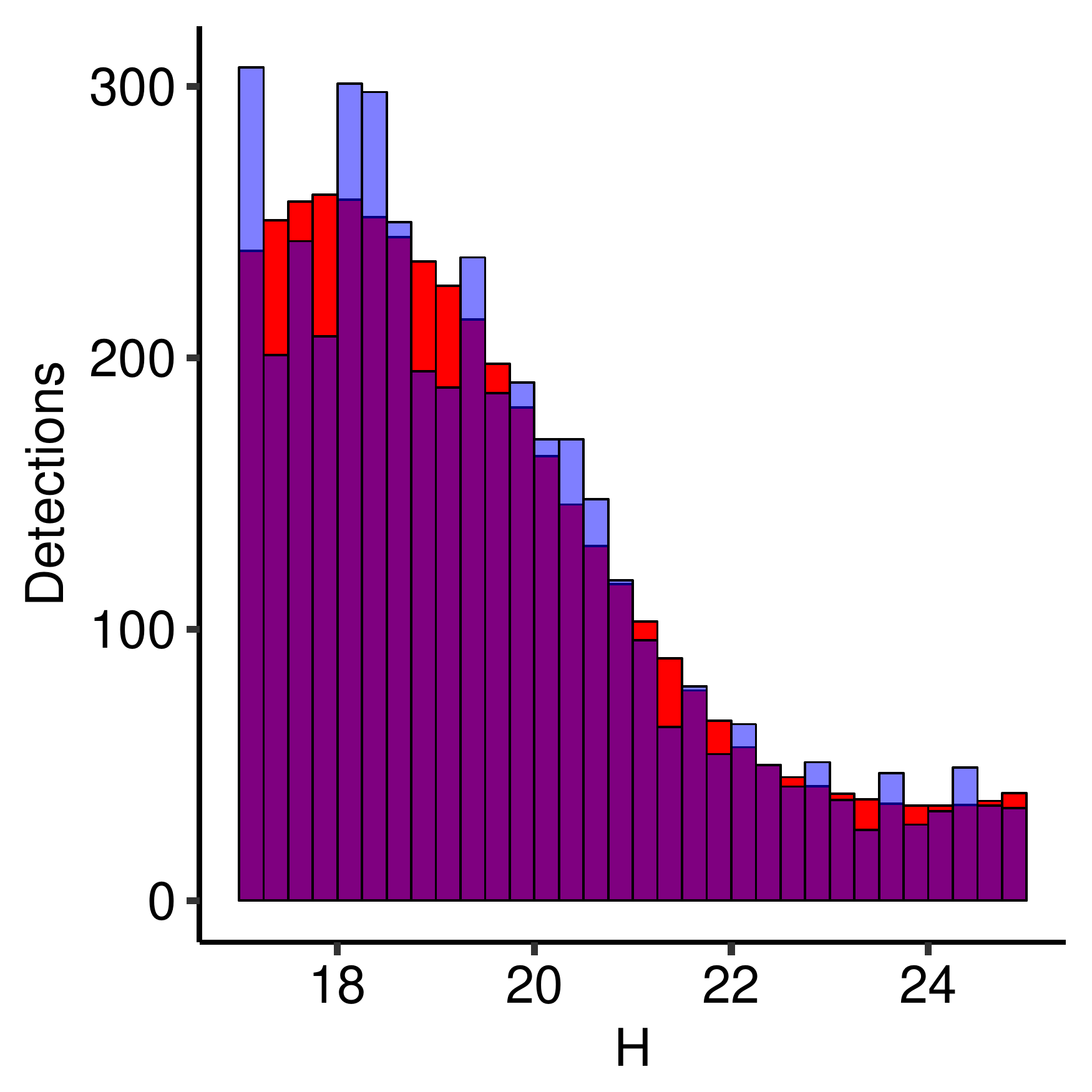} \\
  {\bf \small Prediction for 703 detections using model based on G96 detections} \vspace{1mm} \\
  \includegraphics[width=0.45\columnwidth]{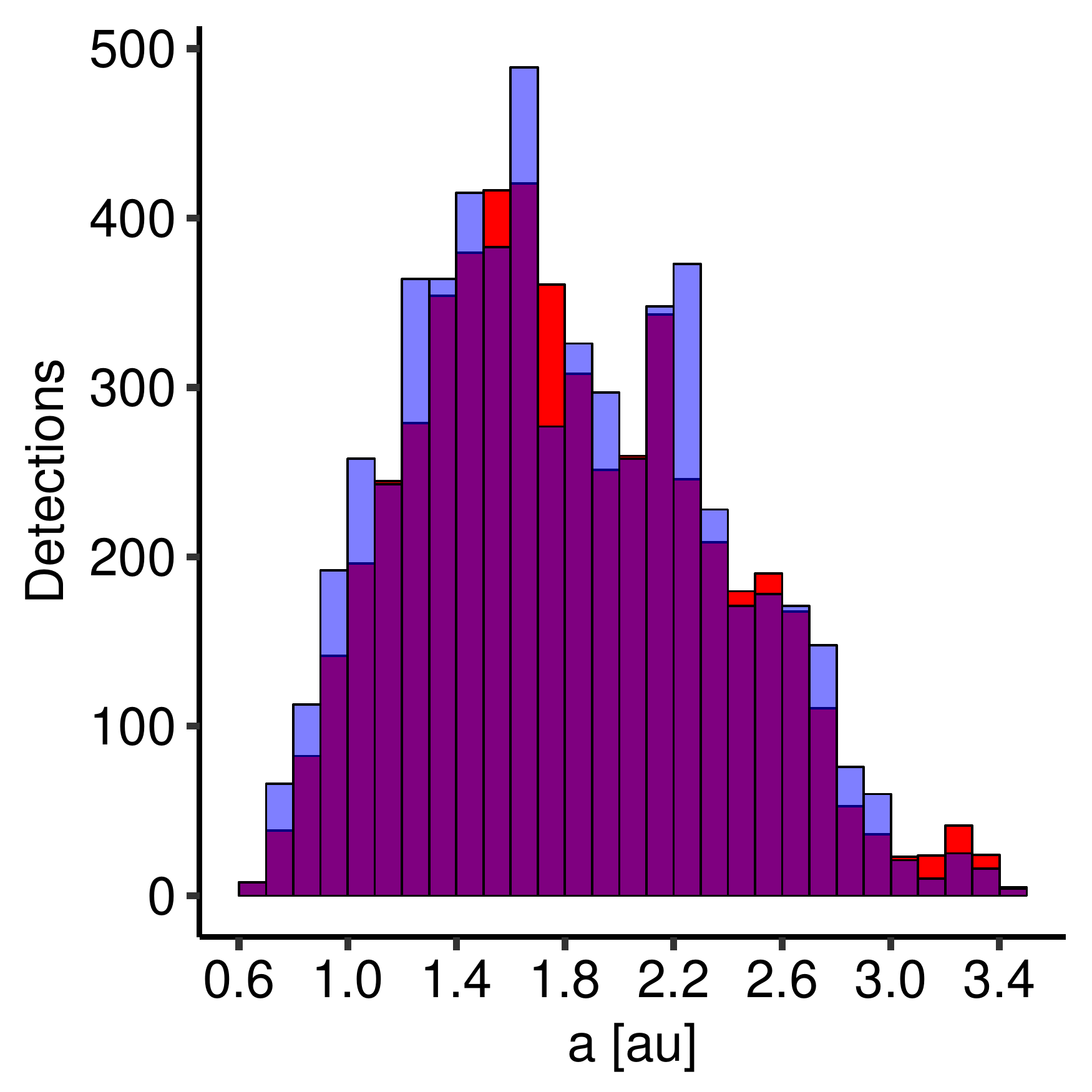} \vspace{2mm}
  \includegraphics[width=0.45\columnwidth]{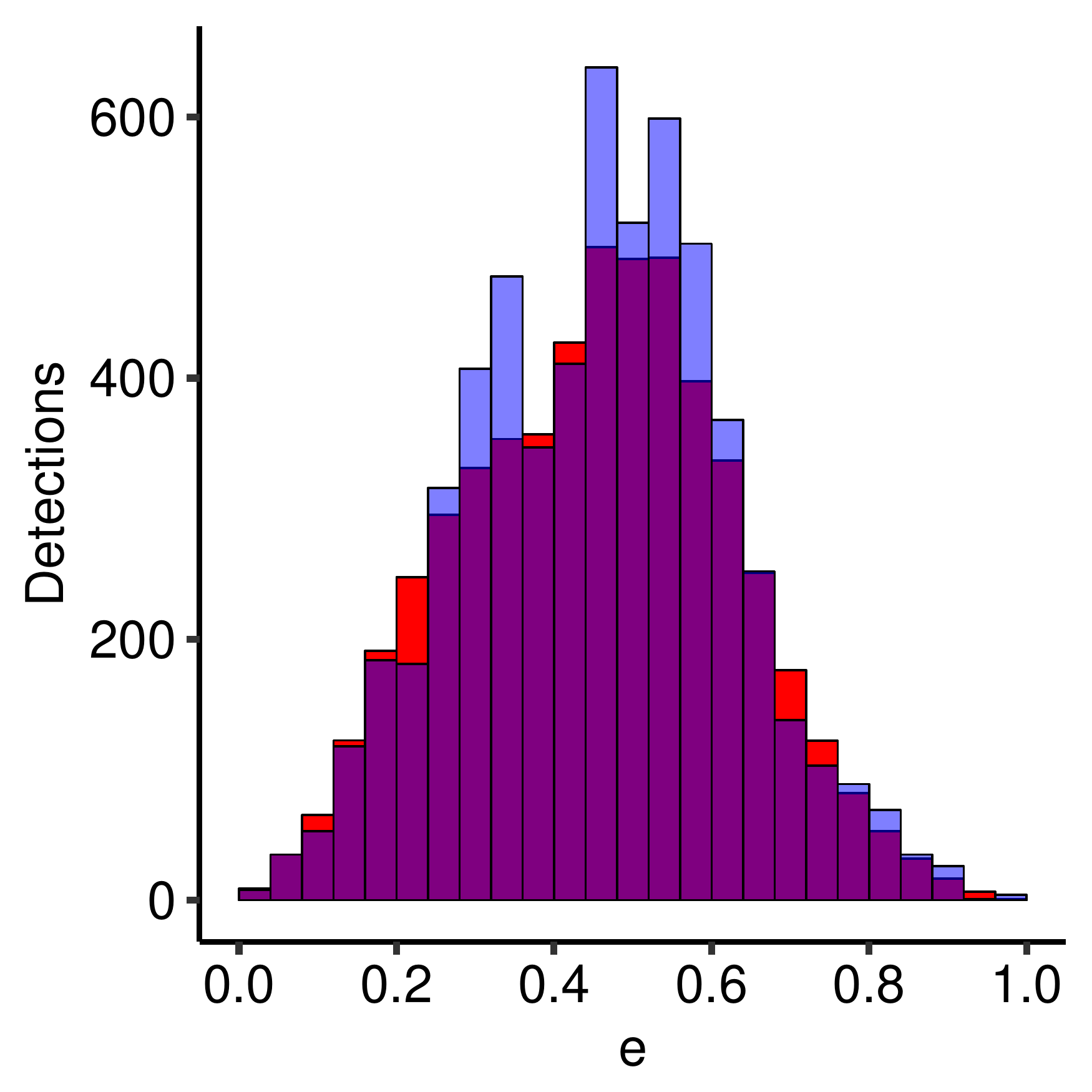}
  \includegraphics[width=0.45\columnwidth]{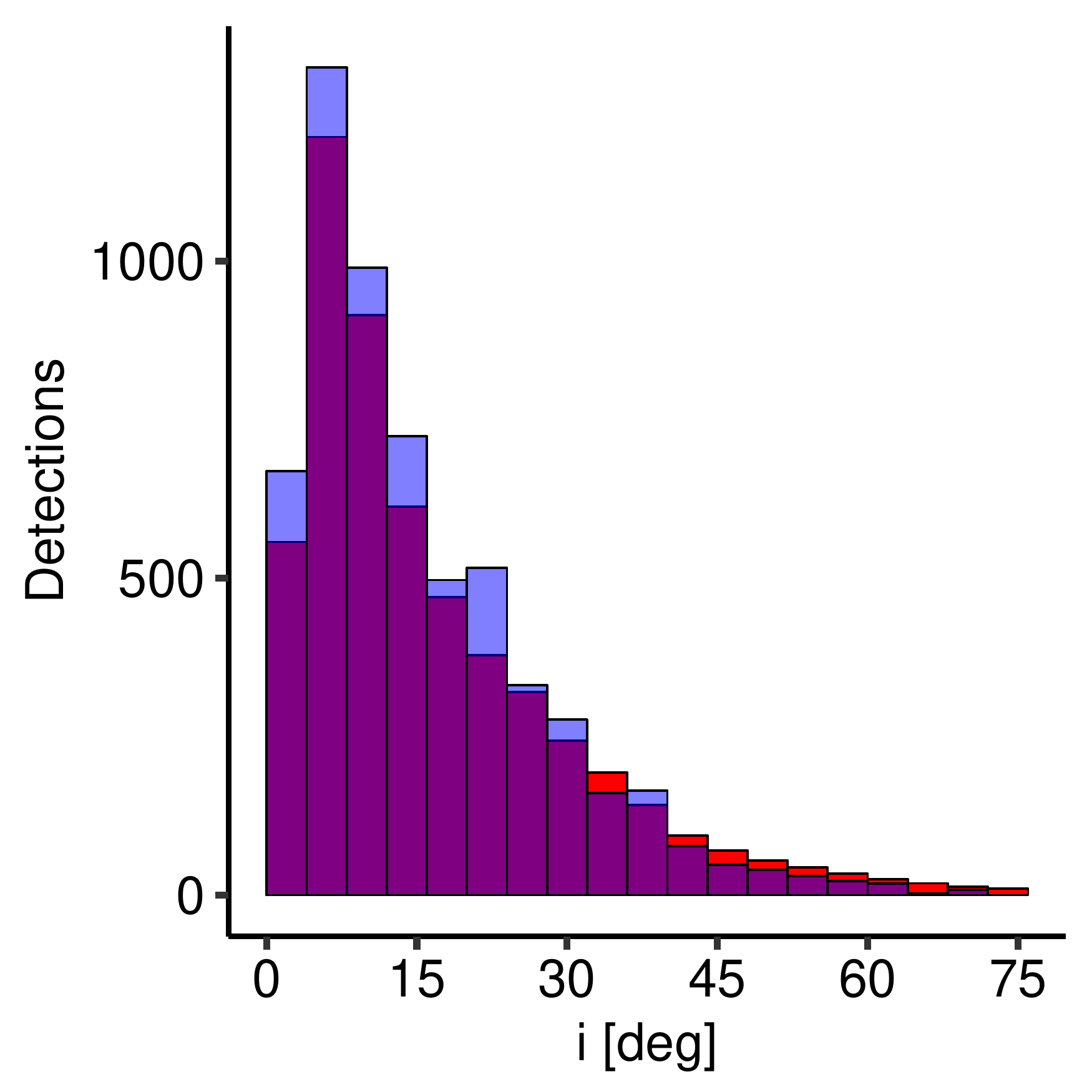}
  \includegraphics[width=0.45\columnwidth]{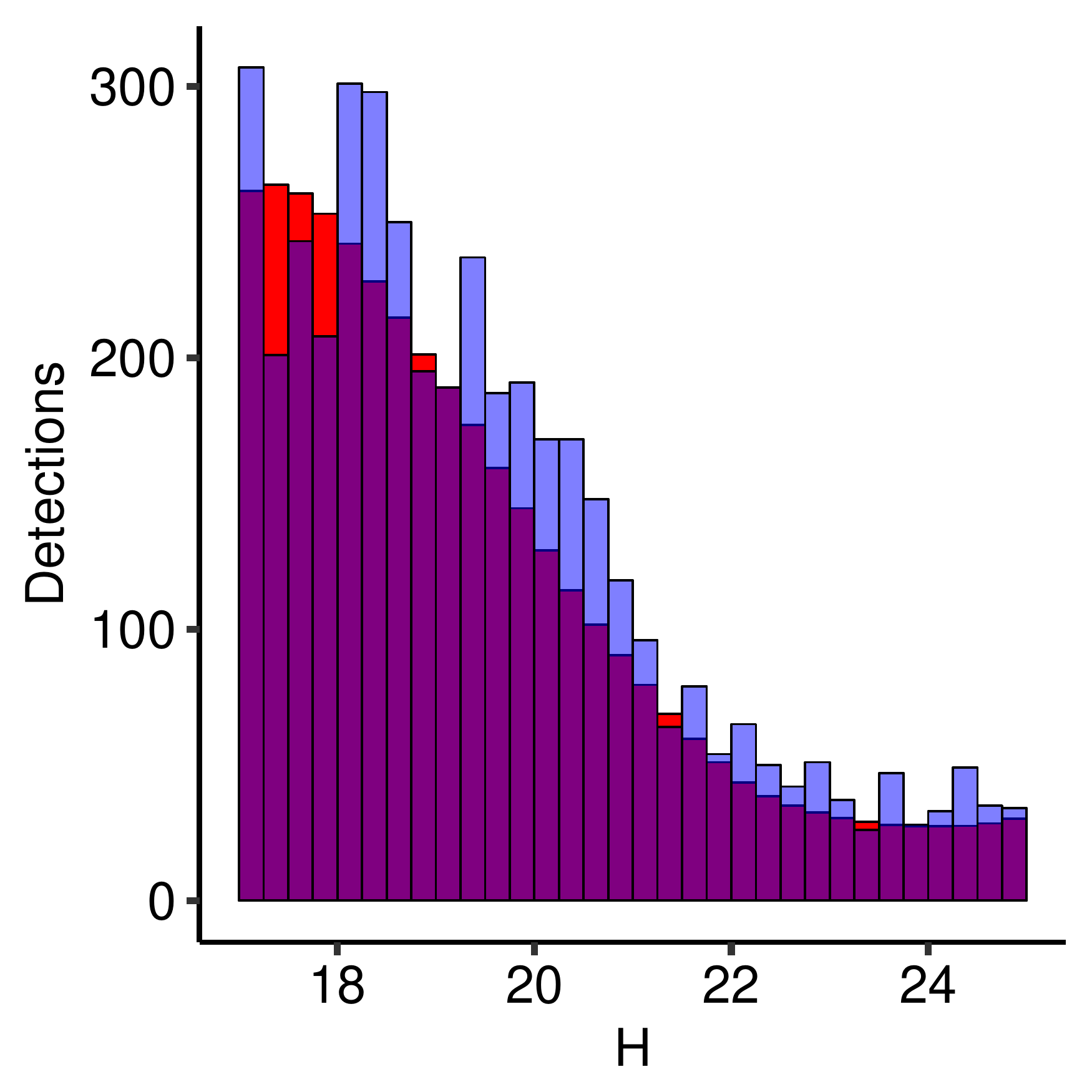} \\
  {\bf \small Prediction for G96 detections using model based on 703 detections} \vspace{1mm} \\
  \includegraphics[width=0.45\columnwidth]{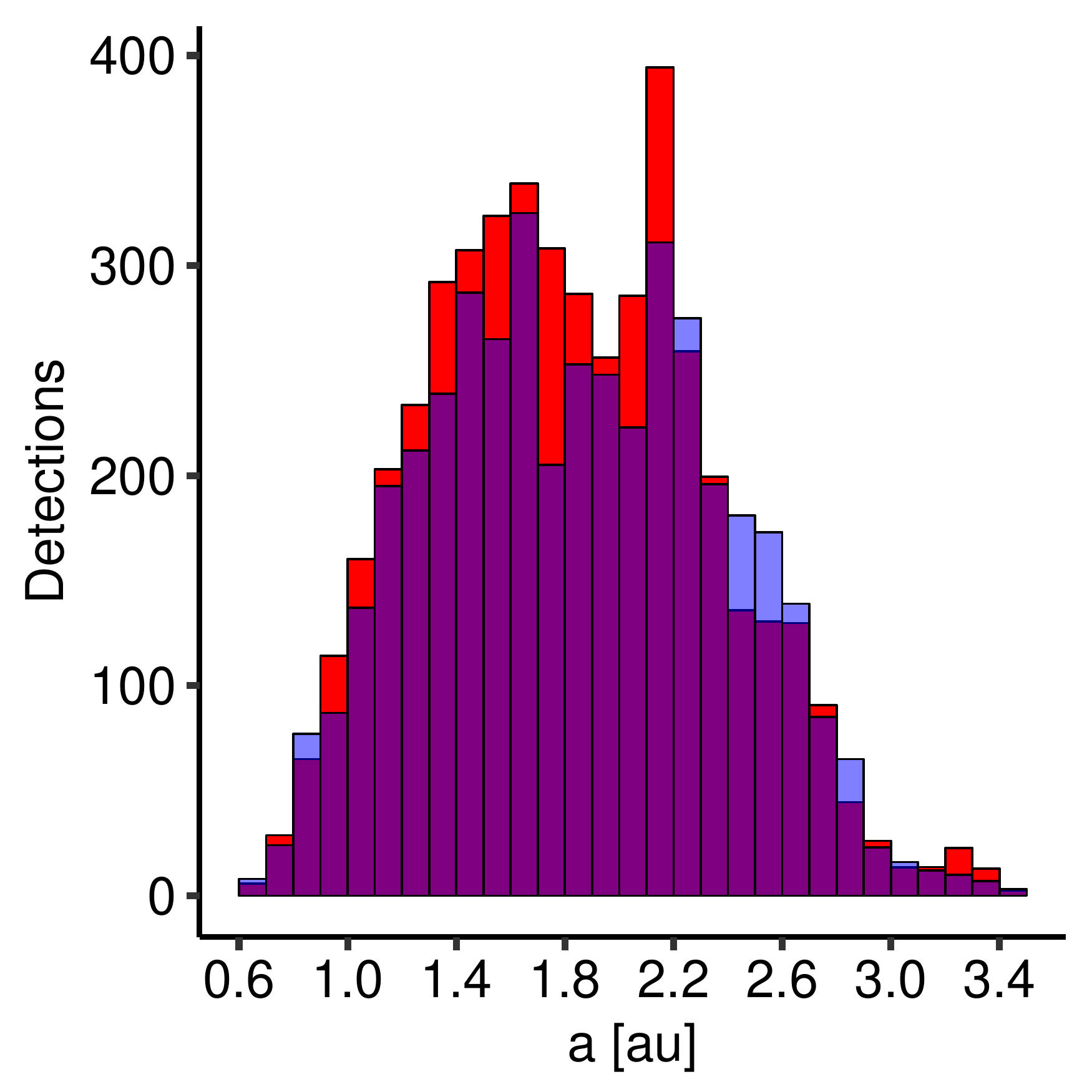} \vspace{2mm}
  \includegraphics[width=0.45\columnwidth]{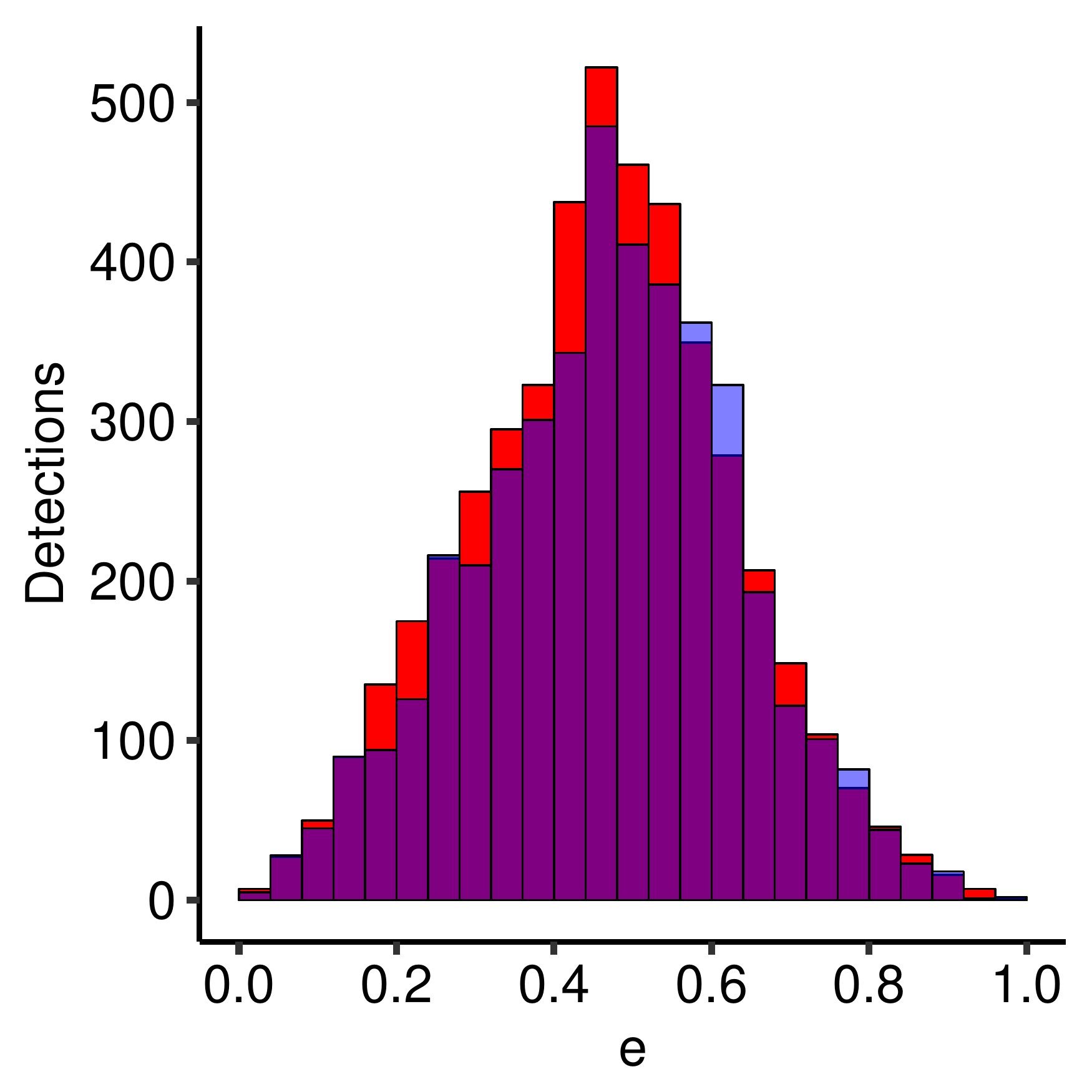}
  \includegraphics[width=0.45\columnwidth]{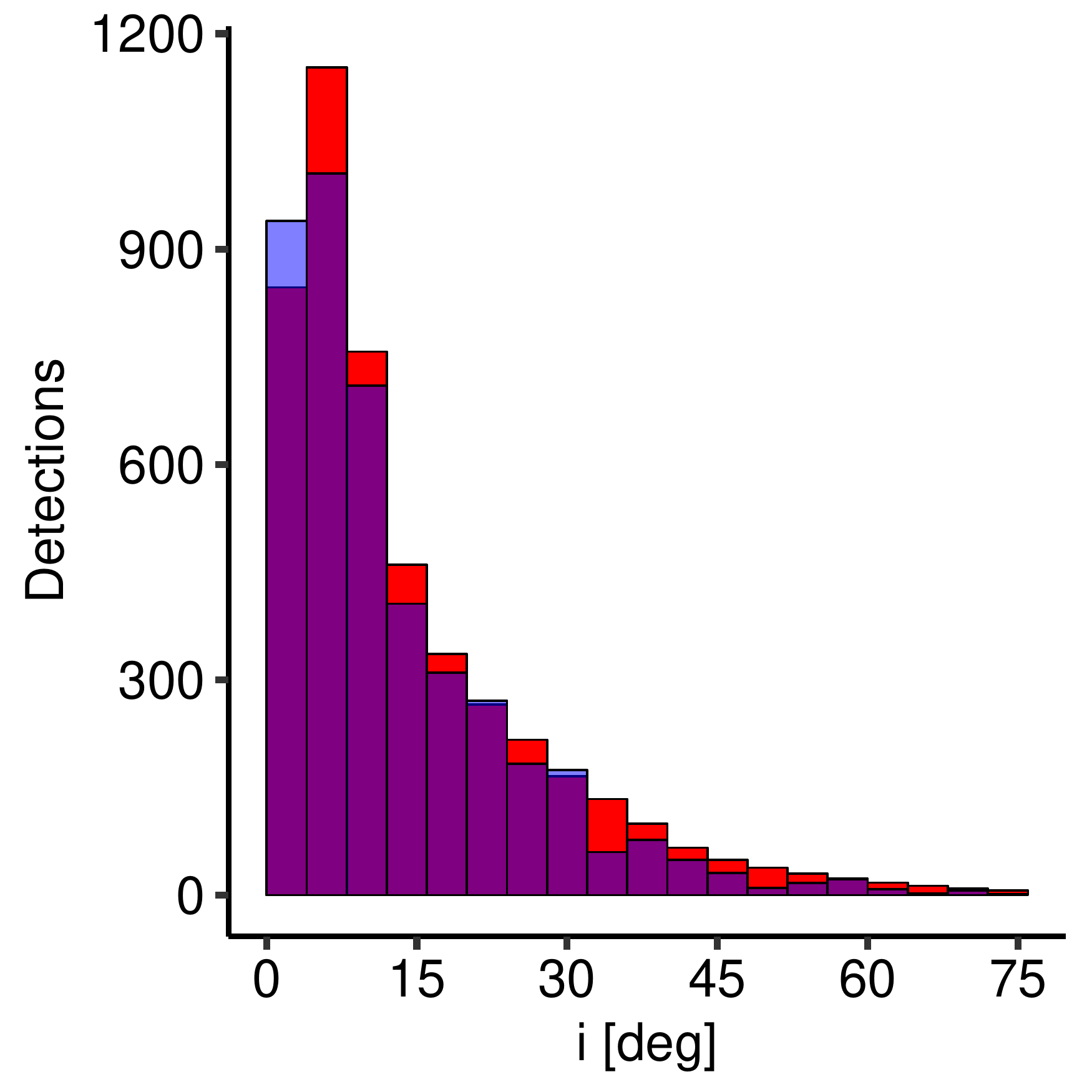}
  \includegraphics[width=0.45\columnwidth]{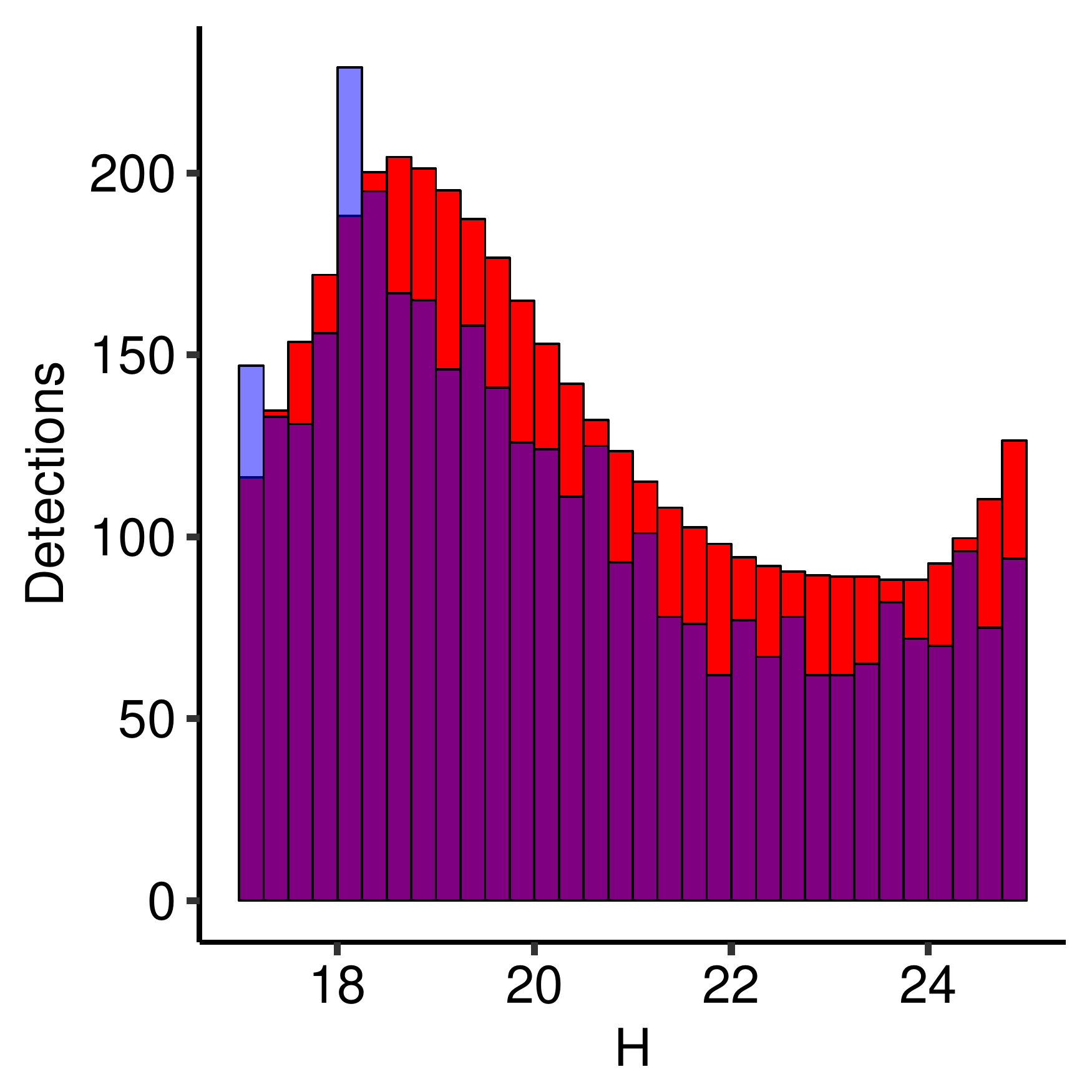} \\
  {\bf \small Prediction for G96 detections using model based on G96 detections} \vspace{1mm} \\
  \includegraphics[width=0.45\columnwidth]{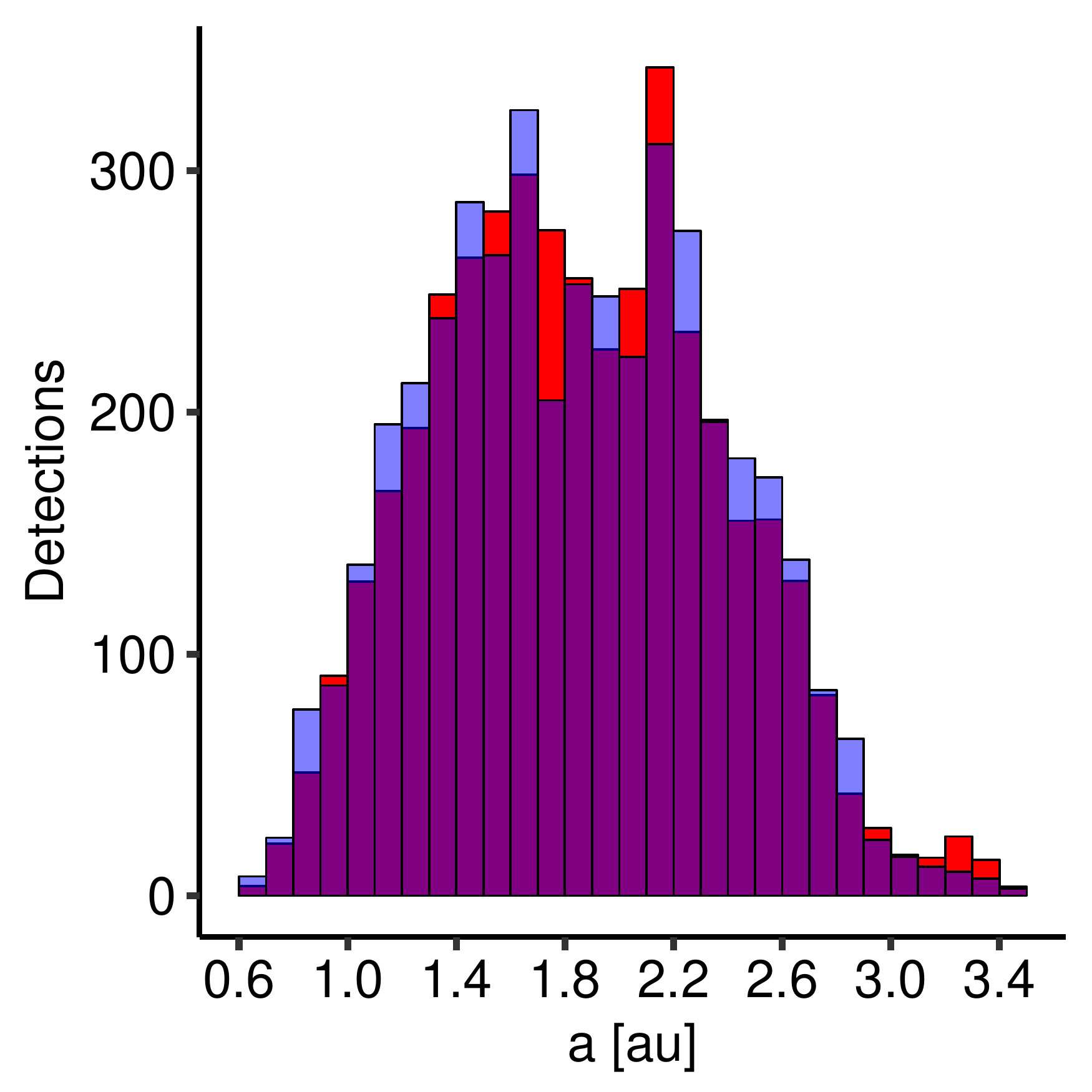}
  \includegraphics[width=0.45\columnwidth]{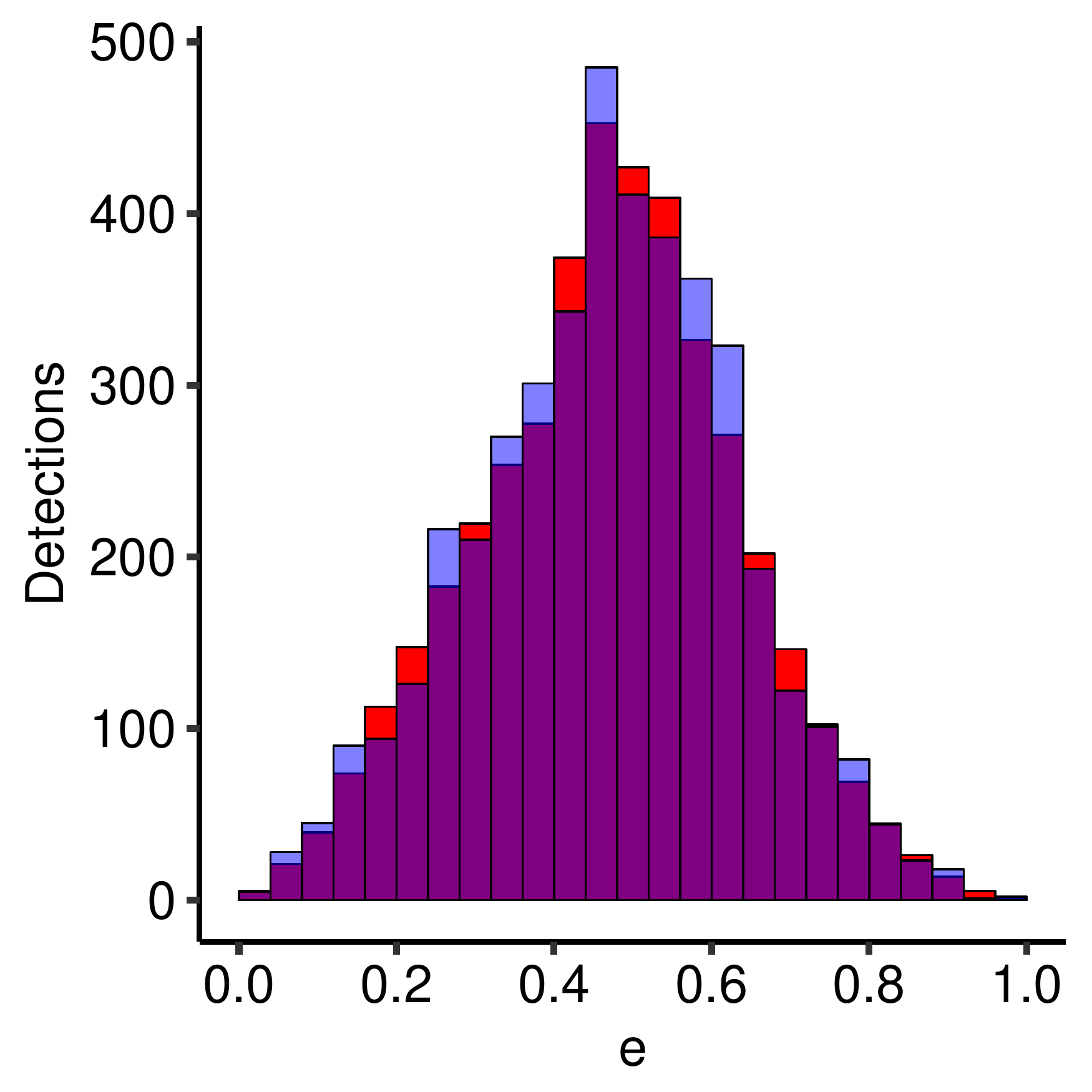}
  \includegraphics[width=0.45\columnwidth]{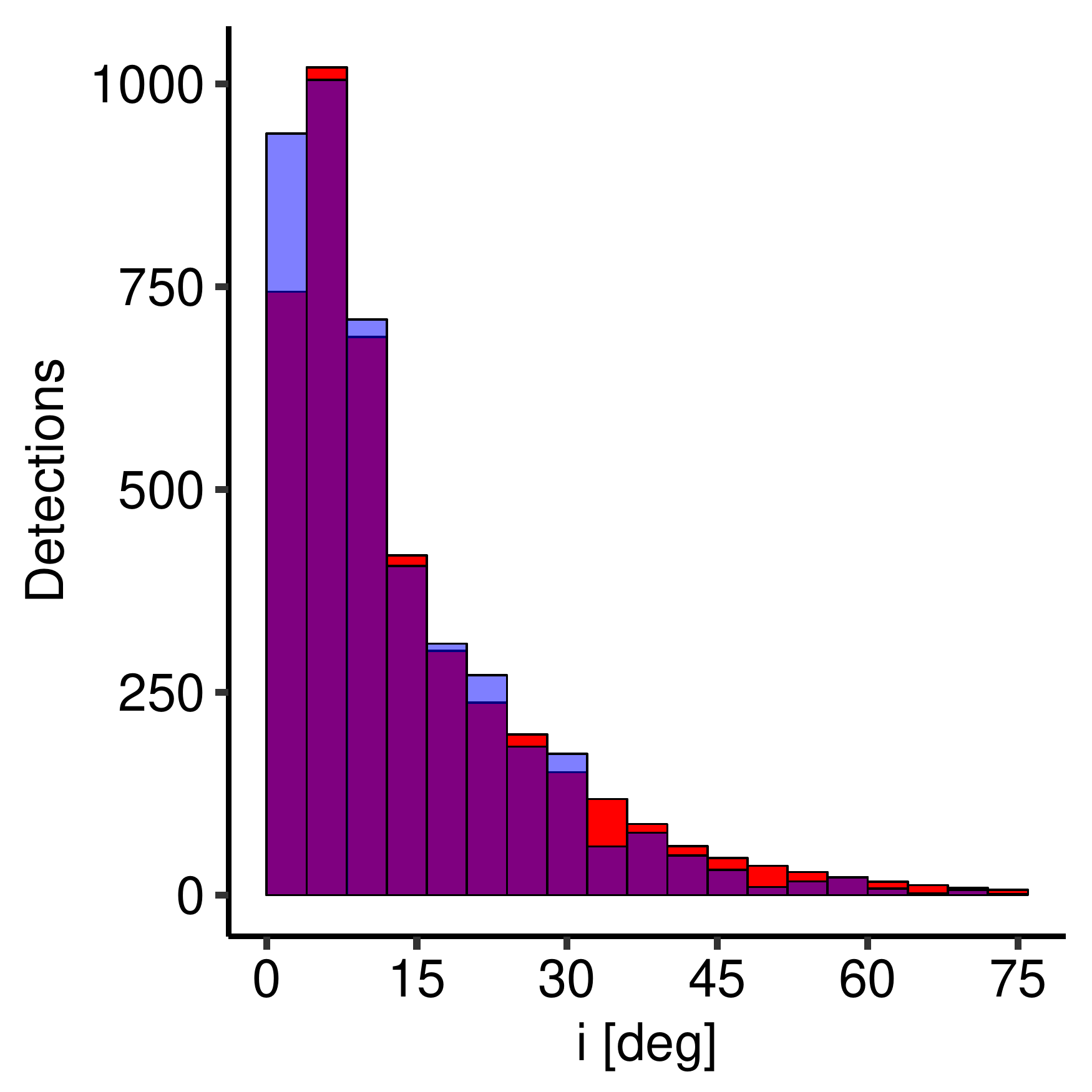}
  \includegraphics[width=0.45\columnwidth]{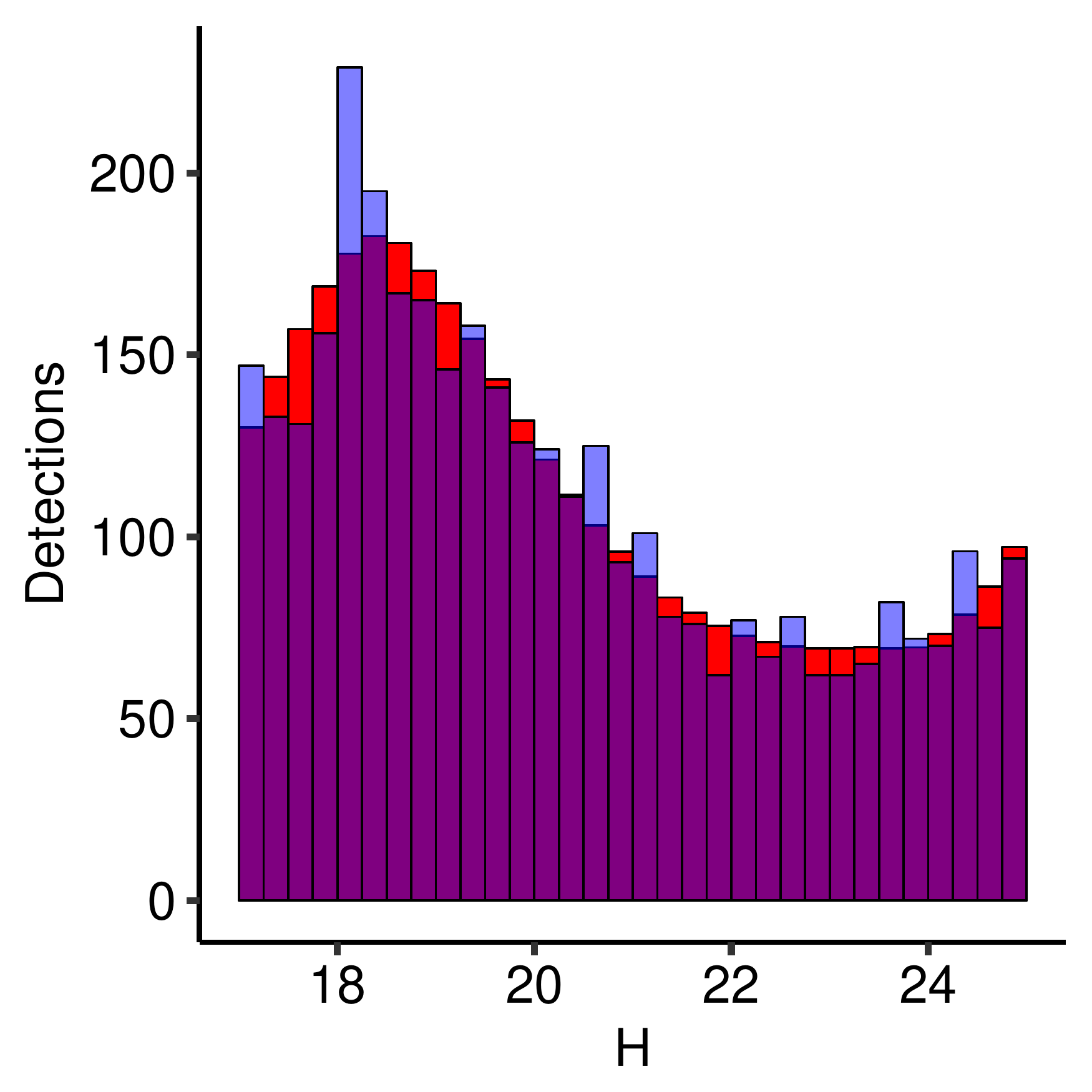}
  \caption{(Top row) Comparison between detections by 703 (blue;
    $n(a,e,i,H)$) and prediction based on the best-fit model using
    data by 703 (red; $\epsilon(a,e,i,H)\,N(a,e,i,H)$).  (Upper middle
    row) Comparison between detections by 703 (blue) and prediction
    based on the best-fit model using data by G96 (red). The purple
    color indicates overlapping distributions. (Lower middle row)
    Comparison between detections by G96 (blue) and prediction based
    on the best-fit model using data by 703 (red). (Bottom row)
    Comparison between detections by G96 (blue) and prediction based
    on the best-fit model using data by G96 (red).}
  \label{fig:singlefits}
\end{figure*}

To assess the sensitivity of the model on the data set used for its
calibration, we first divided the observations into two sets: those
obtained by G96 and those obtained by 703. Then we constructed two
models based on the data sets by using an approach otherwise identical
to that described in Sect.~\ref{sec:nominal7smodel}. The marginal
($a$,$e$,$i$,$H$) distributions of the biased model,
$\epsilon(a,e,i,H)\,N(a,e,i,H)$, compared to the observations show
that both models accurately reproduce the observations from which they
were derived (top and bottom panels in Fig.~\ref{fig:singlefits}).

The critical test is then to predict, based on a model of G96 (703),
what the other survey, 703 (G96), should have observed and compare
this to the actual observations. For example, we should be able to
detect problems with the input inclination distribution or the bias
calculations, because the latitude distribution of NEO detections by
703 is wider than the latitude distribution by G96. It turns out that
the shapes of the model distributions accurately match the observed
distributions, but there are minor systematic offsets in the absolute
scalings of the distributions (upper and lower middle panels in
Fig.~\ref{fig:singlefits}). The model based on detections by 703 only
predicts about 9\% too many detections by G96 whereas the model based
on detections by G96 only predicts about 8\% too few detections by
703. We interpret the discrepancy as a measure of the systematic
uncertainty arising from the data set used for the modeling. Given
that the models are completely independent of each other we consider
the agreement satisfactory.  In fact, none of the models described in
the Introduction \citep[except that of][]{2016Natur.530..303G} have
undergone a similar test of their predictive power. We also stress
that the nominal model combines the 703 and G96 data sets and thus the
systematic error is reduced from the less than 10\% seen here.

\begin{figure}[h]
  \centering
  \includegraphics[width=\columnwidth]{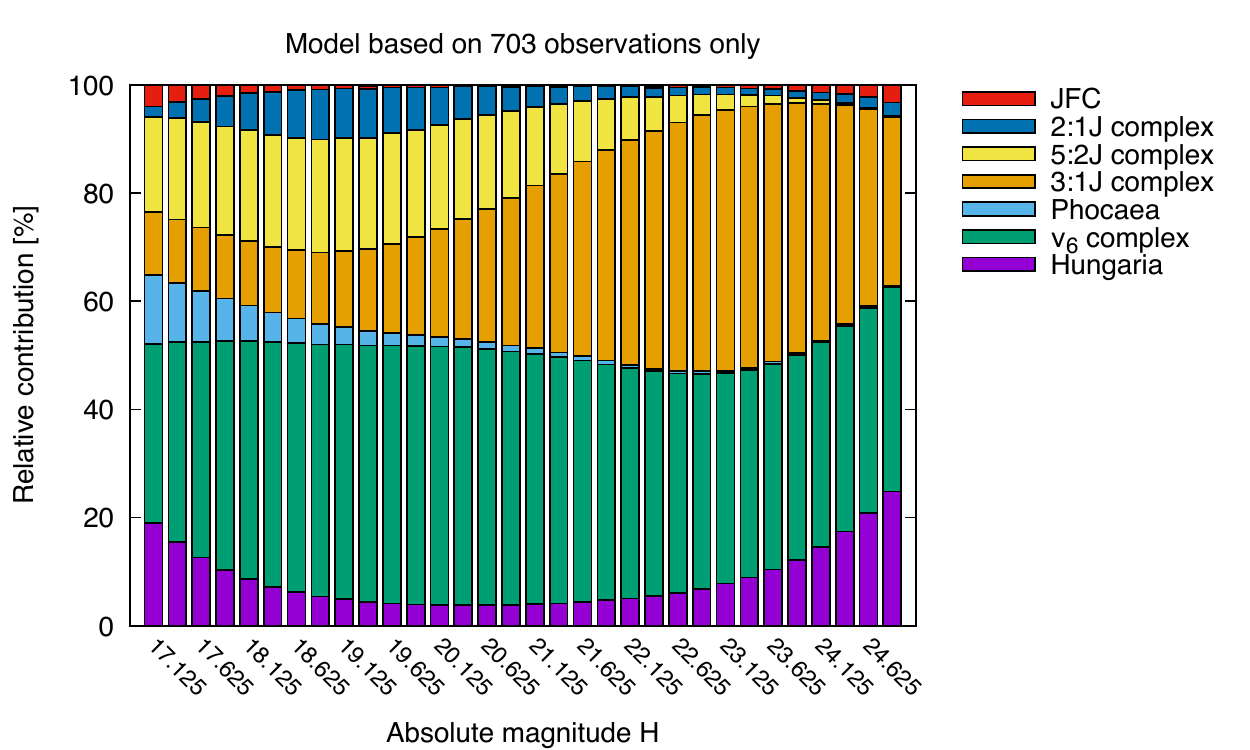}
  \includegraphics[width=\columnwidth]{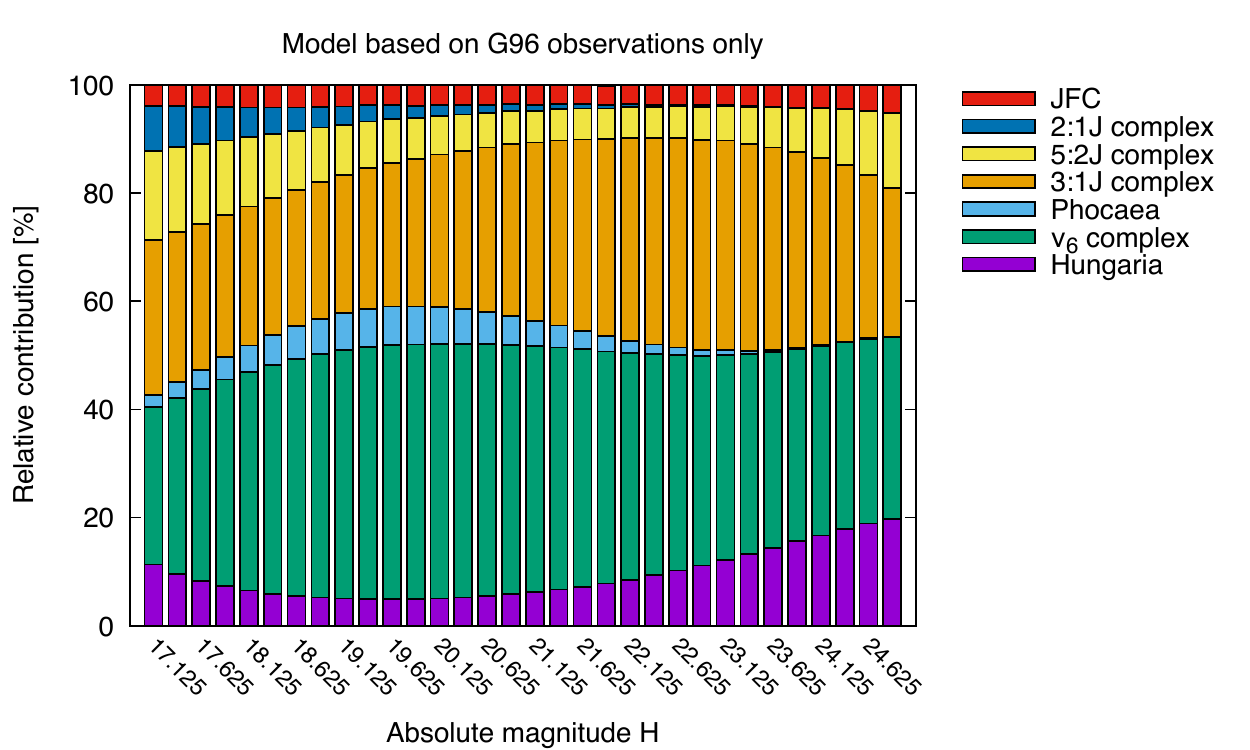}
  \caption{Relative importance of 7 different NEO source regions as a
    function of the absolute magnitude when the model is based only on
    (top) 703 and (bottom) G96.}
  \label{fig:relsourcessing}
\end{figure}

We can take the test further by making predictions about the relative
importance of different source regions as a function of $H$
(Fig.~\ref{fig:relsourcessing}).
The overall picture is consistent with our expectations in that the
$\nu_6$ and 3:1J dominate both models. The difference for $\nu_6$ can
largely be described by a constant such that the G96 model gives is a
systematically smaller (by a few \%-units) importance compared to the
703 model. The Hungarias show similar trends in both models, but the
difference cannot be described as a simple systematic offset. Phocaeas
show a clear difference in that their relative importance peaks at
$H=17$ and $H\sim19.5$ for 703 and G96, respectively. The common
denominator for Phocaeas is that both models give them negligible
importance for $H\gtrsim23$. The 3:1J shows a similar overall shape
with the importance peaking at $H\sim23$ although the 703 model
predicts a lesser importance at $H\lesssim21$ compared to the G96
model. The G96 model gives a similar importance for 5:2J throughout
the $H$ range whereas the 703 model gives a similar overall importance
but limiting it to $H\lesssim23$. The 2:1J has the opposite behaviour
compared to the 5:2J in that its importance in the G96 model is
primarily limited to $H\lesssim21$ whereas its importance in the 703
model gives it a non-negligible importance throughout the $H$
range. Finally, the G96 model gives JFCs a nearly constant (a few
\%-units) importance throughout the $H$ range whereas the 703 model
predicts that their importance is negligible for $19 \lesssim H
\lesssim 23.5$. We stress that this analysis does not account for
uncertainties in model parameters which would make a difference in the
relative importance more difficult to assess, in particular at large
$H$ (see, e.g., Fig.~\ref{fig:neosfds}).

\subsection{Uniqueness}

\begin{figure}[h]
  \centering
  \includegraphics[width=\columnwidth]{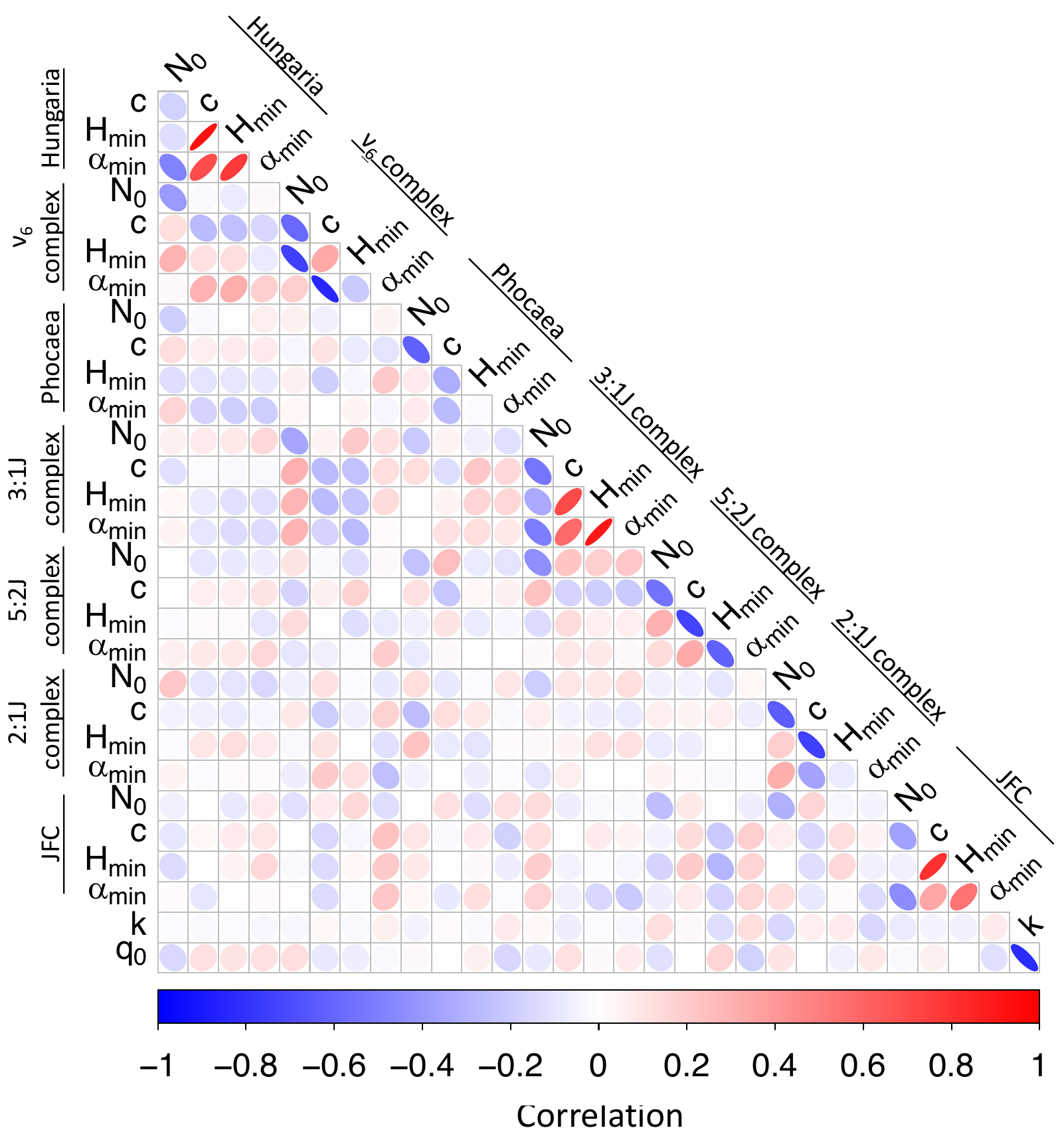}
  \caption{Correlations between model parameters.}
  \label{fig:paramcorr}
\end{figure}

Although the similarity of the solutions based on different data sets
suggest that the overall solution is stable, it does not directly
imply that the parameters are unique. To study the uniqueness of the
solution we computed a correlation matrix based on the parameters of
the best-fit solution and the parameters of the 100 alternate
solutions used for the uncertainty analysis described above. The
correlation matrix shows that the 7 ERs and the penalty function are
largely uncorrelated, but the parameters describing a single
absolute-magnitude distribution or the penalty function are typically
strongly correlated (Fig.~\ref{fig:paramcorr}). This suggests that the
use of 7 ERs does not lead to a degenerate set of model
parameters. Weak correlations are present between the Hungarias and
the $\nu_6$ complex as well as between the $\nu_6$ and the 3:1J
complexes. Both correlations are likely explained by partly
overlapping orbit distributions. Somewhat surprisingly there is hardly
any correlation between the two outer-MAB ERs --- the 5:2J and 2:1J
complexes. This suggests that they are truly independent components in
the model.

A caveat with this analysis is that the samples of synthetic
detections used for the alternative solutions are based on the
steady-state orbit distributions and we used the same orbit
distributions for fitting the model parameters. The small correlation
between ERs might thus be explained by the fact that the synthetic
orbits belong a priori to different ERs. We assume that the
non-negligible overlap between the steady-state orbit distributions
mitigate some or all of these concerns.

\section{Discussion}
\label{sec:discussion}

\subsection{Comparison to other population estimates}

\begin{figure*}[h]
  \centering
  \includegraphics[width=\columnwidth]{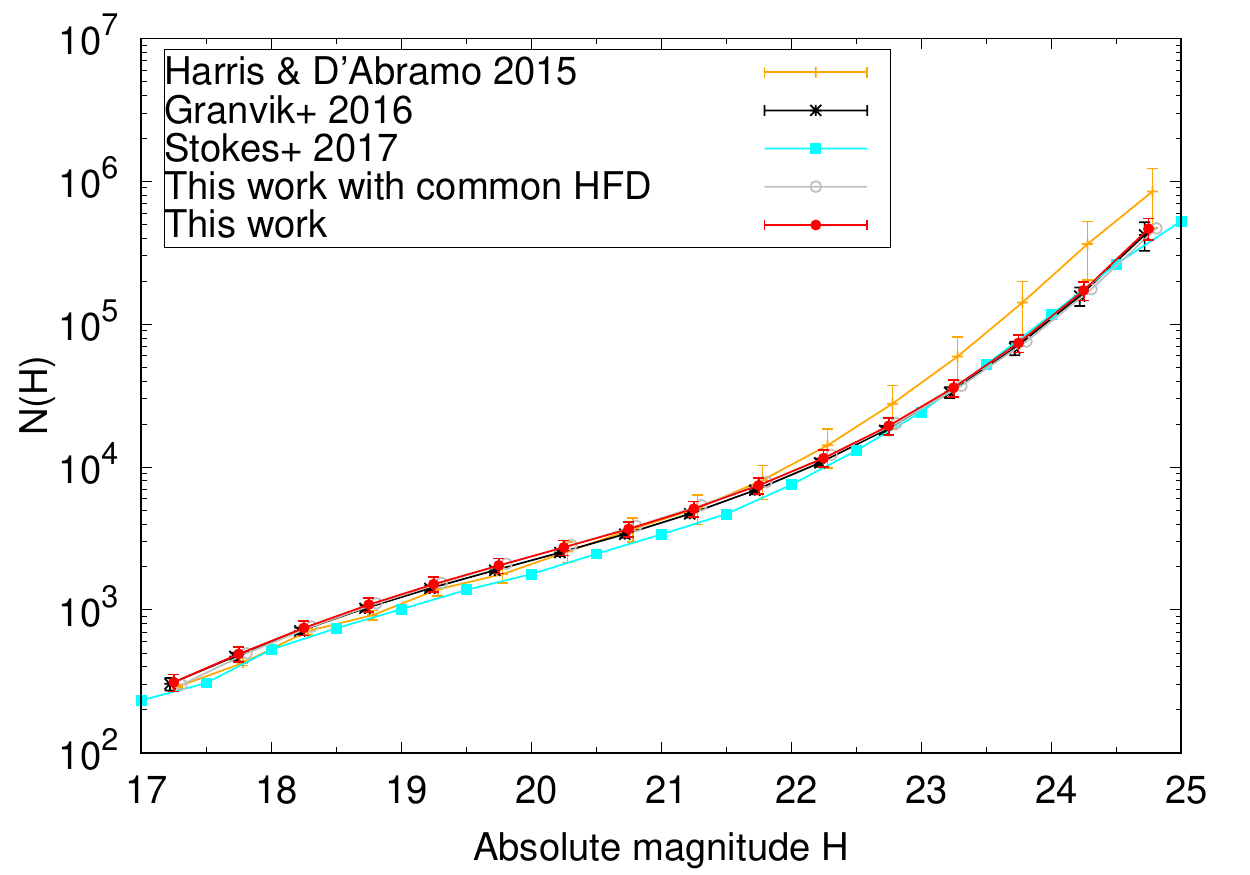}
  \includegraphics[width=\columnwidth]{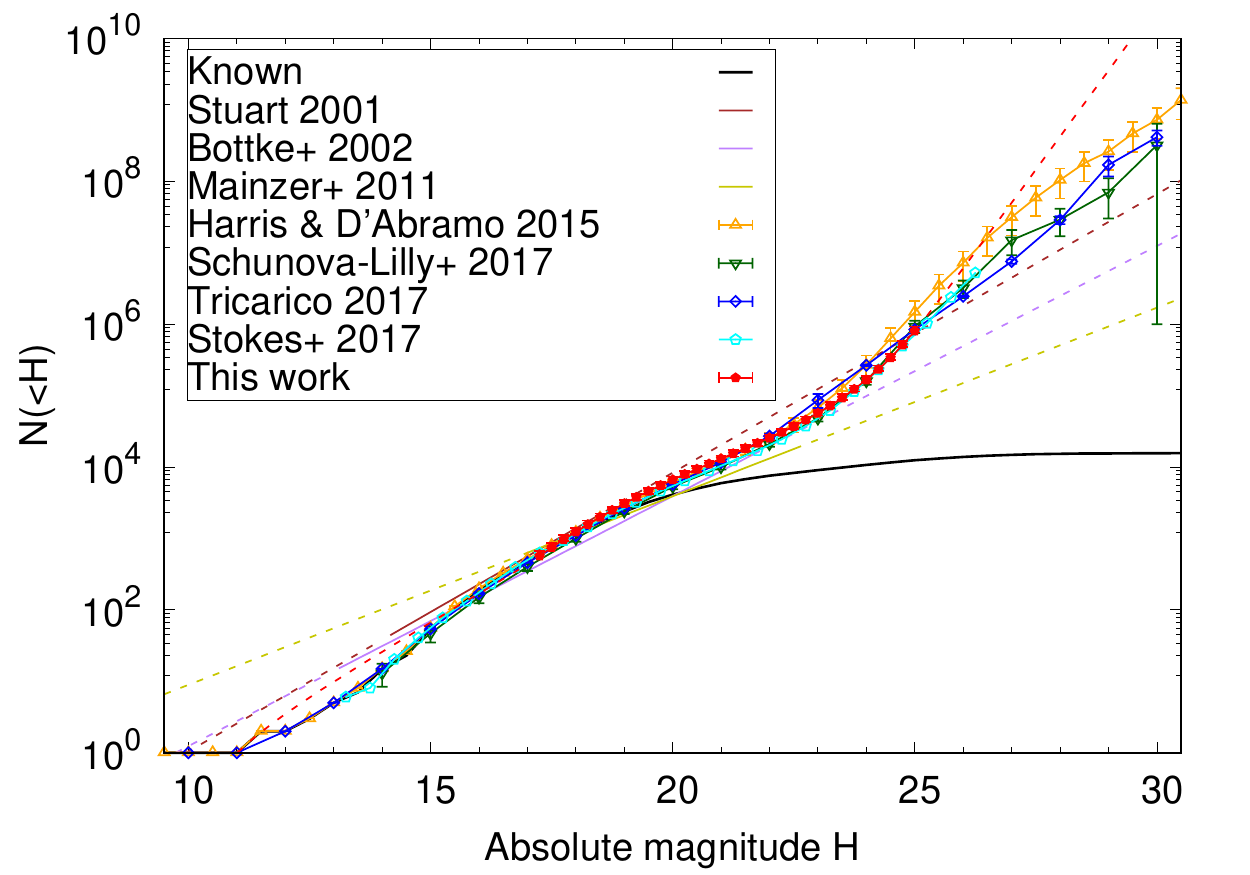}
  \caption{Our estimates for the debiased incremental (left) and
    cumulative (right) absolute-magnitude distributions for NEOs
    compared to published estimates. Extrapolations outside the
    validity regions for the different models are marked with dashed
    lines. The error bars for the model presented in this paper
    include both random and systematic uncertainties. The latter
    component is estimated to be 10\% (Sects.~\ref{sec:sensorb} and
    \ref{sec:sensobs}).}
  \label{fig:hdist}
\end{figure*}

Our estimates for the incremental and cumulative NEO HFDs agree the
most recent estimates published in the literature
(Fig.~\ref{fig:hdist} and Table~\ref{table:hdist}). The relatively
large uncertainties in the cumulative HFD are a reflection of the
large uncertainties in the extrapolation to $H<17$.

Our estimates also both agree and disagree with older estimates such
as that by \citet{bot2002a}, for reasons we explain below.  The
estimate by \citet{bot2002a} is based on the work by
\citet{bot2000a}. The ML technique used by both \citet{bot2000a} and
\citet{bot2002a} was able to characterize the slope of the NEO HFD
where most of their data existed, namely between $17 < H < 22$. The
HFD slope they found, $\gamma = 0.35$, translates into a cumulative
power-law size-distribution slope of -1.75. These values match our
results and those of others \citep[see,
  e.g.,][]{2015Icar..257..302H}. The challenge for \citet{bot2000a}
was setting the absolute calibration of the HFD. Given that the slope
of their HFD was likely robust, they decided to extend it so that it
would coincide with the expected number of NEOs with $13 < H < 15$, a
population that was much closer to completion.  This latter population
would serve as the ca\-lib\-ra\-tion point for the model.  At the time
of the analysis by \citet{bot2000a} there were 53 known NEOs in this
$H$ range and they assumed a completeness ratio of 80\% based on
previous work by \citet{rab2000a}. The expected number of NEOs with
$13 < H < 15$ was therefore assumed to be about 66 whereas today we
know that there are 50 such NEOs. The most likely explanation for the
reduced number of known NEOs with $13<H<15$ over the past 17 years
(from 53 to 50) is that the orbital and absolute-magnitude parameters
of some asteroids have changed (i.e., improved) and they no longer
fall into the category. We also now know that the HFD has a wavy shape
with an inflection point existing near $H = 17$, as shown in our work
(Figs. \ref{fig:neosfds} and \ref{fig:hdist}).  Thus, while the
\citet{bot2002a} estimates for $H < 18$ are on the low side but in
statistical agreement with those provided here (i.e., $960 \pm 120$ $H
< 18$ NEOs for \citet{bot2002a} vs. $\sim 1250$ here), their
population estimate for $H < 16$ is progressively too high.

\begin{table*}[]
    \centering
    \caption{Incremental ($N(H)$) and cumulative ($N(<H)$) HFD in 
    the fitted $H$ interval. The cumulative HFD has been scaled so that
    an extrapolation to smaller $H$ will predict 170 NEOs with $H<16$. 
    The uncertainty estimates only account for the random component.}
    \label{table:hdist}
    \begin{small}
    \begin{tabular}{ccccccc}
    \hline
    $H$ & $N(H)$ & $\sigma_\mathrm{N(H),-}$ & $\sigma_\mathrm{N(H),+}$ & $N(<H)$ & $\sigma_\mathrm{N(<H),-}$ & $\sigma_\mathrm{N(<H),+}$ \\ 
    \hline
17.125  &  1.378e+02  &  -5.900e+00  &  5.400e+00  &  5.695e+02  &  -4.640e+01  &  4.550e+01  \\ 
17.375  &  1.739e+02  &  -5.700e+00  &  5.200e+00  &  7.434e+02  &  -5.110e+01  &  4.960e+01  \\ 
17.625  &  2.187e+02  &  -6.000e+00  &  4.800e+00  &  9.620e+02  &  -5.590e+01  &  5.200e+01  \\ 
17.875  &  2.728e+02  &  -5.800e+00  &  5.200e+00  &  1.235e+03  &  -5.900e+01  &  5.500e+01  \\ 
18.125  &  3.370e+02  &  -6.800e+00  &  5.500e+00  &  1.572e+03  &  -6.100e+01  &  5.600e+01  \\ 
18.375  &  4.113e+02  &  -7.100e+00  &  7.300e+00  &  1.983e+03  &  -6.100e+01  &  5.800e+01  \\ 
18.625  &  4.960e+02  &  -8.600e+00  &  9.200e+00  &  2.479e+03  &  -6.300e+01  &  5.700e+01  \\ 
18.875  &  5.914e+02  &  -1.080e+01  &  1.130e+01  &  3.071e+03  &  -6.600e+01  &  5.500e+01  \\ 
19.125  &  6.979e+02  &  -1.300e+01  &  1.380e+01  &  3.768e+03  &  -7.100e+01  &  5.600e+01  \\ 
19.375  &  8.165e+02  &  -1.680e+01  &  1.480e+01  &  4.585e+03  &  -7.800e+01  &  5.900e+01  \\ 
19.625  &  9.490e+02  &  -2.140e+01  &  1.580e+01  &  5.534e+03  &  -8.600e+01  &  6.800e+01  \\ 
19.875  &  1.098e+03  &  -2.300e+01  &  2.000e+01  &  6.632e+03  &  -9.600e+01  &  8.100e+01  \\ 
20.125  &  1.269e+03  &  -2.800e+01  &  2.400e+01  &  7.901e+03  &  -1.140e+02  &  9.600e+01  \\ 
20.375  &  1.467e+03  &  -3.300e+01  &  2.800e+01  &  9.368e+03  &  -1.390e+02  &  1.130e+02  \\ 
20.625  &  1.701e+03  &  -4.000e+01  &  3.500e+01  &  1.107e+04  &  -1.800e+02  &  1.300e+02  \\ 
20.875  &  1.985e+03  &  -4.700e+01  &  4.500e+01  &  1.305e+04  &  -2.100e+02  &  1.700e+02  \\ 
21.125  &  2.334e+03  &  -5.600e+01  &  5.800e+01  &  1.539e+04  &  -2.500e+02  &  2.100e+02  \\ 
21.375  &  2.775e+03  &  -7.000e+01  &  7.200e+01  &  1.816e+04  &  -2.900e+02  &  2.800e+02  \\ 
21.625  &  3.343e+03  &  -9.500e+01  &  8.800e+01  &  2.151e+04  &  -3.700e+02  &  3.500e+02  \\ 
21.875  &  4.088e+03  &  -1.210e+02  &  1.170e+02  &  2.559e+04  &  -4.400e+02  &  4.700e+02  \\ 
22.125  &  5.088e+03  &  -1.570e+02  &  1.560e+02  &  3.068e+04  &  -5.800e+02  &  5.900e+02  \\ 
22.375  &  6.453e+03  &  -2.020e+02  &  2.170e+02  &  3.713e+04  &  -7.500e+02  &  7.700e+02  \\ 
22.625  &  8.358e+03  &  -2.850e+02  &  2.860e+02  &  4.549e+04  &  -1.010e+03  &  1.000e+03  \\ 
22.875  &  1.107e+04  &  -3.900e+02  &  4.000e+02  &  5.656e+04  &  -1.360e+03  &  1.340e+03  \\ 
23.125  &  1.502e+04  &  -5.700e+02  &  5.400e+02  &  7.158e+04  &  -1.850e+03  &  1.840e+03  \\ 
23.375  &  2.090e+04  &  -8.400e+02  &  7.700e+02  &  9.247e+04  &  -2.630e+03  &  2.480e+03  \\ 
23.625  &  2.988e+04  &  -1.270e+03  &  1.120e+03  &  1.224e+05  &  -3.800e+03  &  3.400e+03  \\ 
23.875  &  4.399e+04  &  -2.030e+03  &  1.690e+03  &  1.663e+05  &  -5.400e+03  &  5.100e+03  \\ 
24.125  &  6.685e+04  &  -3.280e+03  &  2.900e+03  &  2.332e+05  &  -8.400e+03  &  7.500e+03  \\ 
24.375  &  1.051e+05  &  -5.490e+03  &  5.600e+03  &  3.383e+05  &  -1.350e+04  &  1.210e+04  \\ 
24.625  &  1.717e+05  &  -1.040e+04  &  1.160e+04  &  5.101e+05  &  -2.220e+04  &  2.290e+04  \\ 
24.875  &  2.923e+05  &  -1.890e+04  &  2.920e+04  &  8.024e+05  &  -4.240e+04  &  4.790e+04  \\ 
    \hline
    \end{tabular}
    \end{small}
\end{table*}

We confirm that the slope of the HFD reaches a minimum at $H\sim20$
(Fig.~\ref{fig:hdist}) and, since it is also seen for different ERs
(Fig.~\ref{fig:neosfds}), conclude that it cannot be caused by the
combination of differently-sloped and power-law shaped HFDs. Instead
the wavy shape is likely related to the nature of how asteroids
disrupt \citep[see,
  e.g.,][]{1998Icar..135..431D,2005Icar..179...63B,2015aste.book..701B}. Small
asteroids are considered part of the “strength-scaling” regime, where
the fragmentation of the target body is governed by its tensile
strength, while large asteroids are considered part of the “gravity
scaling” regime, where fragmentation is controlled by the self-gravity
of the target. Laboratory experiments and hyd\-ro\-code modeling work
suggest the transition between the two regimes occurs near
100--200-meter-diameter bodies, which corresponds to $21 \lesssim H
\lesssim 22$ assuming a geometric albedo $p_V=0.14$ (smaller $H$ for
higher albedos).  It has also been suggested that it is caused by the
transition from the strong "monolithic" structures to the weaker
"rubble-pile" structures held together by gravitational forces, as
proposed by \citet{2015Icar..257..302H}.  Objects with $21 \lesssim H
\lesssim 22$ fall in between these categories and are more easily
disrupted than larger or smaller objects. An alternative explanation
is that the physical size distribution can be described by a powerlaw
but the albedo distribution varies as a function of size. This could
effectively lead to a dip in the HFD. However, this would probably
require an unrealistically large change in the average albedo as a
function of $H$ \citep{wer2002a}.

\begin{figure}[h]
  \centering
  \includegraphics[width=\columnwidth]{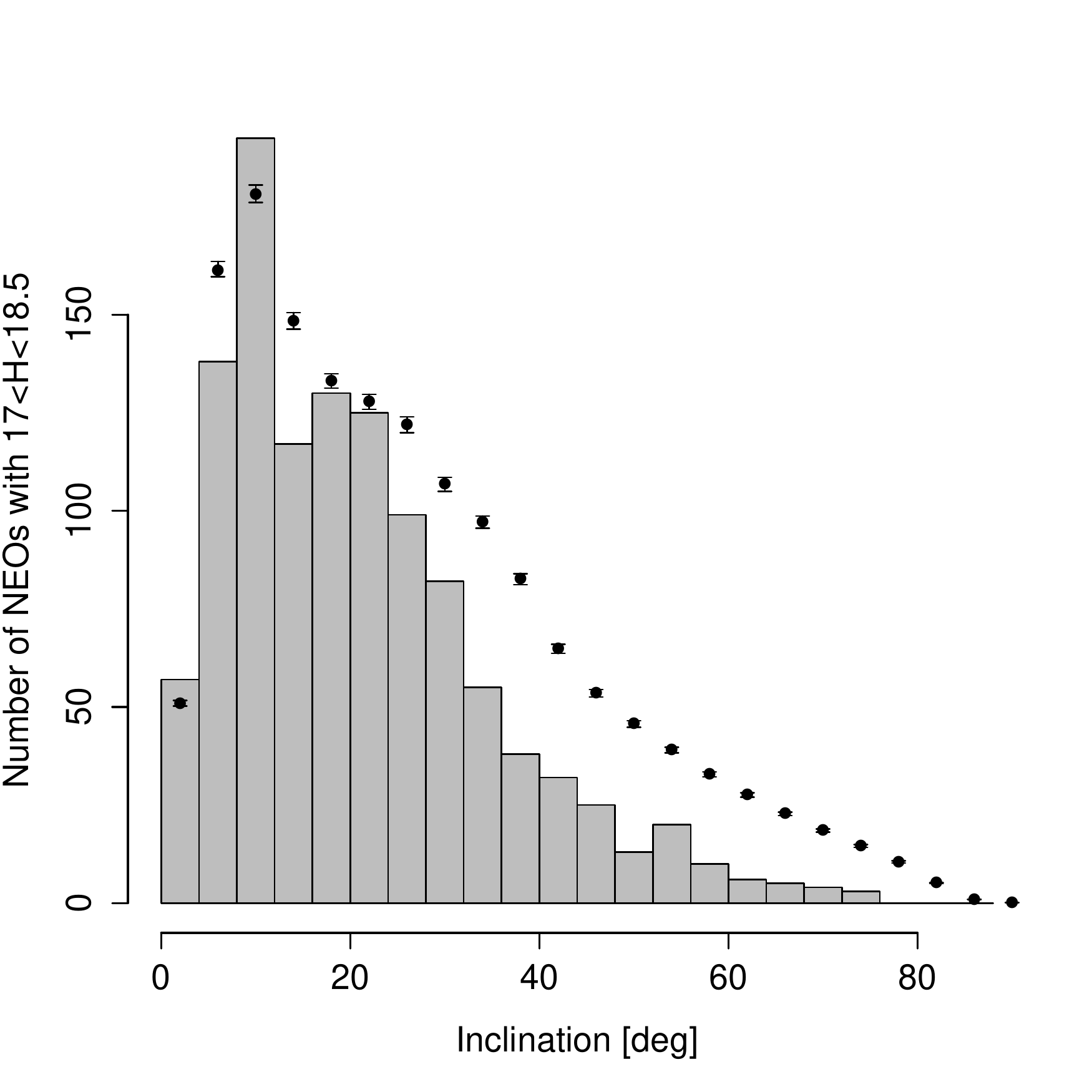}
  \caption{Our estimate for the debiased inclination distribution for
    NEOs with $17<H<18.5$ compared to currently known NEOs in the same
    $H$ range (ASTORB 2018-01-30). The uncertainties were computed as
    single-sided RMS estimates with respect to the nominal model by
    utilizing 100 alternative models as explained in
    Sect.~\ref{sec:nominal7smodel}. The error bars do not account for
    systematic uncertainties.}
  \label{fig:ibump}
\end{figure}

Our results for the orbital distributions also mostly agree with
results found in the literature. For large NEOs ($H<18.5$) we can
compare our debiased marginal ($a$,$e$,$i$) distributions with those
by \citet{2001Sci...294.1691S} and \citet{Tricarico2017}.  Whereas we
find a good agreement for the $a$ and $e$ distributions we do not
confirm the strong peak in the inclination distribution at $20\deg
\lesssim i \lesssim 30\deg$ predicted by \citet{2001Sci...294.1691S}.
However, we do predict that the distribution has a lesser bump at
$20\deg \lesssim i \lesssim 40\deg$ (Fig.~\ref{fig:ibump}) in
agreement with \citet{Tricarico2017}. We also note that the
inclination distribution for the known NEOs shows a similar bump
suggesting that we are looking at a real phenomenon, although the
exact shape of the bump is sensitive to the bin size. The bump was not
predicted by the model by \citet{bot2002a}, because it is caused by
the Hungarias and Phocaeas and those ERs were not included in their
model.

For NEOs with $22<H(<25)$ there are no debiased orbit models available
and hence we cannot compare our results to the literature.

\subsection{Extrapolation to larger and smaller NEOs}

Although the calibration of our model is limited to $17<H<25$, we may
occasionally want to extrapolate to larger and/or smaller objects.
The limitation of such extrapolations is that the orbit distributions,
and hence the relative ratios between the ERs, are fixed to either
$H=17$ or $H=25$, depending on if the extrapolation is towards larger
or smaller NEOs, respectively.

When extrapolating to larger NEOs, that is, to those with $H<17$, we
use a power-law function for the extrapolation and obtain the slope
from our estimate for $17<H<17.5$ NEOs. We also rescale the
extrapolation so that the cumulative number of NEOs with $H<16$
coincides with the current number of known NEOs, that is, 170. An
extrapolation to smaller $H$ values would predict 207 $H<16$ NEOs
without the rescaling. The extrapolation predicts a few too many NEOs
at $12\lesssim H \lesssim15$ but we think that the agreement with the
known population is still reasonable for most purposes
(Fig.~\ref{fig:hdist} right). Considering that the NEO inventory is
virtually complete for $H<17$ one may alternatively choose to simply
append the known population for the $H<17$ part.

An issue, however, is that many larger MBOs escape the MAB out of the
"forest" of weak resonances in the inner MAB
\citep[e.g.,][]{1998AJ....116.3029N,1999Icar..139..295M}. The bigger
they are, the more difficult time they will have reaching one of the
main NEO sources via Yarkovsky drift \citep{2006AREPS..34..157B}. As
discussed above, a potential problem with our model for large
asteroids is that the test asteroids we have chosen to represent them
have such high Yarkovsky semimajor axis drift rates that most will
jump over these weak resonances. Given that most large NEOs are known,
this does not present a problem for our NEO model per se, but a future
NEO model wanting to make predictions about, e.g., planetary impacts
by large NEOs will need to worry about getting the dynamics right for
these bodies. In addition, \citet{2018AJ....155...42N} have recently
shown that the large end of the NEO population does not appear to be
in steady-state which is the assumption behind the modeling carried
out here. Hence it is not surprising that an extrapolation of the
model described here cannot accurately reproduce the large end of the
NEO HFD.

When extrapolating to smaller NEOs, that is, those with $H>25$ we can
estimate the slope from our model for $24.5<H<25$ and use that for the
extrapolation. However, that approach does not lead to a good
agreement with the HFD by \citet{2015Icar..257..302H} for
$H\geq27$. When converted to the rate of Earth impacts,
\citet{2015Icar..257..302H} is in reasonable agreement with the
observed rate of impacts on the Earth
\citep{bro2002a,2013Natur.503..238B}. A more accurate extrapolation,
that is, one that is more in line with the literature, can be obtained
by using a slope found by others
\citep[e.g.,][]{bro2002a,2013Natur.503..238B} for $H>25$.

\subsection{Completeness of the current inventory of NEOs}

The surveys have so far found 905 NEOs with the estimated $D>1\km$
($H<17.75$; ASTORB 2018-01-30). Assuming that the $H<16$ population is
essentially complete and currently includes 170 NEOs, we predict a
population of $962^{+52}_{-56}$ for NEOs with $H<17.75$ (the
uncertainty estimates only account for the random component;
Fig.~\ref{fig:hdist} right and Table~\ref{table:hdist}). This implies
that about 94\% of all NEOs with $H<17.75$ have been found to date.

\begin{figure}[!h]
  \centering
  \includegraphics[width=0.8\columnwidth]{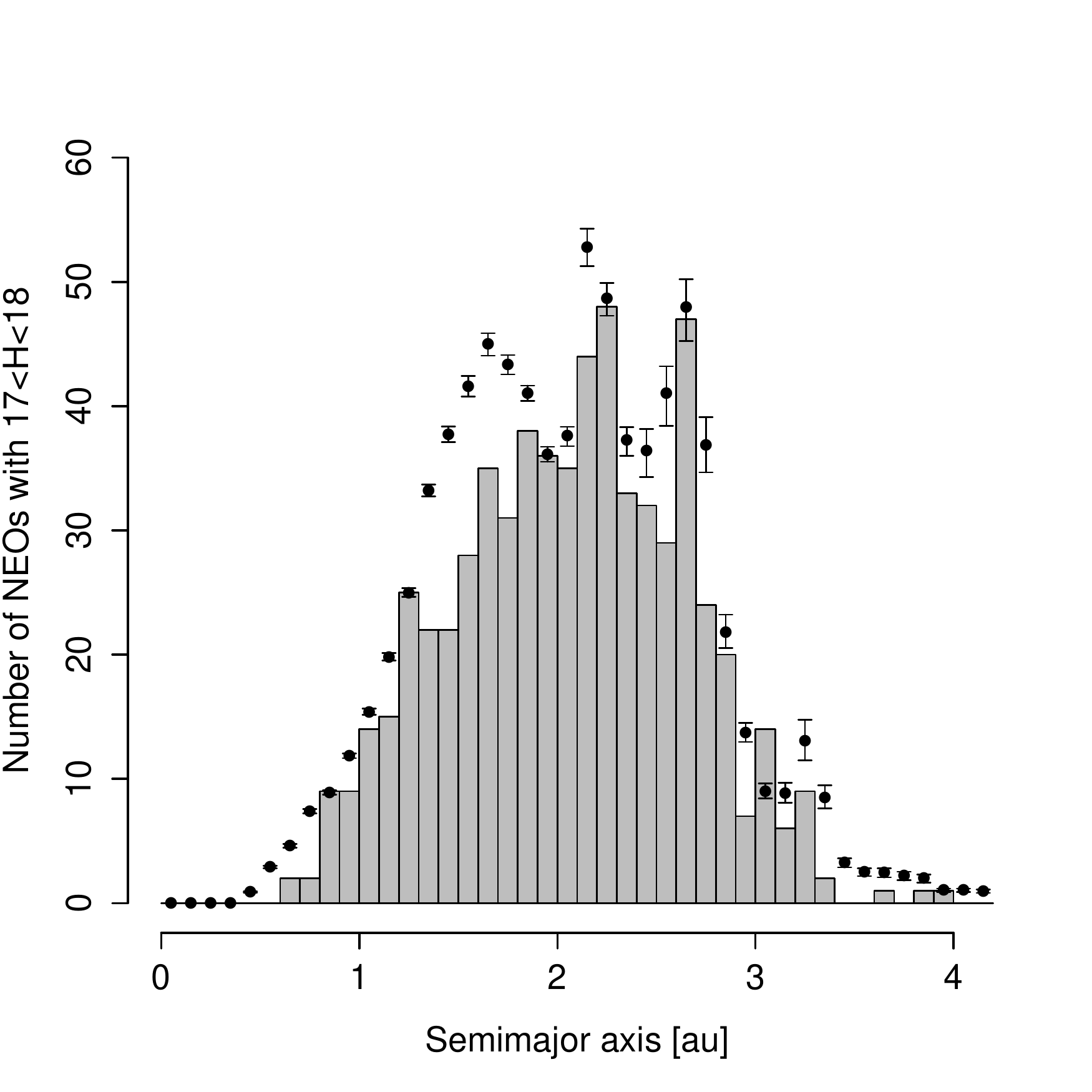}
  \includegraphics[width=0.8\columnwidth]{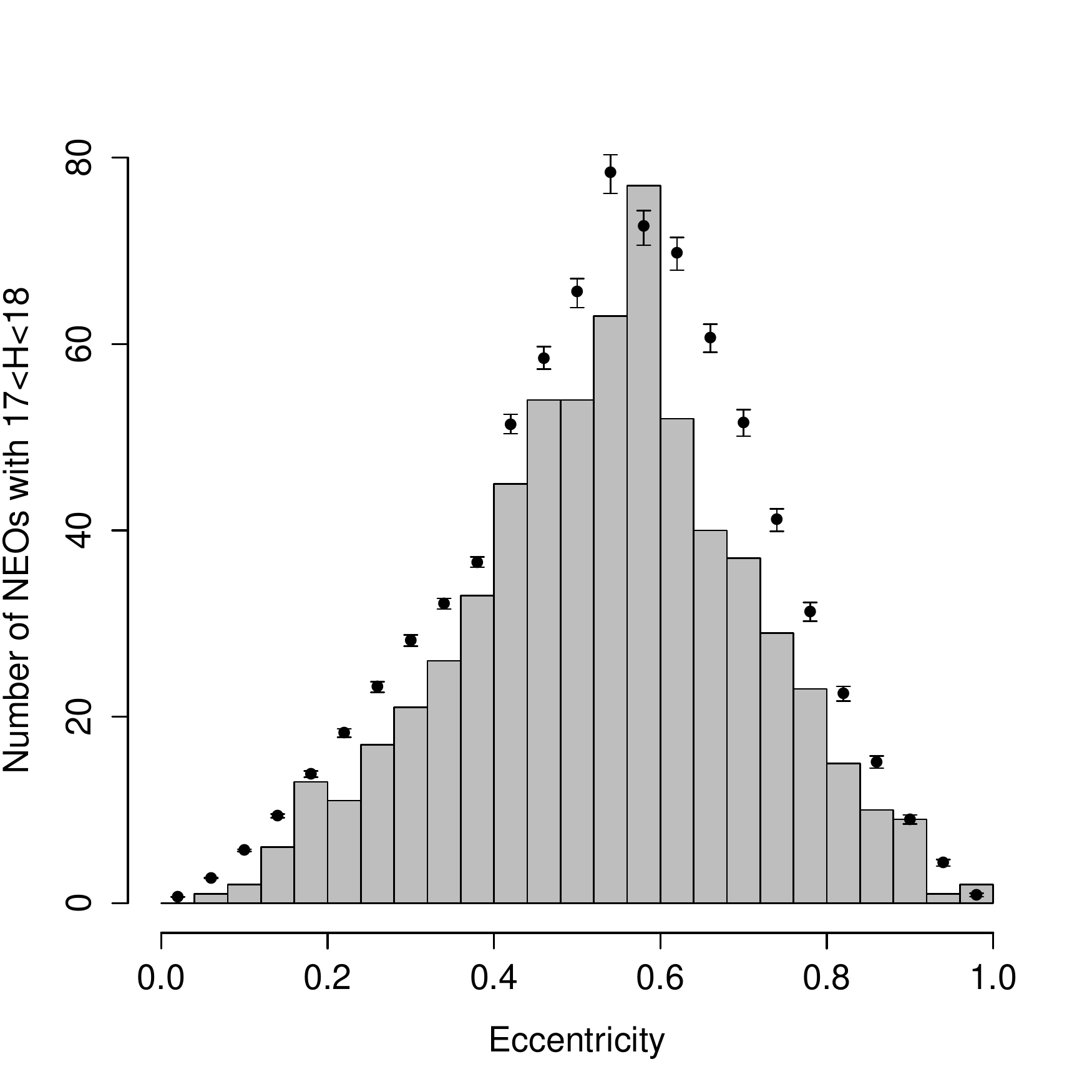}
  \includegraphics[width=0.8\columnwidth]{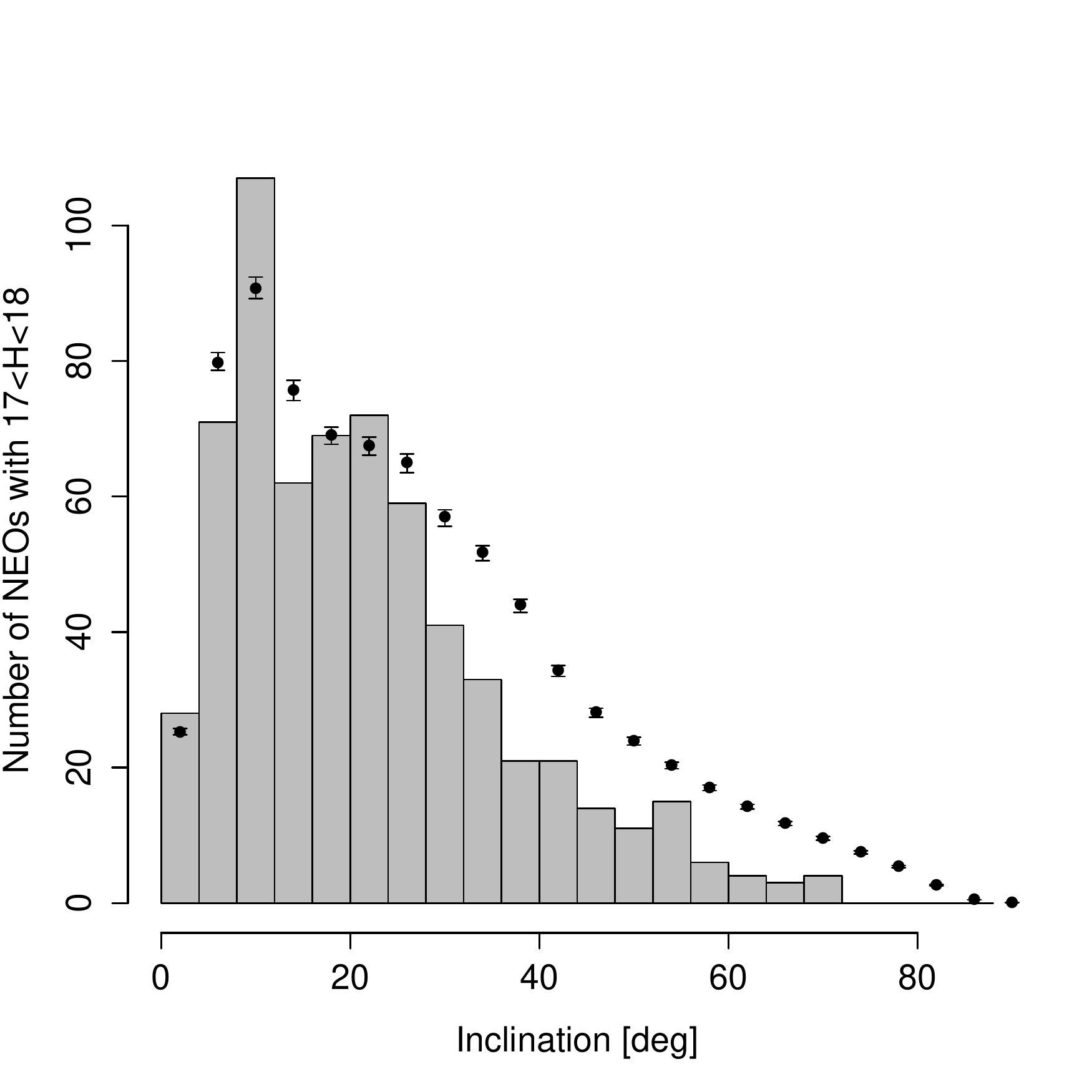}
  \caption{Known (ASTORB 2018-01-30; histogram) and predicted (dot
    with errorbars) marginal orbital-element distributions for NEOs
    with $17<H<18$. The uncertainties were computed as single-sided
    RMS estimates with respect to the nominal model by utilizing 100
    alternative models as explained in
    Sect.~\ref{sec:nominal7smodel}. The error bars do not account for
    systematic uncertainties.}
  \label{fig:1dorbit17h18}
\end{figure}

The orbits of the undiscovered large NEOs are characterized by high
inclinations and relatively small semimajor axes
(Fig.~\ref{fig:1dorbit17h18}). NEOs with such orbital characteristics
are challenging to detected because they can have relatively long
synodic periods and they may be bright enough only at perihelion when
they can be in the southern hemisphere. Finding these NEOs thus
require longer surveys carried out (also) in the southern hemisphere
and/or larger apertures. An example of such a challenging NEO to
discover is 2017~MK$_8$ ($a=2.51\au$, $e=0.67$, $i=31.6\deg$,
$H=16.5$) which was discovered by Pan-STARRS as recently as in June
2017. This particular object crosses the ecliptic approximately at
perihelion when inside the Earth's orbit and at aphelion (at a
distance of about $4\au$ from the Sun).

\begin{figure}[!h]
  \centering
  \includegraphics[width=0.8\columnwidth]{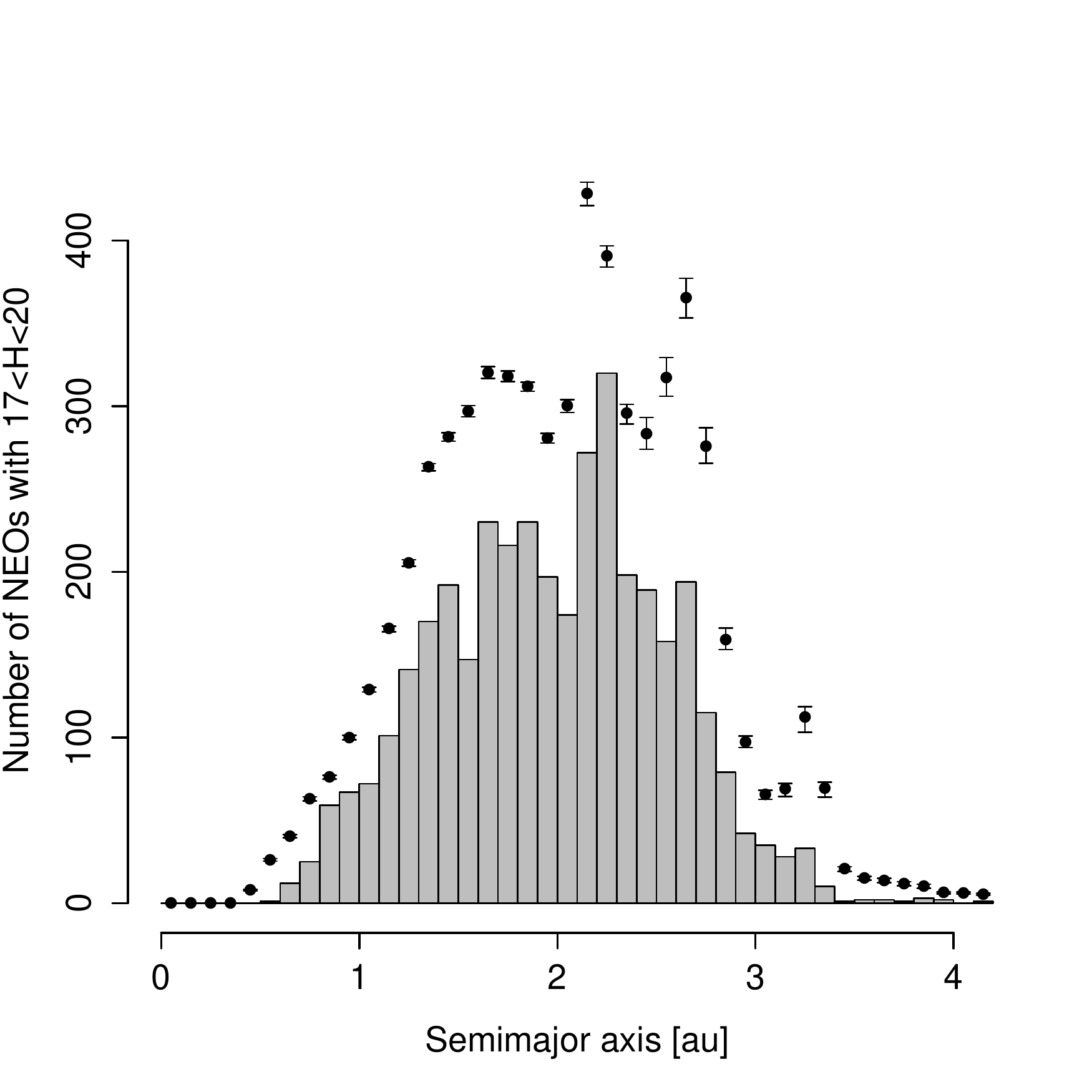}
  \includegraphics[width=0.8\columnwidth]{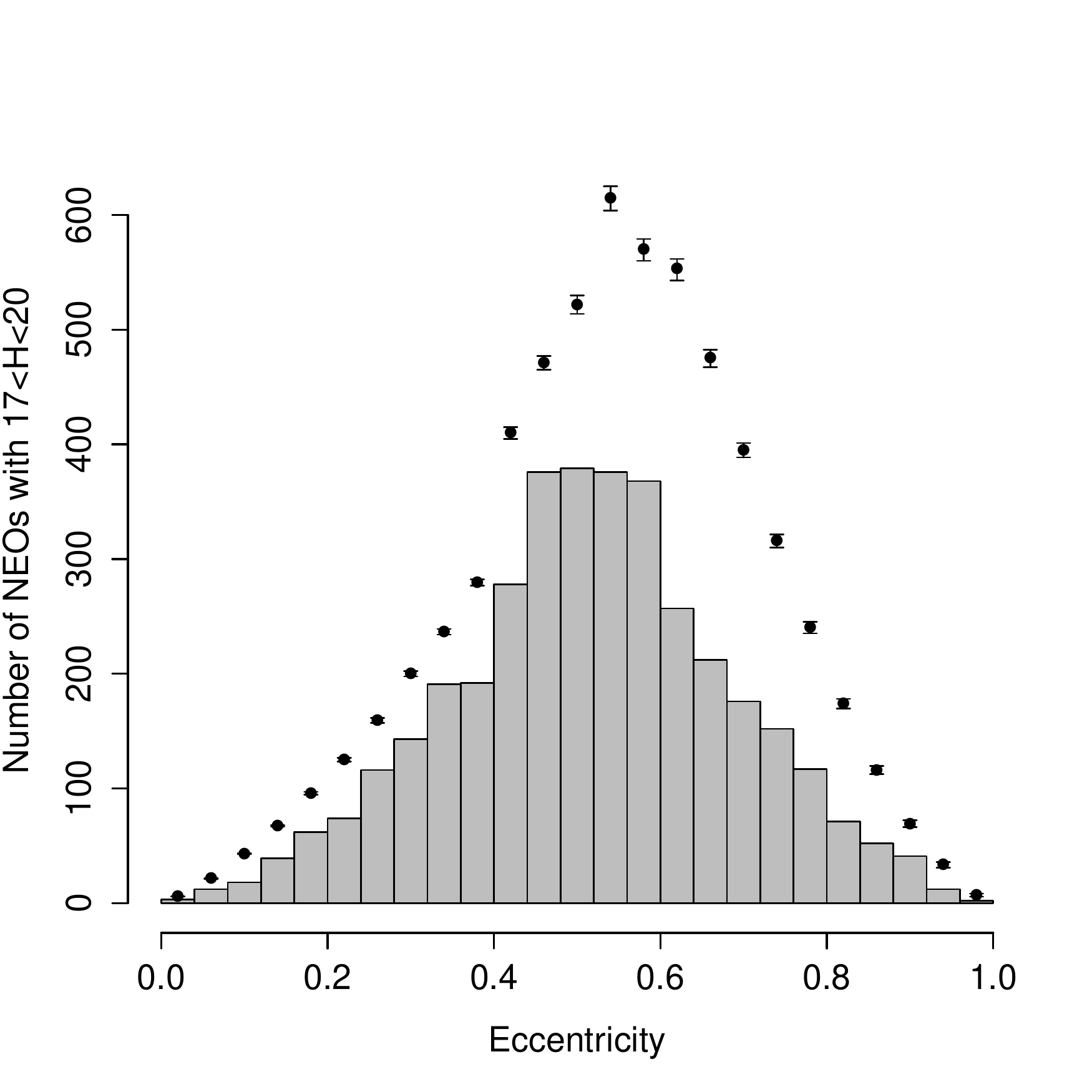}
  \includegraphics[width=0.8\columnwidth]{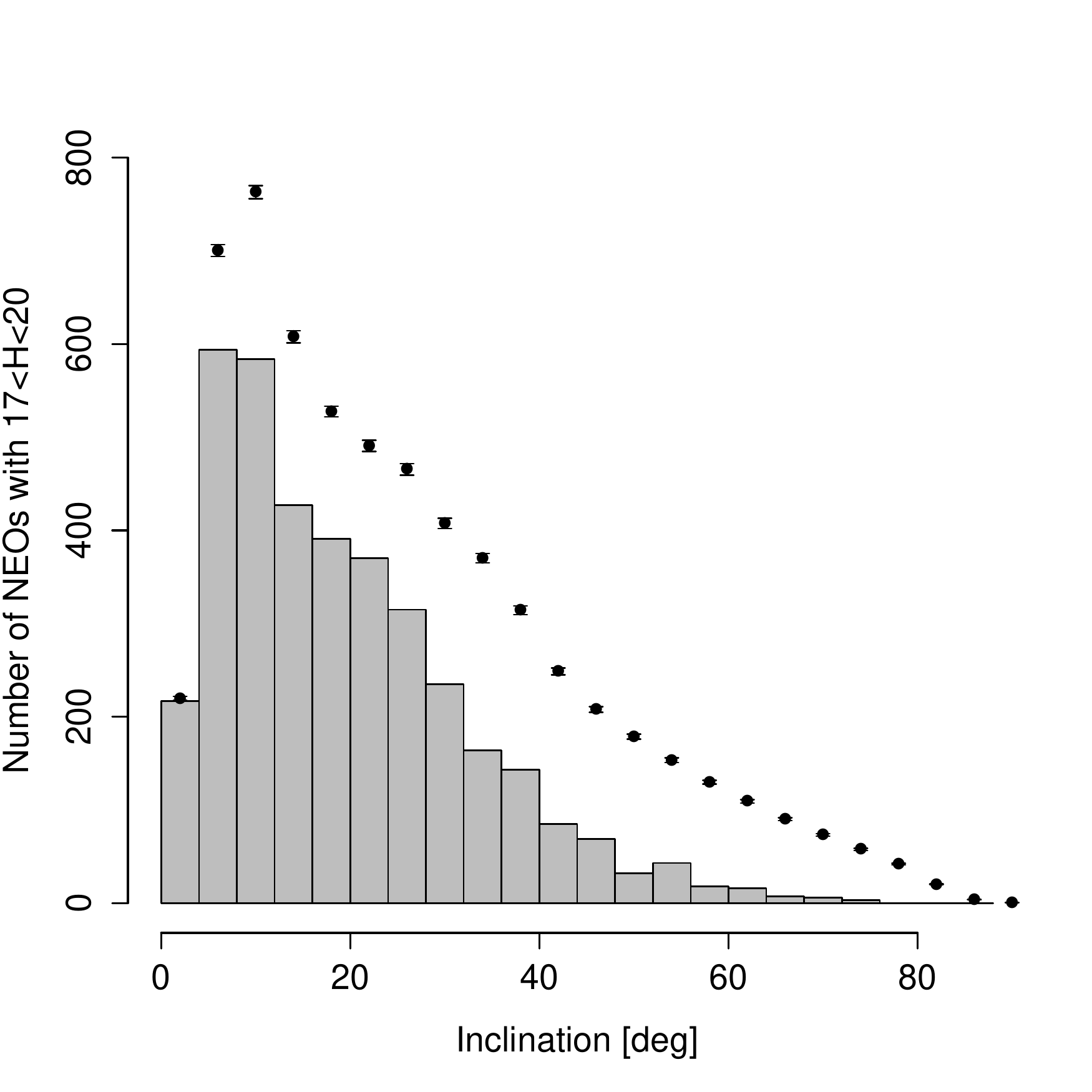}
  \caption{Known (ASTORB 2018-01-30; histogram) and predicted (dot
    with errorbars) marginal orbital-element distributions for NEOs
    with $17<H<20$. The uncertainties were computed as single-sided
    RMS estimates with respect to the nominal model by utilizing 100
    alternative models as explained in
    Sect.~\ref{sec:nominal7smodel}. The error bars do not account for
    systematic uncertainties.}
  \label{fig:1dorbit17h20}
\end{figure}

For smaller objects with $17<H<20$ the need for improved
instrumentation becomes even more urgent as in addition to high-$i$
NEOs also large-$a$ NEOs remain undiscovered
(Fig.~\ref{fig:1dorbit17h20}). Smaller and more distant NEOs are
difficult to detect due to their greater average distances from the
observer and higher rates of motion when close to the Earth.

The main difference between the orbits of undiscovered small
(Fig.~\ref{fig:1dorbit17h20}) and large (Fig.~\ref{fig:1dorbit17h18})
NEOs is that the former are more notably characterized by large
eccentricities. As most of the known high-inclination NEOs have been
discovered prior to, e.g., Pan-STARRS, which is the most prolific
survey telescope currently operating, we find it unlikely that the
remaining large, high-inclination NEOs would be discovered in the next
decade without substantial improvements in observation strategy and/or
instrumentation.

\subsection{NEO flux from different ERs}

The relative flux of asteroids and comets into the NEO population as a
function of ER is strongly size dependent
(Table~\ref{table:parameters}). The number-weighted flux of NEOs in
general ($17<H<25$) is dominated by inner-MAB ERs whereas the
number-weighted flux of $D>100\meter$ ($17<H<22$) NEOs is dominated by
outer-MAB ERs.  The domination of outer-MAB ERs for large NEOs has
been seen before \citep{bot2002a} but the change to domination by
inner-MAB ERs for smaller NEOs has not been shown before.

Recently \citet{gra2017a} estimated the relative flux of asteroids
into the NEO population from different ERs through direct integrations
of MBOs. They found a good agreement with \citep{bot2002a} for
$D=3\km$ objects but unfortunately the smallest diameter considered,
$D=0.1\km$ ($H\sim22.7$), is still fairly close to the "large" group
and hence they do not see the transition to inner-MAB domination.
Instead the relative fluxes for all the diameters considered
($0.1\km$--$3.0\km$) are statistically indistinguishable. Focusing on
the large group only we find that the flux through the 5:2J complex is
the highest (Table~\ref{table:parameters}) followed by the 3:1J, 2:1J
and $\nu_6$ complexes and Phocaeas and Hungarias in descending order.
The relative numbers are remarkably close to those predicted by
\citet{gra2017a} for $D=0.1\km$ asteroids through direct orbital
integrations. The largest relative difference between our estimates
and those by \citet{gra2017a} is found for Hungarias in that our
estimate is a factor of about three higher.

\subsection{NEAs on retrograde orbits}

\begin{figure}[h]
  \centering
  \includegraphics[width=\columnwidth]{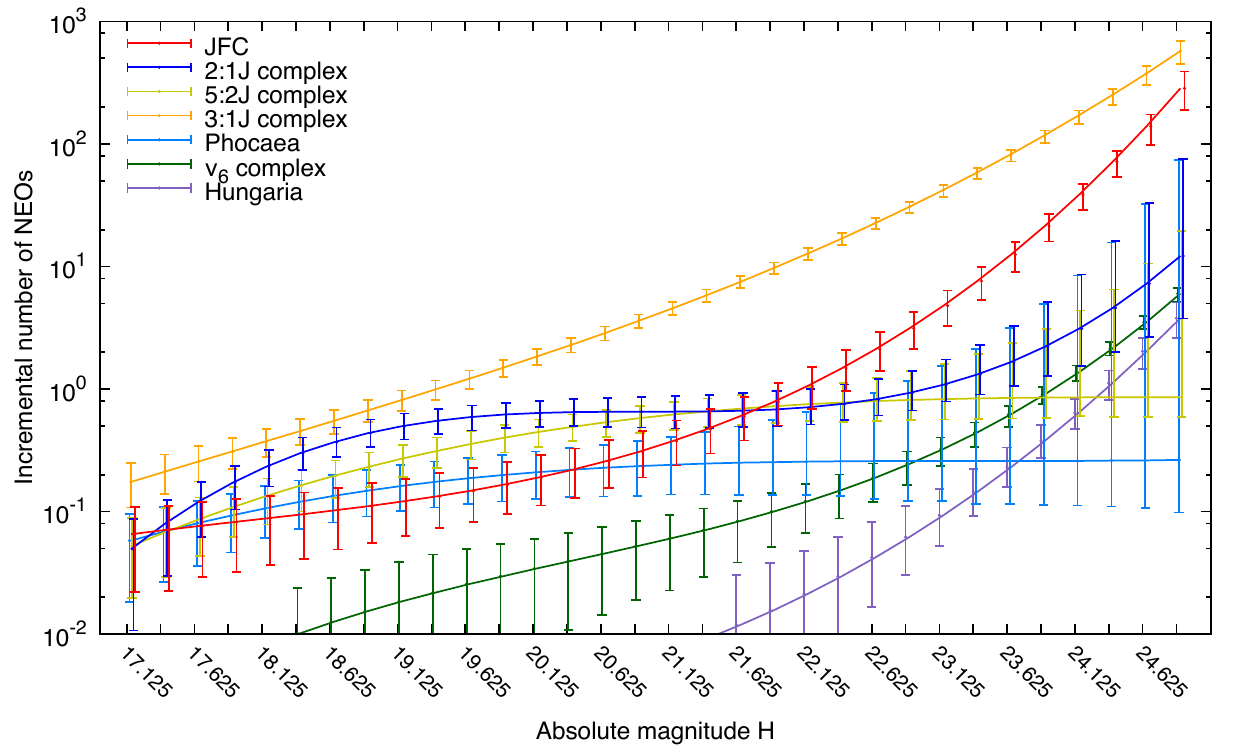}
  \caption{Debiased, incremental absolute-magnitude distributions for
    NEOs on retrograde orbits from different ERs.}
  \label{fig:neoretrosfd}
\end{figure}

We find that the fraction of retrograde objects ranges from about 1\%
to 2.5\% depending on the range in $H$ magnitude and the main ERs are
the 3:1J complex and JFCs (Fig.~\ref{fig:neoretrosfd}). In particular,
when extrapolating our model down to $H=15$ it predicts that there are
0.5 (3.5) NEAs on retrograde orbits with $15<H\le16$
($15<H\le18$). \citet{gre2012b} suggest that the NEO (343158) 2009
HC$_{82}$ is of asteroidal origin despite its retrograde orbit. The
absolute magnitude for (343158) is 16.1 according to the MPC which
agrees with our predicted number of NEAs on retrograde orbits assuming
a 100\% completeness for $H\lesssim16$ NEAs and suggests that (343158)
is the largest NEA on a retrograde orbit. Based on the new model we
predict that there are still 2--3 $H<18$ NEAs on retrograde orbits
left to be discovered. The fraction of unknown retrograde $H<18$ NEOs
may appear surprisingly high considering that about 90\% of all
prograde NEOs in the same $H$ range have been discovered. We note,
however, that the undiscovered large retrograde NEOs likely reside on
nearly polar orbits that make them hard to discover
(cf. Fig~\ref{fig:1dorbit17h18}).

\begin{figure}[h]
  \centering
  \includegraphics[width=\columnwidth]{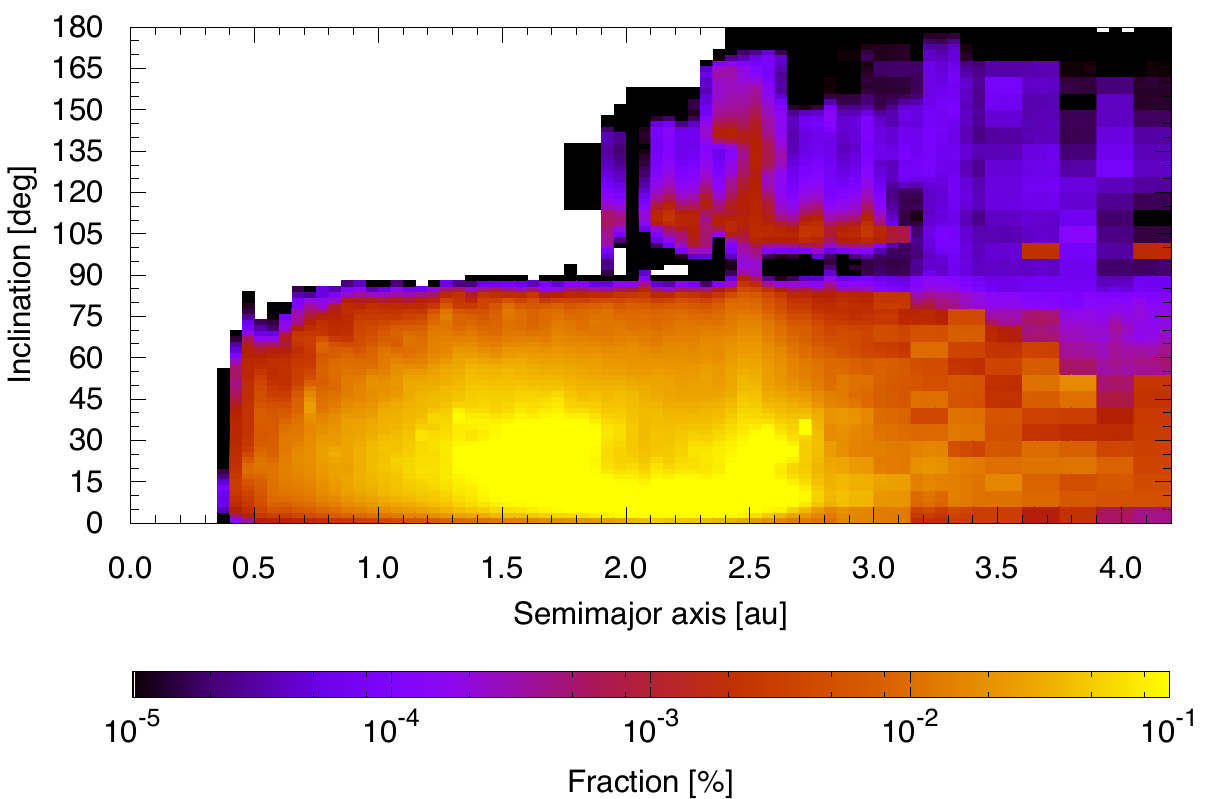}
  \caption{NEO steady-state ($a,i$) distribution for $17<H<25$.}
  \label{fig:neoretro}
\end{figure}

The mechanism producing NEAs on retrograde orbits is primarily related
to the 3:1J MMR as suggested by \citet{gre2012b} because most NEAs
evolve to $i>90\deg$ orbits when their $a\sim2.5\au$
(Fig.~\ref{fig:neoretro}). Note that (343158) 2009 HC$_{82}$ has
$a\sim2.53\au$ which puts it in the 3:1J MMR.

\subsection{Collision rate on terrestrial planets}

In order to compute the collision probabilities of NEOs with Mercury,
Venus and the Earth (we neglect Mars because the latter is also
bombarded by asteroids which have $q>1.3$ au, not included in our
model) we adopted the following procedure. For each cell in orbital
space $(a, e, i)$ considered in our NEO model, we have computed the
collision probability $P_{\rm col} (a,e,i)$ and impact velocity
$v_{\rm col} (a,e,i)$ with the considered planet, using an \"Opik-like
code described in \citet{2012Icar..219..150V} and
\citet{2013Icar..226..682P}. This code is superior to the one
originally described in \cite{1967JGR....72.2429W}, because it
accounts for the fact that the eccentricity and the inclination of an
object oscillate in a coupled manner together with the precession of
the argument of perihelion $\omega$. This oscillation is prominent
when the $z$-component of the angular momentum is small \citep[i.e.,
  the well-known Lidov-Kozai
  oscillations:][]{1962P&SS....9..719L,1962AJ.....67..591K}. Thus, the
code requires that the values of $a, e, i$ of the projectile be
specified as well as the corresponding value of $\omega$. Because
$\omega$ was not tracked in our model, for each cell we considered 10
particles, each with the ($a,e,i$) values corresponding to the center
of the cell, and values of $\omega$ ranging from 0 to 90 degrees, with
steps of 10 degrees. For simplicity we have assumed that each planet
has a null inclination relative to the reference plane, but we used
its actual orbital eccentricity. For each of these 10 particles, the
code outputs a different collision probability $P_{(a,e,i)}(\omega)$
and velocity $v_{(a,e,i)}(\omega)$. The collision probability and
velocity for objects in the cell $P_{\rm col} (a,e,i), v_{\rm col}
(a,e,i)$ are computed as the averages of these quantities. In
computing the averages we recognize the symmetry of the
$\omega$-induced dynamics relative to the axes $\sin{\omega}=0$ and
$\cos{\omega}=0$ and therefore the values of $P_{(a,e,i)},
v_{(a,e,i)}$ for $\omega\ne 0, 90^\circ$ are considered twice while
$P_{(a,e,i)}(0), v_{(a,e,i)}(0), P_{(a,e,i)}(90), v_{(a,e,i)}(90)$
only once. Moreover, in computing the average of $v_{(a,e,i)}$ we
weight with $P_{(a,e,i)}$.

Once the values $P_{\rm col} (a,e,i), v_{\rm col} (a,e,i)$ are
computed for each cell, the total collision probability with the
planet (impacts per year) is computed as
\begin{equation}
 P_{\rm tot}=R_p^2 \sum_{a,e,i} N_{\rm tot}(a,e,i)P_{\rm col}(a,e,i) \left(1+\frac{v_{\rm esc}^2}{v_{\rm col}^2(a,e,i)}\right)\,
\end{equation}
where $R_p$ is the radius of the planet, $v_{\rm esc}$ is the escape
velocity for its surface, and $N_{\rm tot}(a,e,i)=\sum_H N(a,e,i,H)$
is the number of asteroids in each orbital-magnitude cell of our
model. Note that the term in parentheses on the right hand side
implies that our collision probability calculation accounts for
gravitational focusing.

\begin{figure}[h]
  \centering
  \includegraphics[width=\columnwidth]{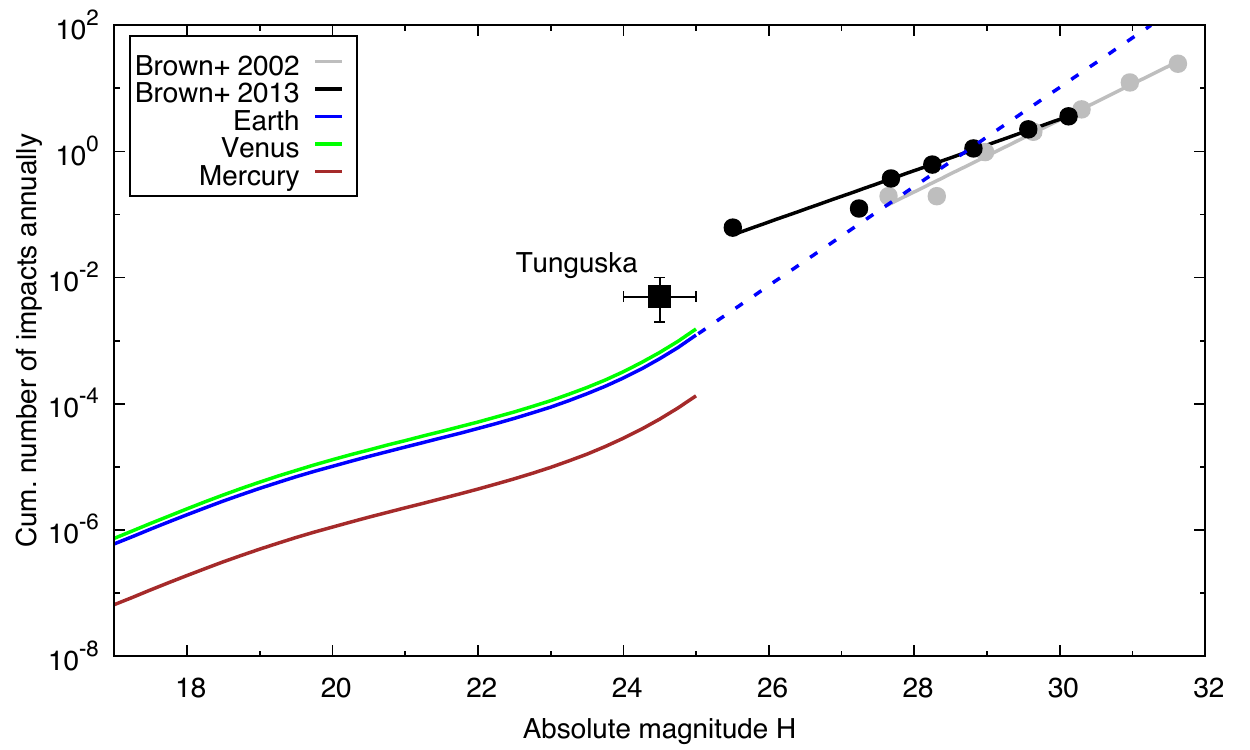}
  \caption{Cumulative annual rate of impacts on terrestrial planets
    predicted by our model and observed rate of bolides on the Earth
    \protect \citep{2013Natur.503..238B, bro2002a}. The dashed line
    marks a linear extrapolation based on our prediction for the slope
    at $24.5<H<25$. The conversion from bolide energy to absolute
    magnitude $H$ assumes a spherical shape, a bulk density of
    $3,000\kg\meter^{-3}$, an average impact speed of
    $20.3\km\second^{-1}$, and a geometric albedo of 0.14.  The error
    bars (and the nominal value) for the Tunguska event are
    approximate assuming that similar events happen every 100--500
    years and that the diameter of the impactor is about 50 meters
    with the geometric albedo ranging from 0.05 to 0.25.}
  \label{fig:planetimp}
\end{figure}

The result is illustrated in Fig.~\ref{fig:planetimp} in a cumulative
form (number of impacts per year on a planet for impactors brighter
than a given magnitude $H$).  The cumulative rate of Earth impacts by
NEOs with $H<25$ is approximately once per millenium
(Fig.~\ref{fig:planetimp}). The results are in very good agreement
with those reported in \citet{2002Icar..158..329M} (Table 5) for
$H<17.3$, $H<19.0$, and $H<20.6$. While our nominal rate is about 3
times smaller than another contemporary estimate for $H<25$
\citep{2015Icar..257..302H}, we stress that this difference is
explained by the difference in the HFDs rather than in the calculation
of the impact rate. The estimates overlap at the 1$\sigma$ level when
accounting for the uncertainties of the HFDs (Fig.~\ref{fig:hdist}).

A linear extrapolation of the cumulative impact rate in the
($H,\log{N(<H)}$) space reproduces the observed rate of
decameter-scale and smaller asteroids and meteoroids to within an
order of magnitude
\citep[Fig.~\ref{fig:planetimp};][]{2013Natur.503..238B,bro2002a}.  A
better match to the observed rate of bolide impacts would require a
steeper slope at $24\lesssim H \lesssim 26$. If the
higher-than-expected rate of large bolides is more than just a
statistical anomaly, the extrapolation suggests that the NEO HFD has a
bump at $24\lesssim H \lesssim 28$ that has not been predicted by NEO
models so far to the best of our knowledge. However, one has to bear
in mind that the largest bolides are single events and therefore their
frequency is uncertain \citep[see, e.g.,][]{boslough2015}.

\begin{figure}[h]
  \centering
  \includegraphics[width=\columnwidth]{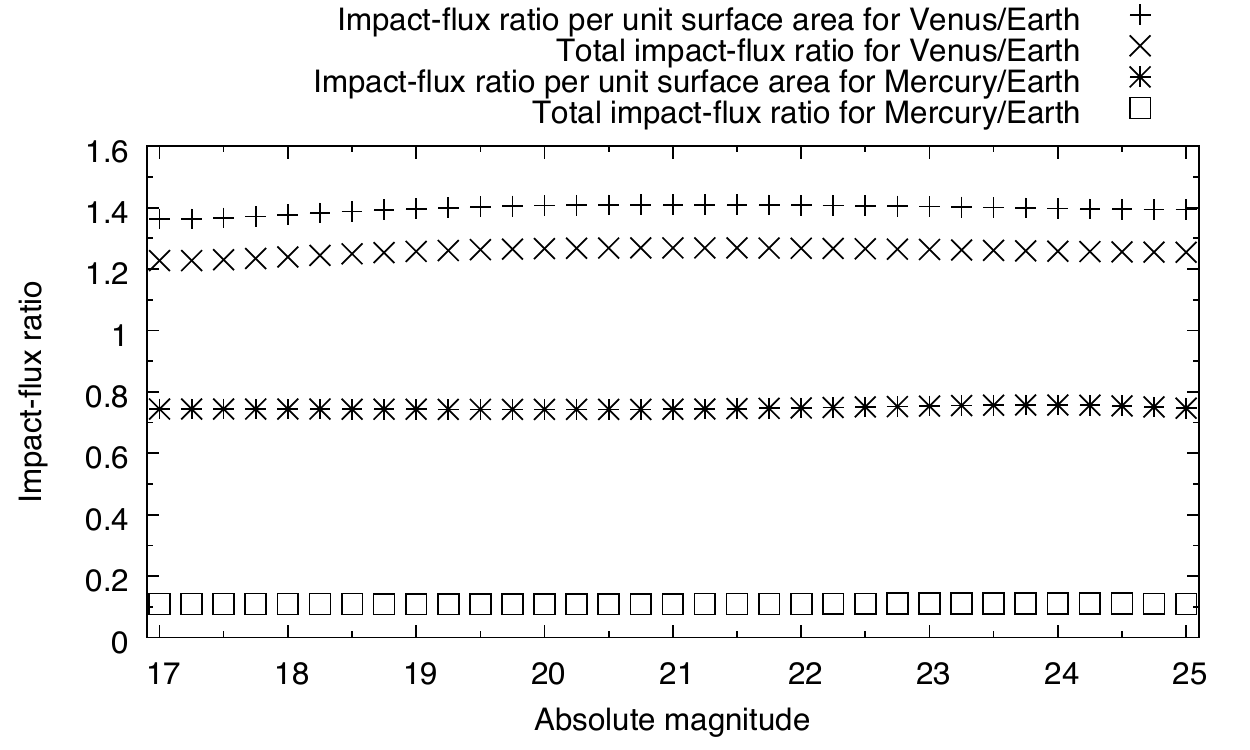}
  \caption{Total and per-surface-area impact-flux ratios Venus/Earth
    and Mercury/Earth as a function of impactor $H$ magnitude. These
    ratios do not account for the destruction of asteroids at small
    $q$ which changes all ratios as a function of $H$ (Table
    \ref{table:sinksmeanlife}).}
  \label{fig:impratios}
\end{figure}

The impact-flux ratios are fairly stable throughout the considered $H$
range (Fig.~\ref{fig:impratios}). The uncertainty on these estimates
is driven by the uncertainty in the orbit distribution and HFD, and
not more than about 10\% based on the discussion in
Sect.~\ref{sec:nominal7smodel}. Our total impact flux ratio for Venus
and Earth ($\sim1.2$) agrees with \citet{2017AJ....153..172V} whereas
our estimates for the impact flux ratios per surface area for Venus
and Earth ($\sim1.4$) and Mercury and Earth ($\sim0.75$) do not agree
with the ones reported in \citet{gre2012a} but are about 20\% higher
and 40\% lower, respectively. Given the rather trivial conversion from
the total impact-flux ratio to the impact-flux ratio per surface area
it seems that also \citet{2017AJ....153..172V} and \citet{gre2012a}
are at odds with each other.

\begin{figure}[h]
  \centering
  \includegraphics[width=\columnwidth]{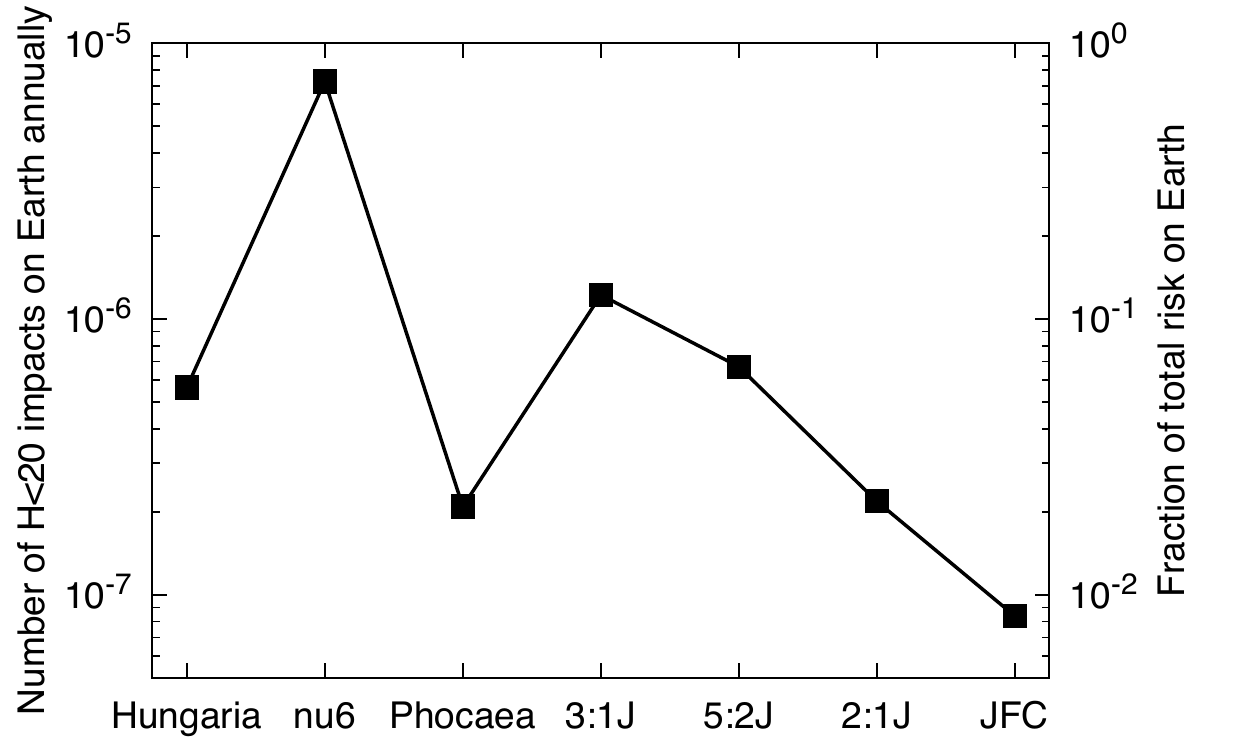}
  \caption{Source regions for Earth impactors with $H<20$.}
  \label{fig:earthimpsource}
\end{figure}

Fig.~\ref{fig:earthimpsource} shows the relative contribution of each
source to the terrestrial impact rate. About 80\% of the impacts come
from the $\nu_6$ SR. Thus, the inner MAB is the predominant source of
impactors. Given that the population of primitive asteroids in the
inner MAB is more than 20\% of the total \citep{2014Natur.505..629D},
this implies that most of the primitive NEOs also come from the
$\nu_6$ SR. This is in agreement with the results of
\citet{2010ApJ...721L..53C,2013AJ....146...26C} and
\citet{2015Icar..247..191B} who investigated the most likely origin of
specific primitive NEOs.

The production rate of $D > 20\km$ craters across the Earth's surface
over the last 100 Myr or so has been estimated from lunar craters to
be $2.5 (\pm 1.1) \times 10^{-15}$~km$^{-2}$~yr$^{-1}$ and from
terrestrial craters to be $2.8 (\pm 1.1) \times
10^{-15}$~km$^{-2}$~yr$^{-1}$
\citep{maz2017a}. \citet{2000MNRAS.317..429H}, using a different
method, estimated the production rate of $D > 22$~km craters across
the Earth's surface over the last 125 Myr to be $3.0 (\pm 0.3) \times
10^{-15}$~km$^{-2}$~yr$^{-1}$.  We can compare these values to
predictions from our model, assuming that the scaling relationship to
turn projectiles into terrestrial craters is a factor of 20
\citep[see, e.g.,][]{1989icgp.book.....M}.  Combining our collision
probability results with 962 km-sized NEOs ($H < 17.75$), 58\%
which are on Earth crossing orbits, yields a model production rate for
terrestrial $D > 20$ km craters of $2.4 \times
10^{-15}$~km$^{-2}$~yr$^{-1}$.  This value is in statistical agreement
with the estimates from \citet{maz2017a} but slightly outside the
error bars for \citet{2000MNRAS.317..429H}. The difference for the
latter, however, could simply suggest we need to slightly modify our
crater scaling laws \citep[see, e.g.,][]{2016LPI....47.2036B}.

\subsection{ESA NEO service and access to realizations of the model}

The model and tools for survey simulations have been made available by
ESA through their Space Situational Awareness website dedicated to
NEOs\footnote{\url{http://neo.ssa.esa.int/neo-population}}. The tool
for generating a realization of the model produces a distribution of
orbital elements and absolute magnitudes in the range $17<H<25$ by
default, but the user has the option to extrapolate to larger and/or
smaller $H$ magnitudes.  The user may either define his/her own slope
for the HFD outside the default range or use one of the predefined
slopes. To improve the statistics it is also possible to provide a
scaling factor, which is unity for the nominal model and, e.g., 10 for
a population 10 times larger than the nominal population for the same
$H$ range.

To simplify access to the model, we also provide direct access to a
realization of the nominal model for the default $H$ range. The flat
file\footnote{\url{http://www.iki.fi/mgranvik/data/Granvik+_2018_Icarus}}
contains orbital elements and absolute magnitudes ($a$,$e$,$i$,$H$)
for 802,000 NEOs with $a<4.2\au$ and $17<H<25$.

\section{Conclusions}

We have developed a four-dimensional model describing the debiased NEO
orbit ($a,e,i$) and absolute-magnitude ($H$) distributions. The free
parameters in our modeling approach describe $H$ distributions for
asteroids and comets that entered the NEO region through 7 different
ERs in the MAB or the cometary region. Our modeling methodology, tools
and results have been carefully vetted through comparing independent
predictions of NEO detections and actual detections by CSS's stations
703 and G96 --- to the best of our knowledge such detailed and
independent quality-control measures have never been employed in the
past during the development of NEO population models.

We see statistically-significant differences in the shapes of the
fitted $H$ distributions for the different ERs. The shapes range from
almost flat (Phocaeas and the 5:2J complex) to simple power-law (3:1J
complex) to increasing slope (Hungarias and JFCs) to waves (the
$\nu_6$ and 2:1J complexes). Understanding the reason behind these
differences is challenging because the shapes of the $H$ distributions
are a convolution of the dynamical mechanisms (such as Yarkovsky and
YORP) that replenish the NEO population and the asteroids' material
properties.

The fitted $H$ distributions also provide direct estimates for the
absolute contributions of NEOs from 7 different ERs. Our predicted
fractional contributions agree with previous estimates by
\citet{bot2002a} in the sense that the $\nu_6$ and 3:1J complexes are
the most significant ERs. Most NEOs thus originate in the inner
MAB. The outer MAB and the JFCs contribute only about 10--25\% of the
steady-state NEO population depending on $H$. The JFCs contribution
alone is about 2--10\% depending on $H$. In addition, our model shows
for the first time that the Hungaria group is an important source for
NEOs whereas the Phocaea group is a less important source. Combined
these high-inclination groups solve the controversy as to (initially)
the existence \citep{2001Sci...294.1691S} and (later) the origin of
high-inclination NEOs.

Our estimate for the number of NEOs on retrograde orbits is in
agreement with \citet{gre2012a,gre2012b}. These retrograde NEOs are
dominated by the asteroids from the 3:1J complex with a lesser
contribution from JFCs. This results in a substantially different
overall shape for the HFD compared to NEOs on prograde orbits. We also
note that these results clearly imply that the 3:1J MMR plays a key
role in the production of NEOs on retrograde orbits.

Our results for the debiased marginal $a$, $e$, $i$ and $H$
distributions for large NEOs generally agree with the most recent
literature with the exception that the inclination distribution is
weakly bimodal due to the contribution from the Hungaria group.
Although the main features of the orbit distribution are fairly stable
across the considered $H$ range due to the substantial contribution
from $\nu_6$ and 3:1J complexes across $H$, there is a clear but
complex fluctuation with $H$ on top of that. Most of the fluctuation
is explained by variation in contributions from the 5:2J complex and
the Hungarias.  In particular, the contribution from Hungarias is
largest at the smallest sizes whereas the opposite is true for the
5:2J complex.

The relative fractions of Amors, Apollos, Atens, Atiras, and Vatiras
are fairly insensitive to $H$. Our estimates for the relative
fractions for $17<H<25$ are markedly different compared to the
estimates by both \citet{gre2012a} and \citet{bot2002a} for $H<18$. A
difference with respect to both \citet{gre2012a} and \citet{bot2002a}
suggests that statistical uncertainties in the ER-specific
steady-state orbit distributions that are used for the orbit models is
not the culprit. Instead we think that the difference is driven by
improved estimates of the relative contributions from different ERs.

Based on our NEO model and \"Opik-like impact analysis we find a good
agreement to earlier estimate for the impact rate on terrestrial
planets in the literature for large NEOs. For smaller NEOs we can
compare a linear extrapolation of our model to smaller sizes with the
observed rate of bolides on the Earth. The agreement is reasonably
good apart for $24<H<28$ where the frequency of Tunguska-sized impacts
remains an unsettled issue. We also find a good agreement between our
prediction and lunar cratering records.

Finally, our work suggests that the NEO population is in a steady
state, at least for $H\geq 17$, because our model is based on that
assumption and it accurately reproduces the observed population.

\section*{Acknowlegdements}

We thank Pasquale Tricarico and Alan Harris for their detailed 
reviews that helped improve the manuscript. M.G.\ was funded by 
grants 137853 and 299543 from the Academy of
Finland, and D.V.\ by grant GA13-01308S of the Czech Science
Foundation. W.F.B.\ thanks NASA's Near Earth Object Observation
program for supporting his work in this project. We acknowledge
support by ESA via contract AO/1-7015/11/NL/LvH (SGNEOP). CSC --- IT
Centre for Science, Finland, the Finnish Grid Infrastructure and the
mesocentre SIGAMM at Observatoire de la C\^ote d'Azur provided
computational resources.



\section*{References}



\bibliographystyle{model2-names.bst}\biboptions{authoryear}

\bibliography{asteroid}

\end{document}